\documentclass[12pt,draftclsnofoot,onecolumn]{IEEEtran}

\ifodd 0

\else

\fi

\ifodd 0

\else

\fi

\ifx\UseOption\undefined
\def\UseOption{opt1} 
\fi

\usepackage[T1]{fontenc}

\immediate\write18{%
	makeindex -s nomencl.ist -o \jobname.nls -t \jobname.nlg \jobname.nlo%
}

\usepackage{nomencl}
\makenomenclature

\usepackage{etoolbox}
\renewcommand\nomgroup[1]{%
	\item[\it
	\ifstrequal{#1}{A}{Acronyms}{%
		\ifstrequal{#1}{B}{Variables}{%
			\ifstrequal{#1}{C}{Random variables}{%
				\ifstrequal{#1}{D}{Parameters/constants}{%
					\ifstrequal{#1}{P}{Symbols of payoffs}{}}}}}%
	]}

\usepackage{cite}

\ifCLASSINFOpdf
\usepackage[pdftex]{graphicx}
\else
\fi
\usepackage{subfigure}

\usepackage{amsmath,amsthm,bm,amssymb}
\newtheorem{defi}{Definition}
\newtheorem{prop}{Proposition}
\newtheorem{lemma}{Lemma}
\newtheorem{thm}{Theorem}
\newtheorem{ass}{Assumption}
\usepackage{algorithm}  
\usepackage{algorithmic}

\usepackage{array}

\usepackage[usenames]{color}
\usepackage{xspace,colortbl}
\usepackage{url}
\usepackage{caption}

\usepackage[T1]{fontenc}

\begin{document}
\title{Storage or No Storage: Duopoly Competition Between Renewable Energy Suppliers in a Local Energy Market}

\author{Dongwei~Zhao,~\IEEEmembership{Student Member,~IEEE,}
	Hao~Wang,~\IEEEmembership{Member,~IEEE,}\\
	Jianwei~Huang,~\IEEEmembership{Fellow,~IEEE,}
	and~Xiaojun~Lin,~\IEEEmembership{Fellow,~IEEE}
\thanks{This work is supported by the Presidential Fund from the Chinese University of Hong Kong, Shenzhen, China, the Shenzhen Institute of Artificial Intelligence and Robotics for Society (AIRS), and in part by the NSF award ECCS-1509536.  Part of the results have appeared in IEEE ICC 2019 \cite{Dongwei2}.}
\thanks{Dongwei Zhao is with the Department of Information Engineering, The Chinese University of Hong Kong, Hong Kong, China (e-mail: zd015@ie.cuhk.edu.hk). Hao Wang is with the Department of Civil and Environmental Engineering and the Stanford Sustainable Systems Lab, Stanford University, CA 94305 USA (e-mail: haowang6@stanford.edu). Jianwei Huang is with the School of Science and Engineering, The Chinese University of Hong Kong, Shenzhen, China, the Shenzhen Institute of Artificial Intelligence and Robotics for Society (AIRS), and the Department of Information Engineering, The Chinese University of Hong Kong, Hong Kong, China (e-mail: jianweihuang@cuhk.edu.cn).  Xiaojun Lin is with the School of Electrical and Computer Engineering, Purdue University, West Lafayette, IN 47907, USA (e-mail: linx@ecn.purdue.edu).}	}

\maketitle

\begin{abstract}
	Renewable energy generations and energy storage are playing  increasingly important roles in serving consumers in power systems. This paper studies the market competition between renewable energy suppliers with or without energy storage in a local energy market. The storage investment   brings the  benefits of  stabilizing renewable energy suppliers' outputs, but it also leads to substantial investment costs as well as some  surprising changes in the market outcome. To study the equilibrium decisions of storage investment  in the renewable energy suppliers' competition, we model the interactions between suppliers and consumers using a three-stage game-theoretic model. In Stage \uppercase\expandafter{\romannumeral1}, at the beginning of the investment horizon  (containing many days), suppliers decide  whether to invest in storage.   Once such decisions have been made (once), in the day-ahead market of each day, suppliers decide on their bidding prices and quantities in Stage \uppercase\expandafter{\romannumeral2}, based on which consumers  decide the electricity quantity purchased from each supplier in Stage \uppercase\expandafter{\romannumeral3}. In the real-time market,  a supplier is penalized if his actual generation falls short of his commitment. We characterize a price-quantity competition equilibrium of Stage \uppercase\expandafter{\romannumeral2} in the local energy market,  and we further  characterize a storage-investment equilibrium in Stage \uppercase\expandafter{\romannumeral1} incorporating electricity-selling revenue and storage cost. Counter-intuitively,  we show that the uncertainty of renewable energy without storage investment can lead to higher supplier profits  compared with the stable generations with  storage investment due to the reduced market competition  under random energy generation. Simulations further illustrate  results due to the market competition. For example, a higher penalty  for not meeting the commitment, a higher storage cost, or a   lower consumer demand can sometimes increase a supplier's  profit. We also show that although storage investment can increase a supplier 's profit, the first-mover supplier who invests in storage  may  benefit less than the  free-rider competitor who chooses not to invest.
\end{abstract}


\IEEEpeerreviewmaketitle

\printnomenclature[4em]

\nomenclature[A,01]{${\text{S}_1\text{S}_1}$}{the case where both suppliers invest in storage}
\nomenclature[A,02]{${\text{S}_0\text{S}_0}$}{the case where neither supplier invests in storage}
\nomenclature[A,03]{${\text{S}_1\text{S}_0}$}{the case where one supplier invests in storage and the other does not}
\nomenclature[B,01]{$\varphi_i$}{storage investment decision of supplier $i$}
\nomenclature[B,02]{$p_i^{d,t}$ }{bidding price of supplier $i$ at hour $t$ of day $d$}
\nomenclature[B,03]{$y_i^{d,t}$}{bidding quantity of supplier $i$ at hour $t$ of day $d$}
\nomenclature[B,04]{$x_i^{d,t}$}{electricity quantity that consumers purchase from supplier $i$ at hour $t$ of day $d$ }
\nomenclature[C,01]{$X_i^{d,t}$}{generation amount  of supplier $i$ at hour $t$ of day $d$ }
\nomenclature[C,02]{$CD_i^{d,t}$}{ charge and discharge power of supplier $i$ at hour $t$ of day $d$}
\nomenclature[D,01]{$\lambda$}{ penalty price for the supply shortage}
\nomenclature[D,02]{$\bar{p}$ }{ price cap for the bidding price}
\nomenclature[D,03]{$D^{d,t}$}{ demand of consumers at hour $t$ of day $d$}
\nomenclature[D,04]{$c_i$}{unit storage investment cost of supplier $i$ over the investment horizon}
\nomenclature[D,05]{$\kappa_i$}{scaling factor of supplier $i$}
\nomenclature[D,06]{$S_i$}{storage capacity  of supplier $i$}	
\nomenclature[D,07]{$C_i$}{storage investment cost (scaled in one hour) of supplier $i$ }
\nomenclature[P,01]{$\pi_i^{R,d,t}$}{supplier $i$'s revenue at hour $t$ of day $d$}
\nomenclature[P,02]{$\pi_i^{RE,d,t}$ }{supplier $i$'s  equilibrium revenue at hour $t$ of day $d$}
\nomenclature[P,03]{$\pi_i^{\text{S}_1\text{S}_1}$}{supplier $i$'s expected equilibrium revenue in  $\text{S}_1\text{S}_1$  case over the investment horizon}
\nomenclature[P,04]{$\pi_i^{\text{S}_0\text{S}_0}$}{supplier $i$'s expected equilibrium revenue in  $\text{S}_0\text{S}_0$ case over the investment horizon}
\nomenclature[P,05]{$\pi_i^{\text{S}_1\text{S}_0\mid \text{Y}}$}{with-storage  supplier $i$'s expected equilibrium revenue in $\text{S}_1\text{S}_0$ case over the investment horizon}
\nomenclature[P,06]{$\pi_i^{\text{S}_1\text{S}_0\mid \text{N}}$}{without-storage  supplier $i$'s expected equilibrium revenue in $\text{S}_1\text{S}_0$ case over the investment horizon}
\nomenclature[P,07]{$\Pi_i$}{supplier $i$'s profit over the investment horizon}

\section{Introduction}
\subsection{Background and motivation}

Renewable energy, as a clean and sustainable energy source,  is playing an increasingly important role in power systems \cite{renewstatus}. For example, from the year 2007 to 2017, the global installed capacity of solar panels  has increased from 8 Gigawatts to 402 Gigawatts, and the wind power capacity has increased from 94 Gigawatts to 539 Gigawatts\cite{renewstatus}. 
Compared with traditional larger-scale generators, renewable energy sources can be more spatially distributed across the power system, e.g.,  at the distribution level near residential consumers\cite{renewstatus}.   Due to the distributed nature of renewable energy generations, there has been growing interest in forming local energy markets for renewable energy suppliers and consumers to trade electricity at the distribution level  \cite{dismarket}.  Such local energy markets will allow consumers to purchase electricity from the least costly sources locally\cite{localmarketover}, and allow suppliers to compete in selling electricity directly to consumers  (instead of dealing with the utility companies).

However, many types of renewable energy are inherently random, due to factors such as weather conditions that are difficult to predict and control. Under current multi-settlement  energy market structures with day-ahead and real-time bidding rules (which are mostly designed for controllable generations)\cite{fundamentals}, renewable energy suppliers  face a severe disadvantage in the competition by making forward commitment (in the day-ahead market) that they may not be able to deliver in real time. For example, suppliers   are often subject to  a penalty cost if their real-time delivery  deviates from the commitment in the day-ahead market\cite{dt2014bid_renwable_deviation}.

Energy storage has been considered  as an important type of flexible resources for renewable energy suppliers to  stabilize their outputs\cite{overview1}. Investing in storage can potentially improve the renewable energy suppliers' position in these energy markets. However, investing in storage incurs substantial investment  costs. Furthermore, the return of storage investment depends on the outcome of the market, which in turn depends on how suppliers  with  or without storage compete for the demand. Therefore, it remains an open problem regarding \textit{whether competing renewable energy suppliers should invest in energy storage in the market competition  and what  economic benefits the storage can bring to the suppliers.}

\subsection{Main results and contributions}
In this paper, we formulate a three-stage game-theoretic model to study the market equilibrium for both storage investment as well as price and quantity bidding of  competing renewable energy suppliers.  In Stage \uppercase\expandafter{\romannumeral1}, at the beginning of the investment horizon, each supplier decides  whether to invest in storage. We formulate a storage-investment  game between  two suppliers in Stage  \uppercase\expandafter{\romannumeral1}, which is based on a bimatrix game to model suppliers' storage-investment decisions for maximizing profits\cite{mangasarian1964equilibrium}.  Given the storage-investment decisions in Stage \uppercase\expandafter{\romannumeral1},  competing suppliers decide  the bidding price and bidding quantity in the (daily) local energy market in Stage \uppercase\expandafter{\romannumeral2}. We formulate a price-quantity competition game between suppliers using the Bertrand-Edgeworth  model \cite{betrand3} (which models  price competition with capacity constraints) in Stage  \uppercase\expandafter{\romannumeral2}.  Given suppliers' bidding strategies,  consumers  decide the electricity quantity purchased  from each supplier in Stage \uppercase\expandafter{\romannumeral3}.   To the best of our knowledge, our work is the first to study the storage-investment equilibrium between competing renewable energy suppliers in the two-settlement  energy market. This problem is quite nontrivial due to the penalty cost on the  random generations of a general probability distribution.

By studying this three-stage model, we reveal a number of new and surprising insights that are against the prevailing wisdom in the literature  on the renewable energy suppliers' revenues in such  a two-settlement market  \cite{dt2014bid_renwable_deviation,bringwind} and on  the economic benefits of storage  supplementing in  renewable energy sources \cite{renewsto3,connolly2012technical}.
\begin{itemize}
	\item First, \emph{the uncertainty of the renewable generation can be favorable to suppliers}.  Note that the prevailing wisdom is that storage investment (especially when the storage cost is low) will improve suppliers' revenue by stabilizing  their outputs  \cite{renewsto3,connolly2012technical}. In contrast, we find that the opposite may be  true  when considering market competition. Specifically, without storage, suppliers with random generations  always have strictly positive revenues when facing any positive consumer  demand. However, if both suppliers invest in storage and stabilize their renewable outputs, their revenues reduce to zero once the consumer demand is below a threshold, which is due to the increased market competition after   storage investment.
	\item Second, \textit{a higher penalty and a higher  storage cost can also be favorable to the suppliers}.  Note that the common wisdom is that a higher penalty\cite{bringwind} and  a higher storage cost\cite{renewsto3} will decrease suppliers' profit. However, when considering market competition, the opposite may be true. With a higher penalty  for not meeting the commitment,  renewable energy suppliers become more conservative in their bidding quantities, which can decrease market competition and increase their profits. Furthermore, a higher storage cost  may change one supplier's storage-investment  decision, which can benefit the other supplier. 
	\item Third, \emph{the  first-mover supplier who invests in energy storage can be at the disadvantage  in terms of profit increase}, which is contrary to the first-mover advantage gained by  early  investment of resources or new technologies  \cite{grant2016contemporary}.  We find that although investing in storage can increase one supplier's profit,  it may benefit himself less than his competitor (who does not invest in storage). This is because  the  later mover becomes a free rider, who may benefit from the changed price equilibrium in the energy market (due to the storage investment of the other supplier) but does not need to bear the investment cost.
\end{itemize}

In addition to these surprising and new insights, a key technical  contribution of our work is the solution to the game-theoretic model for the price-quantity competition, which involves a general penalty cost  due to  random generations of a general probability distribution. Note that such  a    price-quantity competition  with the Bertrand-Edgeworth model has been studied in literature under quite different conditions from ours. The works in \cite{betrand1,betrand2,betrand4} studied a general competition  between suppliers with strictly convex production costs. They focused on the analysis of pure strategy equilibrium without characterizing the mixed strategy equilibrium.  The study in \cite{capacityprice} characterized both pure and mixed strategy equilibrium between suppliers with deterministic supply. However, this work considered  zero cost related to the production (i.e., no production cost or possible penalty cost).      In electricity markets, the works in   \cite{capacitypricerenew}  and \cite{bertrandrenew} also used Bertrand-Edgeworth model  to analyze the competition among renewable energy suppliers with random generations. However, both  \cite{capacitypricerenew}  and \cite{bertrandrenew} considered the suppliers' electricity-selling competition in a single-settlement energy market, and suppliers  deliver  random generations in real time. These studies did not  consider day-ahead bidding strategies and any deviation penalty cost.  In particular, the two-settlement markets with deviation penalty have been essential for ensuring  the reliable  operation of power systems.   Our work is the first to consider the two-settlement energy market, characterizing both pure and  mixed strategy equilibrium based on the Bertrand-Edgeworth model.  Such a setting is nontrivial due to the penalty cost caused by  the suppliers' random production of a general probability distribution.

The remainder  of the paper is organized as follows. First, we introduce the system model  in Section \ref{section:model}, as well as   the three-stage game-theoretic formulation  between suppliers and consumers in  Section \ref{section:stage}. Then, we solve the three-stage problem through backward induction.  We first  characterize the  consumers' optimal purchase decision of Stage \uppercase\expandafter{\romannumeral3} in Section \ref{section:stage3}.  Then, we characterize the  price-quantity equilibrium of Stage \uppercase\expandafter{\romannumeral2} and the storage-investment  equilibrium of Stage \uppercase\expandafter{\romannumeral1} in Sections \ref{section:stage2} and \ref{section:stage1}, respectively. We propose a probability-based method to compute the storage capacity in Section \ref{section:capacity}. Furthermore, in Section \ref{section:extenstion}, we extend some of the theoretical results and insights from the duopoly case  to the  oligopoly case.  Finally, we present the simulation results in Section \ref{section:sim} and conclude this paper in  Section \ref{section:con}.

\section{System Model}\label{section:model}
\begin{figure}[ht]
	\centering
	\includegraphics[width=2.6in]{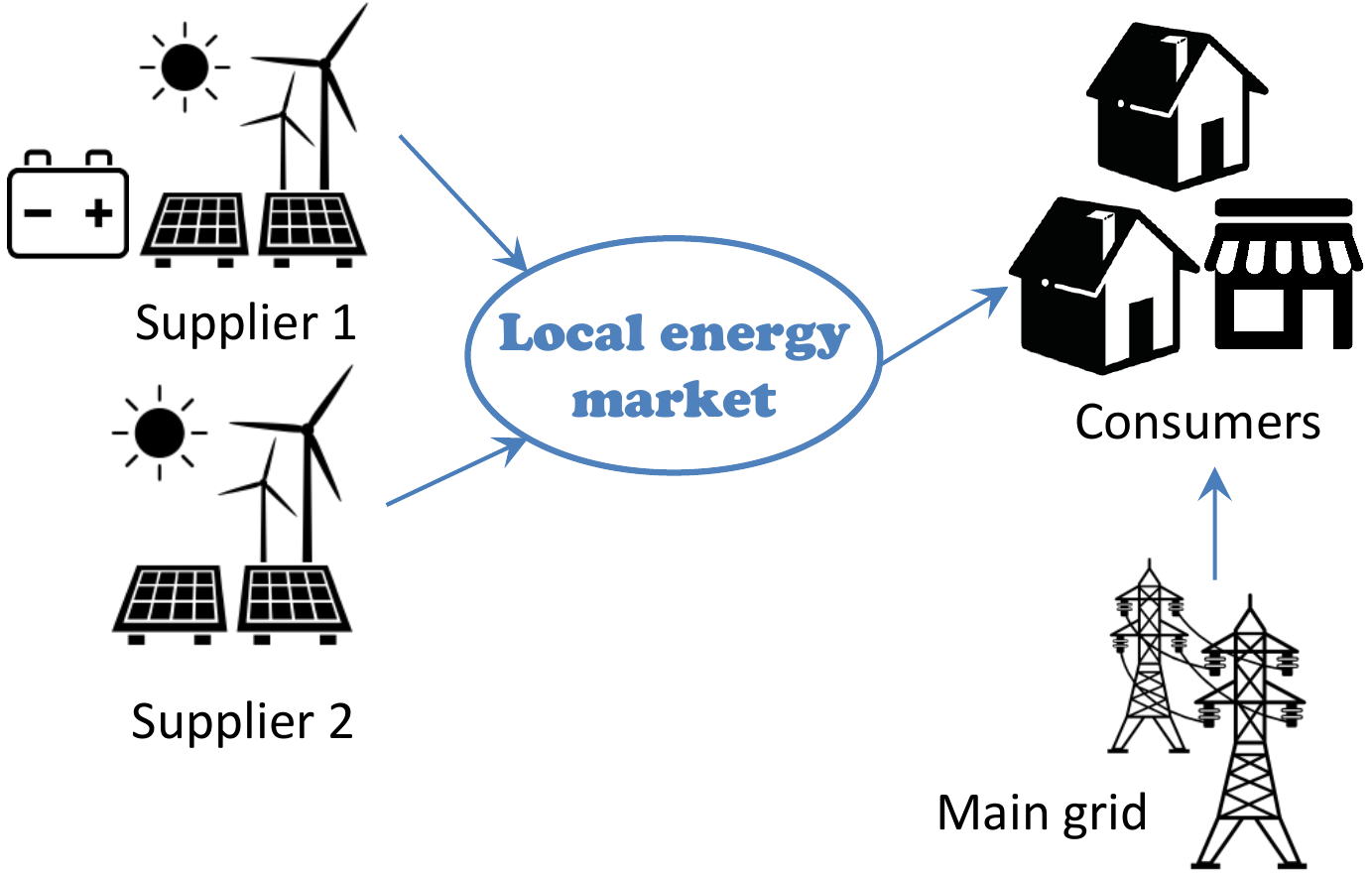}
	\vspace{-1mm}
	\caption{System structure.}
	\label{fig_sim}	
	\vspace{-3mm}
\end{figure}

We consider a  local energy market at the distribution level as shown in Figure \ref{fig_sim}.  Consumers can purchase energy  from both the main grid and  local renewable energy suppliers.  To achieve a positive revenue, the  renewable energy suppliers (simply called suppliers in the rest of the paper)  need to set their prices no greater than the grid price, and they will compete for the market share. Furthermore, suppliers can choose to invest in energy storage  to stabilize their renewable outputs and reduce the uncertainty in their delivery. Next, we will introduce the detailed models  of timescales,  suppliers and consumers, and characterize their interactions in the two-settlement local energy market.

\subsection{Timescale}
We consider two timescales of decision-making. One is  the investment horizon $\mathcal{D}\hspace{-1mm}=\hspace{-1mm}\{1,2,...,D_s\}$ of $D_s$ days (e.g., $D_s$ corresponding to the total number of  days  for the storage investment horizon).  Suppliers can decide (once) whether to invest in energy storage at the beginning of  the investment horizon. The investment horizon is divided into many  operational horizons  (many days), and each $d\in \mathcal{D}$ corresponds to the daily operation of the energy market, consisting of many time slots $\mathcal{T}\hspace{-1mm}=\hspace{-1mm}\{1,2,...,T\}$ (e.g., 24 hours of each day).  In the day-ahead market on day $d-1$, suppliers decide the electricity price and quantity to consumers for each hour $t\in \mathcal{T}$ of the next day $d\in \mathcal{D}$. We will introduce the market structure in detail later in Section \ref{section:model}.D.

\subsection{Suppliers}
In Sections \ref{section:stage3}-\ref{section:stage1}, we  focus on the duopoly case of two suppliers in our analysis. Later in Section \ref{section:extenstion}, we further generalize to the oligopoly case with more than two suppliers. The reason for focusing on the duopoly case is twofold.  First, our work focuses on a local energy market that is  much smaller than a traditional wholesale energy market. The number of suppliers serving one local area is also expected to be limited \cite{locallimit}, compared with  thousands of suppliers in the wholesale energy market \cite{pjmdata}.  In such a small local energy market, a few large suppliers may dominate the market\cite{ilas2018renewable}. Second, we consider two suppliers for analytical tractability,  which is without losing key insights and can effectively capture the impact of competition among suppliers considering the  storage investment. For example, we show that in the duopoly case, the uncertainty of renewable generation can be beneficial to suppliers. Such an insight is still valid in the oligopoly case.

We denote $\mathcal{I}=\{1,2\}$ as the set of two   suppliers. For  hour $t$ of day $d$, the  renewable output   of  supplier $i\in \mathcal {I}$ is denoted as a random variable $X_i^{d,t}$, which is bounded in  $[0,\bar{X}_i^{d,t}]$. We assume that the random generation $X_i^{d,t}$ has a  continuous cumulative distribution function (CDF) $F_i^{d,t}$ with the probability  density  function (PDF) $f_i^{d,t}$. {The distribution of wind or solar power  can be characterized using the historical data, which is known to the renewable energy suppliers.\footnote{In Section \ref{section:sim} of simulations, we use historical data to model the empirical CDF of renewable generations, which is explained in detail in Appendix.\ref{appendix:sim}. }} 
As  renewables usually have extremely low marginal production costs compared with traditional generators, we assume zero marginal production costs for the  suppliers\cite{capacitypricerenew}  \cite{bertrandrenew}.

\subsection{Consumers}
We consider the aggregate consumer population,  and we denote the total consumer demand at hour $t$ of day $d$  as $D^{d,t}>0$.  Note that consumers in one local area usually face the same electricity price from the same utility. Thus, if the local market's  electricity price is lower than  the grid price, all the consumers will first purchase electricity from local suppliers.  From the perspective of suppliers, they only care about the total demand of consumers and how much electricity they can sell to consumers. 

Furthermore,   our work conforms to the current energy market practice that suppliers make decisions in the day-ahead market based on the predicted demand. Thus, for the demand $D^{d,t}$, we consider it as a deterministic (predicted) demand  in our model.\footnote{The day-ahead prediction of consumers' aggregated demand can be fairly accurate\cite{sevlian2014loadforecast}.  We assume that the demand and supply mismatch due to the demand forecast error will be regulated by the operator in the real-time market. }
Since the electricity demand is usually inelastic  \cite{fundamentals},  we also assume the following.
\begin{ass}
	Consumers' demand is perfectly inelastic in the electricity price.
\end{ass}
\noindent Consumers must purchase their demand $D^{d,t}$ either from the main grid (at a fixed unit price  $P_g$) or from the local renewable suppliers (with prices to be discussed later).\footnote{We do not consider demand response for the consumers. }

\subsection{Two-settlement local energy market}
We consider a two-settlement local energy market, which consists of a day-ahead  market and a real-time  market\cite{fundamentals}. In such an energy market, suppliers have market power and can strategically decide their selling prices.\footnote{This price  model is  different from the usual practice of the  wholesale energy market, where the market usually sets a uniform clearing price for all the suppliers through market clearing \cite{fundamentals}.}   Consumers have the flexibility to choose  suppliers by comparing prices \cite{localmarketover}. We explain the  two-settlement  energy market in detail as follows.

\begin{itemize}
	\item In the day-ahead market on day $d-1$ (e.g., suppliers' bids are cleared around 12:30pm of day $d-1$, one day ahead of the delivery day $d$\cite{nord}), supplier $i\in \mathcal{I}$  decides the bidding price $p_i^{d,t}$ and the bidding quantity $y_i^{d,t}$ for  each future hour $t\in \mathcal{T}$ of the delivery day $d$. Based on suppliers' bidding strategies,  consumers decide the electricity quantity $x_i^{d,t}~(\leq y_i^{d,t})$ purchased from supplier $i$. Supplier $i$ will get the revenue of   $p_i^{d,t} x_i^{d,t}$ in the day-ahead market  by committing  the delivery quantity $x_i^{d,t}$ to consumers. Thus, the day-ahead market is cleared  through matching supply and demand. Any excessive demand from the consumers will be satisfied through energy purchase from the main grid.
	\item In the real-time market at each hour on the next day $d$, if supplier $i$'s actual generation falls short of the committed quantity $x_i^{d,t}$ (i.e., $x_i^{d,t}>X_i^{d,t}$), he needs to pay the penalty $\lambda( x_i^{d,t}-X_i^{d,t})$  in the real-time market, which is proportional to the shortfall with a unit penalty price $\lambda$.  For the consumers, although  suppliers may not  deliver the committed electricity to them, the shortage part can  be still satisfied by the system operator using  reserve resources. The cost of reserve resources can be covered by the penalty cost on the suppliers. 
\end{itemize}
{Note that the suppliers  and consumers  make  decisions only in the day-ahead market.  No active decisions are made in the real-time market, but there may be penalty cost on the delivery shortage.} 

To facilitate the analysis, we further make several  assumptions of this local energy market as follows. First, for the excessive amount of generations (i.e.,  $x_i^{d,t}<X_i^{d,t}$),  we assume the following.
\begin{ass}
	Suppliers can curtail any excessive renewable energy generation (beyond any specific given level).
\end{ass}	
\noindent  Assumption 2 implies that we do not need to consider the possible penalty or reward on the excessive renewable generations in real time.\footnote{There are different policies to deal with the surplus feed-in energy of renewables. In some European countries, the   energy markets give rewards to the surplus energy \cite{Chakraborty2018renew}. In the US, some  markets  deal with the surplus energy using the real-time imbalance price that can be either penalties or rewards \cite{bringwind}. }

Second,  the local energy market is much smaller compared with the wholesale energy market.  Thus, the suppliers are usually small and hence may focus on serving local consumers. It is less likely for them to trade in the wholesale energy market.  This is summarized in the following assumption. 
\begin{ass}
	Suppliers only participate in the local energy market and  serve local consumers. They do not participate in the wholesale  energy market.
\end{ass}
\noindent Third, for the bidding price  ${p_i}$ and penalty price  $\lambda$, we impose the following bounds. 
\begin{ass}\label{a:cap}
	Each supplier $i$'s bidding price $p_i$  has a cap  $\bar{p}$ that satisfies $p_i\leq \bar{p} < P_g$.
\end{ass}
\begin{ass} \label{a:penalty}
	The  penalty price satisfies $\lambda>\bar{p}$.
\end{ass}

\noindent Assumption \ref{a:cap} is without loss of generality, since no supplier will  bid a price higher than $P_g$; otherwise, consumers will purchase  from the main grid.\footnote{We avoid the case $\bar{p}= P_g$ as it may bring ambiguity  to the local energy market if the bidding price is equal to the main grid price  $P_g$, in which case it is not clear  whether consumers purchase energy from the local energy market or from the main grid.} Assumption \ref{a:penalty} ensures that the penalty is high enough to discourage suppliers from bidding higher quantities  beyond their capability.
Note that  price cap $\bar{p}$ and the penalty $\lambda$ are exogenous fixed parameters in our model. Next,  we introduce how suppliers invest in the energy storage to stabilize their outputs.

\subsection{Storage investment}
Each supplier decides whether to invest in storage at the beginning of the investment horizon. We denote supplier $i$'s storage-investment decision variable as $\varphi_i$, where $\varphi_i=1$ means investing in storage and $\varphi_i=0$ means not investing.  If supplier $i$ invests in storage, we assume the following. 
\begin{ass} \label{A:storage}
	The with-storage supplier will utilize the storage to completely smooth out his power output at the mean value of renewable generations.
\end{ass}
\noindent Thus, supplier $i$ with the renewable generation $X_i^{d,t}$ will charge and discharge his storage\footnote{There can be different ways to deal with the randomness of renewable generations,  including  the curtailment of renewable energy and the use of additional fossil generators to provide additional energy.  It is  interesting to combine energy storage with other mechanisms (such as renewable energy curtailment), which we will explore in the future work.} to stabilize the power output  at the mean value $\mathbb{ E }[X_i^{d,t}]$. The charge and discharge power $CD_i^{d,t}$ is as follows.
\vspace{-2mm}
\begin{align}
CD_i^{d,t}=X_i^{d,t}-\mathbb{ E }[X_i^{d,t}],\label{eq:chdis}
\end{align}\par \vspace{-1mm}
\noindent where  $CD_i^{d,t}>0$ means charging the storage and $CD_i^{d,t}<0$ means discharging the storage.
Note that $\mathbb{ E }_{X_i^{d,t}}[CD_i^{d,t}]=0$, which implies the long-term average power that the suppliers need  to charge or discharge his storage is zero. Next, we introduce how to characterize the storage capacity and the storage  cost.

First, based on the charge and discharge random variable $CD_i^{d,t}$,  we propose a simple yet effective  probability-based method to characterize the storage capacity $S_i$  using  historical data  of renewable generation $X_i^{d,t}$. In particular, we set a probability threshold, and then aim to find a minimum storage capacity $S_i$ such that the energy level in the storage  exceeds the  capacity with a probability no greater than the probability threshold. We will explain this methodology  in  Section \ref{section:capacity}.

Second, we calculate the storage cost of suppliers over the investment horizon (scaled into one hour) as  $C_i=c_i \kappa_i S_i$, where $c_i$ is the unit capacity cost over the investment horizon and $\kappa_i$ is the scaling factor that scales the investment cost over years to one hour. The factor $\kappa_i$ is calculated as follows. We first calculate the present value of an annuity (a series of equal annual cash flows) with the  annual interest rate $r_i$ (e.g., $r_i=5\%$), and then we divide the annuity equally to each hour. This leads to the formulation of the factor $\kappa_i$  as follows \cite{NearSitSizeSto}.
\begin{align}
\kappa_i=\frac{r_i(1+r_i)^{y_i}}{(1+r_i)^{y_i}-1}\cdot \frac{1}{Y_d},\label{eq:factor}
\end{align}
where  $y_i$ is the number of years over the investment horizon (e.g., $y_i=15$ for Li-ion battery that can last for 15 years), and $Y_d$ is the total hours in one year (e.g., $Y_d= 365\times 24$).  

Therefore, given the parameter $c_i$ and  $\kappa_i$ as well as the probability distribution of random generation,  the storage capacity and storage cost can be regarded as the fixed values for the supplier who invests in storage. Note that a higher storage capacity leads to a higher storage investment cost, which can further affect the storage-investment decisions in the suppliers' competition. Next, in the Section \ref{section:stage}, we will introduce the three-stage model between suppliers and consumers in detail.

\section{Three-stage game-theoretic model}\label{section:stage}
We build a three-stage model between suppliers and consumers. In Stage \uppercase\expandafter{\romannumeral1}, at the beginning of the investment horizon, each supplier decides whether  to invest in storage. In the day-ahead energy market, for each  hour of the next  day, suppliers decide the bidding prices and quantities in  Stage \uppercase\expandafter{\romannumeral2}, and consumers make the purchase decision in Stage \uppercase\expandafter{\romannumeral3}.  Next,  we first introduce the types of  renewable-generation distributions for computing suppliers' electricity-selling revenues over the investment horizon, and then we explain the three stages respectively in detail.
\vspace{-2mm}
\subsection{Type of renewable-generation distributions}
{We cluster the distribution of  renewable generation into several types. Note that  suppliers' revenues depend on the distribution of renewable generations. We use historical data of renewable energy to model the generation distribution. Specifically, for the renewable generations at hour $t$ of all the days over the investment horizon, we cluster the empirical distribution  into $M$ types, e.g., $M=12$ for 12 months considering the seasonal effect. In this case, each type $m\in\mathcal{M}=\{1,2,\ldots,M\}$ occurs with a probability $\rho^m=\frac{1}{12}$ considering 12 months.\footnote{There can be other types of clustering with unequal probabilities.}  We use the  data of renewable energy of all days in month $m$ at hour $t$ to approximate the distribution of renewable generation at hour $t$ for all the days in this month $m$.  Then, to study the interactions between consumers and suppliers in the local energy market,  we will assume that the renewable generation of day  $d$ follows a  random type (month) $m$, uniformly chosen from $m\in\mathcal{M}$. For  notation convenience, we replace all the superscripts $d,t$ into $m,t$.

	\subsection{The three-stage model} 
	
	We illustrate the  three-stage model  between  suppliers and  consumers  in Figure 	\ref{fig_stage}.
	\begin{itemize}
		\item Stage \uppercase\expandafter{\romannumeral1}: at the beginning of  the investment horizon, each supplier $i\in\{1,2\}$ decides the storage-investment decisions $\varphi_{i}\in\{0,1\}$.
		\item Stage \uppercase\expandafter{\romannumeral2}: in the day-ahead market, for each hour $t$ of the next day, each supplier $i$ decides his bidding price $p_i^{m,t}$ and bidding quantity $y_i^{m,t}$ based on suppliers' storage-investment decisions, assuming that the renewable-generation distribution is of month $m$.
		\item Stage \uppercase\expandafter{\romannumeral3}: in the day-ahead market, for each hour $t$ of the next day, consumers decide  the electricity quantity $x_i^{m,t}$  purchased from each supplier $i$ based on each supplier's bidding price and  quantity, assuming that the renewable-generation distribution is of month $m$.
	\end{itemize}
	
	\begin{figure}[t]
		\centering
		\includegraphics[width=2.9in]{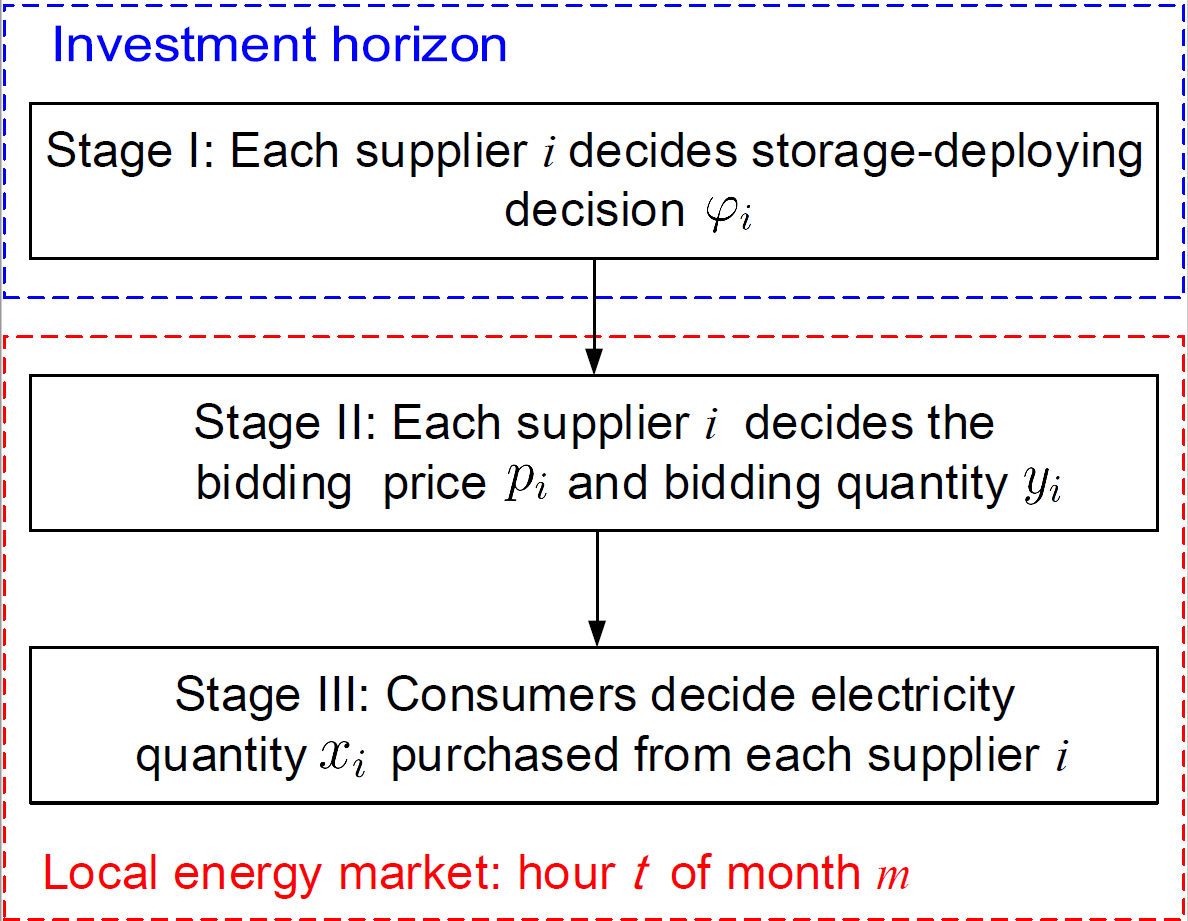}
		\caption{Three-stage model.}
		\label{fig_stage}	
	\end{figure}
	
	This three-stage problem is a dynamic game.  The solution concept of a dynamic game is known as Subgame Perfect Equilibrium, which can be derived through backward induction\cite{gamex}. Therefore, in the following, we will explain the three stages in detail in the order of Stage  \uppercase\expandafter{\romannumeral3}, Stage  \uppercase\expandafter{\romannumeral2}, and Stage  \uppercase\expandafter{\romannumeral1}, respectively.
	
	\subsubsection{Stage   \uppercase\expandafter{\romannumeral3}}
	At hour $t$ of month  $m$, given the bidding price $(p_1^{m,t}, p_2^{m,t})$ and bidding quantity $(y_1^{m,t}, y_2^{m,t})$  of both suppliers in Stage \uppercase\expandafter{\romannumeral2},   consumers decide the electricity quantity $(x_1^{m,t},x_2^{m,t})$ purchased from supplier 1 and supplier 2, respectively. The objective of  consumers is to maximize the cost saving of purchasing energy from local  suppliers compared with purchasing from the main grid only. We denote such cost saving as follows: 
	\vspace{-2mm}
	\begin{align}
	\pi_c^{m,t}(x_1^{m,t}, x_2^{m,t})=(P_g-p_1^{m,t})x_1^{m,t}+(P_g-p_2^{m,t})x_2^{m,t}.
	\end{align} \par \vspace{-1mm}
	\noindent {Recall that we model the collective purchase decision of the entire consumer population together. Consumers must satisfy their demand either from the local energy market or from the main grid (at the fixed price $P_g$). The total cost of satisfying the entire demand by the main grid is  fixed. Therefore, minimizing the total energy cost is equivalent to  maximizing the cost savings  in the local energy market.} We present consumers' {optimal purchase problem}  as follows.
	
	\noindent  \textbf{Stage \uppercase\expandafter{\romannumeral3}: Consumers' Cost Saving Maximization Problem}
	\vspace{-1mm}
	\begin{subequations}\label{eq:consumer}
		\vspace{-1mm}
		\begin{align}
		\max_{x_1^{m,t},x_2^{m,t}}~ & (P_g-p_1^{m,t})x_1^{m,t}+(P_g-p_2^{m,t})x_2^{m,t}, \label{sg2:ob}\\
		\text{s.t.} ~~&x_1^{m,t}+x_2^{m,t}\leq D^{m,t}, \label{sg2:c1}\\
		~~&0\leq x_i^{m,t} \leq y_i^{m,t},i=1,2. ~\label{sg2:c2}
		\end{align}
	\end{subequations}\par \vspace{-1mm}
	\noindent Constraint \eqref{sg2:c1} states that  the total purchased quantity $x_1^{m,t}+x_2^{m,t}$ is no greater than the  demand $D^{m,t}$. Constraints \eqref{sg2:c2} states that the quantity purchased from  supplier $i$ is no greater than his bidding quantity  $y_i^{m,t}$. This problem is  a linear programming and  can be easily solved, which we show in Section \ref{section:stage3}.  We denote the optimal solution to Problem \eqref{eq:consumer} as a function of suppliers' bidding prices and quantities  $(\bm{p}^{m,t},\bm{y}^{m,t})$, i.e., $x_i^{m,t*}(\bm{p}^{m,t},\bm{y}^{m,t}),~\forall i=1,2$, where $\bm{p}^{m,t}=(p_1^{m,t},p_2^{m,t})$ and $\bm{y}^{m,t}=(y_1^{m,t},y_2^{m,t})$.
	
	\subsubsection{Stage  \uppercase\expandafter{\romannumeral2}}
	Given the storage-investment decision $\bm{\varphi}=(\varphi_1,\varphi_2)$ in Stage \uppercase\expandafter{\romannumeral1}, both suppliers decide the bidding price $\bm{p}^{m,t}$ and bidding quantity $\bm{y}^{m,t}$ to maximize their revenues in Stage \uppercase\expandafter{\romannumeral2}. We denote supplier $i$'s electricity-selling  revenue as $\pi_i^{R,m,t}$, which consists of two parts:  the commitment revenue $p_i^{m,t}x_i^{m,t*}(\bm{p}^{m,t},\bm{y}^{m,t})$ from committing the delivery quantity  in the day-ahead market, and the penalty cost in the  real-time market. Supplier $i$ who invests in storage (i.e., $\varphi_{i}=1$) will be penalized if the committed quantity $x_i^{m,t*}(\bm{p}^{m,t},\bm{y}^{m,t})$ is larger than his stable generation $\mathbb{ E }[{X}_i^{m,t}]$.   Supplier $i$  who does not invest in storage (i.e., $\varphi_{i}=0$) will be penalized  if the  commitment $x_i^{m,t*}(\bm{p}^{m,t},\bm{y}^{m,t})$ is larger than his actual random generation ${X}_i^{m,t}$. 
	
	Note that  the decisions of two suppliers are coupled with each other. If one supplier bids a   lower  quantity or a higher price, it is highly possible that consumers will purchase more electricity from  the other supplier. We formulate a price-quantity competition game between suppliers given storage-investment decisions $\bm{\varphi}$ as follows. 
	
	\textbf{Stage  \uppercase\expandafter{\romannumeral2}: Price-quantity competition game} 
	\begin{itemize}
		\item Players: supplier $i\in\{1,2\}$.
		\item Strategies:  bidding quantity $ y_i^{m,t}\geq 0$ and bidding price $p_i^{m,t}\in [0,\bar{p}]$ of each supplier $i$.
		
		\item Payoffs:   supplier $i$'s revenue at hour $t$ of month $m$ is 
		\
		\begin{equation}
		\begin{aligned}
		&\hspace{-4mm}\pi_i^{R,m,t}\left({p}_i^{m,t},x_i^{m,t*}(\bm{p}^{m,t},\bm{y}^{m,t}),\bm{\varphi}\right)\\&\hspace{-7mm}=\left \{
		\begin{aligned}
		&\hspace{-0mm}p_i^{m,t} x_i^{m,t*}(\bm{p}^{m,t},\bm{y}^{m,t})-\lambda (x_i^{m,t*}(\bm{p}^{m,t},\bm{y}^{m,t})-\mathbb{E}[{X}_i^{m,t}])^+,\\	&\hspace{60mm}~\text{if}~\varphi_i=1;\\
		&\hspace{-0mm}p_i^{m,t} x_i^{m,t*}(\bm{p}^{m,t},\bm{y}^{m,t})-\lambda \mathbb{E}_{X_i^{m,t}}\left[(x_i^{m,t*}(\bm{p}^{m,t},\bm{y}^{m,t})- X_i^{m,t})^+\right],\\&\hspace{60mm}~\text{if}~\varphi_i=0,\\
		\end{aligned}
		\right.
		\end{aligned}\label{eq:revenue}
		\end{equation}
		\text{where we define} $(g)^+=\max (g,0).$
	\end{itemize}
	If both suppliers invest in storage (i.e., $\sum_i\varphi_i=2$), the equilibrium has been characterized in \cite{capacityprice}. However, if there is at least one supplier who does not invest in storage (i.e., $\sum_i\varphi_i\leq 1$), characterizing the equilibrium is quite non-trivial due to the penalty cost on the random generation of a general probability distribution. We will discuss how to characterize the equilibrium in detail in Section \ref{section:stage2}. We denote the equilibrium revenue of supplier $i$ as $\pi_i^{RE,m,t}(\bm{\varphi})$.
	
	\subsubsection{Stage  \uppercase\expandafter{\romannumeral1}} At the beginning of the investment horizon, each supplier decides whether to invest in  storage  to maximize his expected profit. We denote supplier $i$'s expected profit as $\Pi_i$, which incorporates the expected revenue in the local energy market and the possible storage investment cost. As one supplier varies his storage-investment decisions, it leads to a different price-quantity subgame, which will affect both suppliers' profits.  Thus, suppliers' storage-investment decisions are coupled and  we formulate a storage-investment game between suppliers as follows.
	
	\textbf{Stage  \uppercase\expandafter{\romannumeral1}: Storage-investment game} 
	\begin{itemize}
		\item Players: supplier $i\in\{1,2\}$.
		\item Strategies:  whether investing in storage $\varphi_i\in \{0,1\}$.
		\item Payoffs:  supplier $i$'s expected profit (scaled in one hour) is
		\vspace{-2mm}
		\begin{align}
		& \Pi_i\left(\bm{\varphi}\right)=\mathbb{ E }_{m,t}[\pi_i^{RE,m,t}(\bm{\varphi}) ]-\varphi_i C_i.
		\end{align}\par \vspace{-1mm}
		
	\end{itemize}
	This storage-investment game is a $2\times 2$ bimatrix game where each supplier has two strategies.  Although the Nash equilibrium of  $2\times 2$ bimatrix game can be easily solved  numerically,  the close-form equilibrium does not exist  in all  subgames of Stage \uppercase\expandafter{\romannumeral2}. It is  challenging  to analyze the storage-investment equilibrium  with respect  to the  parameters, e.g., demand and storage cost, and we discuss it in detail in Section \ref{section:stage1}.
	
	We solve this three-stage problem through backward induction. We first analyze the solution in Stage III given the bidding prices and bidding quantities in Stage II. Then, we incorporate the solution in Stage III to analyze the  price and quantity equilibrium in Stage II, given (arbitrary)  storage-investment decisions in Stage I.  Finally, we incorporate the equilibrium of Stage II into Stage I to solve the storage-investment  equilibrium. In the next three sections of Section \ref{section:stage3}, Section \ref{section:stage2}, and Section \ref{section:stage1}, we will analyze  the three stages in the order of  Stage  \uppercase\expandafter{\romannumeral3}, Stage  \uppercase\expandafter{\romannumeral2}, and Stage  \uppercase\expandafter{\romannumeral1}, respectively.

	\section{Solution of  Stage \uppercase\expandafter{\romannumeral3}}\label{section:stage3}
	In this section, we characterize consumers' optimal purchase solution to Problem \eqref{eq:consumer} in Stage \uppercase\expandafter{\romannumeral3}. We use subscript  $i\in \{1,2\} $ to denote supplier $i$ and we use $-i$ to denote the other supplier. Note that in Stage \uppercase\expandafter{\romannumeral3}, the decisions are made independently  for each hour of each day. For notation simplicity, we omit the superscript $m,t$ in the corresponding variables and  parameters.
	
	Given the bidding price $\bm{p}$ and bidding quantity $\bm{y}$  of  suppliers, we characterize in Proposition \ref{prop:stage3} consumers' optimal decision  $\boldsymbol{x}^*(\boldsymbol{p},\boldsymbol{y}) = (x_i^*(\boldsymbol{p},\boldsymbol{y}),~ i=1,2)$  in Stage \uppercase\expandafter{\romannumeral3}.   Recall that we assume that the bidding price in the local energy market is lower than the main grid price (i.e., $\bar{p}< P_g$).
	
	\begin{prop}[{optimal purchase 	$\boldsymbol{x}^*(\boldsymbol{p},\boldsymbol{y})$ in Stage \uppercase\expandafter{\romannumeral3}}]\mbox{}\label{prop:stage3}
		\begin{itemize}
			\item If $p_i<p_{-i}$  for  some $i\in\{1,2\}$, then
			${x}_i^*(\boldsymbol{p},\boldsymbol{y})= \min \left(D, y_i\right)$ and $
			{x}_{-i}^*(\boldsymbol{p},\boldsymbol{y})=\min \left(D-\min \left(D, y_i\right),y_{-i}\right).$
			
			\item {If $p_1=p_2$}, then the optimal purchase solution can be any element in the following set. 
			\vspace{-1.5mm}
			\begin{align*}
			\mathcal{X}^{opt}\hspace{-0.5mm}=\hspace{-0.5mm}\{\boldsymbol{x}^*(\boldsymbol{p},\boldsymbol{y})\hspace{-0.5mm}: \hspace{-0.5mm}\sum_{i=1}^2 x_i^*(\boldsymbol{p},\boldsymbol{y}) = \min(D, \sum_{i=1}^{2} y_i),\\
			0\leq x_i \leq y_i,~ i=1,2\}.
			\end{align*}
			We assume that the demand will be allocated to the suppliers according to the condition either $p_1<p_2$ or  $p_2<p_1$.  The condition $p_1<p_2$ or  $p_2<p_1$ is selected based on maximizing the two suppliers' total revenue.\footnote{If there is no difference between $p_1<p_2$ and  $p_2<p_1$,  the demand will be allocated by either $p_1 < p_2$ or $p_2 < p_1$ with equal probabilities.}
		\end{itemize}
	\end{prop}
	
	Proposition  \ref{prop:stage3} shows that the consumers will first purchase the electricity  from the supplier who sets a lower price. If there is remaining demand, then they will   purchase  from the other supplier. Furthermore, if consumers' demand cannot be fully satisfied  by  the local suppliers, they will purchase the remaining demand  from the main  grid. We show the proof of 	Proposition  \ref{prop:stage3} in Appendix.\ref{appendix:proofstage3}. Next we analyze the strategic bidding of suppliers in Stage \uppercase\expandafter{\romannumeral2}} by incorporating consumers' optimal purchase decisions $\bm{x}^*(\boldsymbol{p},\boldsymbol{y})$.

\section{Equilibrium analysis of  Stage \uppercase\expandafter{\romannumeral2}}\label{section:stage2}
In this section, we will characterize the bidding strategies of suppliers for the price-quantity competition subgame in Stage \uppercase\expandafter{\romannumeral2}, given the storage-investment decision in Stage \uppercase\expandafter{\romannumeral1}. Note that, depending on the storage-investment decisions in Stage  \uppercase\expandafter{\romannumeral1},  there are three types of subgames: (i) the both-investing-storage ({$\text{S}_1\text{S}_1$}) case, (ii) the mixed-investing-storage ($\text{S}_1\text{S}_0$) case, where one invests in storage and one does not, and (iii) the neither-investing-storage ($\text{S}_0\text{S}_0$) case. The competition-equilibrium characterization between suppliers is highly non-trivial, due to the general distribution of renewable generations and the penalty cost. In particular, the pure price equilibrium may not exist, which requires the characterization of the mixed price equilibrium.  Next, we first show that each supplier's equilibrium bidding quantity is actually a weakly dominant strategy that does not depend on the other supplier's decision, based on which we further derive the suppliers' bidding prices at the equilibrium for each subgame. Note that in Stage \uppercase\expandafter{\romannumeral2}, the decisions are made independently  for each hour of each day. For notation simplicity, we omit the superscript $m,t$ in the corresponding variables and  parameters.

\subsection{Weakly-dominant strategy for  bidding quantity} 
We show that given the bidding price $\boldsymbol{p}$, each supplier has a weakly dominant strategy for  the bidding  quantity that does not depend on the other supplier's quantity or price choice. This is rather surprising, and it will help  reduce the two-dimensional bidding process (involving both quantity and price) into a one-dimensional bidding process (involving only price). Deriving the weakly dominant  strategy is nontrivial due to the  penalty cost on the  renewable generation of a general probability distribution faced by  the without-storage supplier.

We first  define the weakly dominant  strategy for the bidding quantity $y_i^*$ in Definition \ref{def:quantity}, which enables a supplier to obtain a revenue
at least as high as any other bidding quantity $y_i$, no matter what is the other supplier's decision.

\begin{defi}[weakly dominant strategy]\label{def:quantity}
	Given price $\bm{p}$ and storage-investment decision $\bm{\varphi}$, a bidding quantity ${y}_i^*$ is a weakly dominant strategy for supplier $i$ if 
	\vspace{-2mm}
	\begin{align*}
	\pi_i^R(p_i, x_i^*(\bm{p},({y}_i^*,y_{-i})),\bm{\varphi})\geq  \pi_i^R(p_i, x_i^*(\bm{p},({y}_i,y_{-i})),\bm{\varphi}),
	\end{align*}\par \vspace{-2.5mm}
	\noindent \text{for any} $y_{-i}$ and  $y_i\neq y_i^*$.
\end{defi}

We then characterize suppliers' weakly dominant strategy $\bm{y}^*(\bm{p},\bm{\varphi})$  for the bidding quantity in Theorem \ref{thm:quantity}.

\begin{thm}[weakly dominant strategy for the bidding quantity]\label{thm:quantity}
	The weakly dominant strategy  $\bm{y}^*(\bm{p},\bm{\varphi})$ is given by
	\vspace{-1mm}
	\begin{equation}
	y_i^*(p_i,\varphi_i)=\left \{
	\begin{aligned}
	&\mathbb{ E }[{X}_i],~\text{if}~\varphi_i=1,\\
	&F_i^{-1}\left(\frac{p_i}{\lambda}\right),~\text{if}~\varphi_i=0,
	\end{aligned}
	\right.
	\end{equation}\par \vspace{-0.5mm}
	\noindent where $F_i^{-1}$ is the inverse function of the CDF $F_i$  of supplier $i$'s random generation.	
\end{thm}

Theorem \ref{thm:quantity} shows that a with-storage supplier $i$ (i.e., $\varphi_i=1$) should bid the quantity at the stable  production level $\mathbb{ E }[{X}_i]$ (independent of price $\boldsymbol{p}$) so that he can attract the most demand but do not face any  penalty risk in the real-time market.  For a without-storage supplier $i$ (i.e., $\varphi_i=0$), however, he has to trade off between his bidding quantity and the penalty cost  incurred by the random  generation. His weakly dominant strategy $y_i^*(p_i,\varphi_i)$  depends on his own bidding price $p_i$, but does not depend on the other supplier $-i$'s bidding price $p_{-i}$  or bidding quantity $y_{-i}$.  Note that when price  $p_i\hspace{-0.5mm}=\hspace{-0.5mm}0$, the bidding quantity $y_i^*(0,\varphi_i)\hspace{-1mm}=\hspace{-1mm}F_i^{-1}\left(0\right)\hspace{-1mm}=\hspace{-1mm}0$. Furthermore, the bidding quantity $y_i^*(p_i,\varphi_i)$  increases in price $p_i$, which shows that the without-storage supplier $i$ should bid  more quantities when he bids a higher price.  When price  $p_i\hspace{-0.7mm}=\hspace{-0.6mm}\bar{p}$, the bidding quantity satisfies $y_i^*(\bar{p},\varphi_i )\hspace{-0.6mm}=\hspace{-0.6mm}F_i^{-1}\left(\frac{\bar{p}}{\lambda}\right)\hspace{-0.6mm}<\hspace{-0.6mm}\bar{X}_i$ (i.e., the maximum generation amount) since we assume $\bar{p}\hspace{-0.5mm}<\hspace{-0.5mm}\lambda$.

\subsection{Equilibrium price-bidding strategy: pure equilibrium } 
We will further analyze the price equilibrium between suppliers based on the weakly dominant strategies for the bidding quantities in Theorem \ref{thm:quantity}. We characterize the price equilibrium with respect to the demand that can affect the competition level between suppliers. For the $\text{S}_1\text{S}_0$ and $\text{S}_0\text{S}_0$ cases, we show that a pure price equilibrium exists when the
demand $D$ is higher than a  threshold (characterized in the later analysis). However, when the demand $D$ is lower than the threshold, there exists no pure price  equilibrium due to the competition for the limited demand. For the $\text{S}_1\text{S}_1$ case, the equilibrium structure is characterized by two thresholds of the demand  (characterized in the later analysis). A pure price equilibrium will exist when the
demand $D$ is higher than the larger  threshold  or lower than the other smaller threshold. However, when the demand $D$ is in the middle of the two thresholds, there exists no pure price equilibrium.

We first define the pure price equilibrium of suppliers in Definition  \ref{def:pureprice},  where no supplier can increase his revenue through unilateral price deviation.  

\begin{defi}[pure price equilibrium]\label{def:pureprice}
	Given the storage-investment decision $\bm{\varphi}$, a price vector $\bm{p}^*$ is a pure price equilibrium if for both $i=1,2$,
	\vspace{-2mm}
	\begin{align}
	&\pi_i^R\left(p_i^*, x_i^*(\bm{p}^*,\bm{y}^*(\bm{p}^*,\bm{\varphi})),\bm{\varphi}\right)\notag\\&~~~~~~~~~~~~~~\geq \pi_i^R\left(p_i, x_i^*\left((p_i,p_{-i}^*),\bm{y}^*((p_i,p_{-i}^*),\varphi_{i})\right),\bm{\varphi}\right),
	\end{align}\par\vspace{-2mm}
	\noindent \text{for all} ~$0\leq p_i\leq \bar{p}$, where $\bm{y}^*$ is the weakly dominant strategies derived in Theorem \ref{thm:quantity}. 
\end{defi}

Then, we show the existence of the pure
price equilibrium  in Proposition \ref{prop:pureprice}.
\begin{prop}[existence of the pure price equilibrium]\mbox{}\label{prop:pureprice}
	\begin{itemize}
		\item Subgames of type $\text{S}_1\text{S}_0$ and type $\text{S}_0\text{S}_0$ (i.e., when $\sum_i\varphi_i<2$):		
		\begin{itemize}
			\item If  $D \geq \sum_i y_i^*(\bar{p},\varphi_i)$, there exists a pure  price equilibrium $p_i^*=\bar{p}$, with equilibrium revenue $\pi_i^{RE}=\lambda \int_{0}^{F_i^{-1}(\bar{p}/\lambda)}xf_i(x)dx$, for any $i=1,2$.
			\item If  $0<D<\sum_i y_i^*(\bar{p},\varphi_i)$,
			there is no  pure price  equilibrium.
		\end{itemize}
		\item Subgame of type $\text{S}_1\text{S}_1$  ($\sum_i\varphi_i=2$):
		\begin{itemize}
			\item If  $D \geq \sum_i y_i^*(\bar{p},\varphi_i)$, there exists a pure  price equilibrium $p_i^*=\bar{p}$,  with equilibrium revenue $\pi_i^{RE}=\bar{p}\mathbb{ E }[X_i]$, for any $i=1,2$.
			\item If  $D\leq \min_i y_i^*(\bar{p},\varphi_i)$, there exists a pure  price equilibrium $p_i^*=0$,  with equilibrium revenue $\pi_i^{RE}=0$, for any $i=1,2$.
			\item  If  $\min_i y_i^*(\bar{p},\varphi_i)<D< \sum_i y_i^*(p_i,\varphi_i)$,    there is no  pure price  equilibrium.
		\end{itemize}
	\end{itemize}
\end{prop}

We summarize  the existence of  pure price equilibrium and the weakly dominant strategy of bidding quantity in Table \ref{table:price}. 
\begin{table*}[ht] \normalsize
	\centering
	\renewcommand\arraystretch{1.1}
	\begin{tabular}{|p{1.8cm}<{\centering}|p{3cm}<{\centering}|p{5.5cm}<{\centering}|p{5cm}<{\centering}|}
		\hline
		Subgame & Weakly dominant strategy of bidding quantity & Existence of pure price equilibrium  &  Non-existence of pure price equilibrium\\
		\hline
		$\text{S}_1\text{S}_1$  & $y_i^*(p_i,\varphi_i)$, $\forall i=1,2$ & (a) $D \geq \sum_i y_i^*(\bar{p},\varphi_i)$:  $p_i^*=\bar{p}$,  $\forall i=1,2$ (b) $D\leq \min_i y_i^*(\bar{p},\varphi_i)$:  $p_i^*=0$,  $\forall i=1,2$  &
		$\min_i y_i^*(\bar{p},\varphi_i)<D< \sum_i y_i^*(\bar{p}_i,\varphi_i)$:   no  pure price  equilibrium\\
		\hline
		$\text{S}_1\text{S}_0$, $\text{S}_0\text{S}_0$   & $y_i^*(p_i,\varphi_i)$, $\forall i=1,2$  & $D \geq \sum_i y_i^*(\bar{p},\varphi_i)$:  $p_i^*=\bar{p}$,  $\forall i=1,2$.
		& $0<D<\sum_i y_i^*(\bar{p},\varphi_i)$:  no  pure price  equilibrium\\
		\hline
	\end{tabular}
	\caption{Weakly dominant strategy of bidding quantity as well as the conditions for the existence of pure price equilibrium.}
	\label{table:price}
	\vspace{-1mm}
\end{table*}

According to Proposition \ref{prop:pureprice},  for all the types of subgames, when the demand $D$ is higher than the summation of the suppliers' maximum  bidding quantities   (i.e., $D\geq \sum_iy_i^*(\bar{p},\varphi_i)$), both suppliers will bid the highest price $\bar{p}$. The reason is that both suppliers' bidding quantities will be fully sold out in this case, and the highest price will give the highest revenue to each supplier. Basically there is no impact of market competition in this case.  However, for the $\text{S}_1\text{S}_0$ and $\text{S}_0\text{S}_0$ subgames, if the demand $D$ is lower than the threshold $\sum_i y_i^*(\bar{p},\varphi_i)$,  there exists no pure price equilibrium. In contrast, for the $\text{S}_1\text{S}_1$ subgame, it is also  possible that when the demand $D$ is smaller than a threshold (i.e., $D<\min_i y_i^*(\bar{p},\varphi_i)$), both suppliers have to bid zero price and get zero revenue. The intuition is that the competition level of the $\text{S}_1\text{S}_1$ subgame is higher than that of the $\text{S}_1\text{S}_0$  and $\text{S}_0\text{S}_0$ subgames due to both suppliers' stable outputs, which leads to zero bidding prices  if the demand is low. The result of the  subagame $\text{S}_1\text{S}_1$ has been proved in \cite{capacityprice}. We present the proofs of subgames of type $\text{S}_1\text{S}_0$ and  type $\text{S}_0\text{S}_0$ in Appendix.\ref{appendix:proofstage2}.

\subsection{Equilibrium price-bidding strategy: mixed  equilibrium }

When the demand is such a level that there is no pure price equilibrium as shown in Proposition \ref{prop:pureprice}, we characterize the mixed price equilibrium between suppliers. 

First, we define the mixed price equilibrium  under the weakly dominant strategy $\bm{y}^*(\bm{p},\bm{\varphi})$ in Definition \ref{def:mix}, where $\mu$  denotes a probability measure\footnote{A probability measure  is  a real-valued function  that assigns a probability to each event in a probability space.} of the price over $[0,\bar{p}]$\cite{capacityprice}.

\begin{defi}[mixed price equilibrium]\label{def:mix}
	A vector of probability measures $(\mu_1^*, \mu_2^*)$ is a mixed price equilibrium if, for both $i=1,2$, 
	\vspace{-2mm}
	\begin{align*}
	&\int_{{[0,\bar{p}]}^{2}}\hspace{-0.6mm}\pi_i^R \left(p_i,  {x}_i^*\left((p_i, {p}_{-i}),\bm{y}^*(p_i, p_{-i})),\bm{\varphi}\right) d (\mu_i^{*}(p_i) \hspace{-0.6mm}\times\hspace{-0.6mm} {\mu}_{-i}^{*}({p}_{-i}) \right)\\
	\geq &\int_{{[0,\bar{p}]}^{2}}\hspace{-0.6mm}	\pi_i^R \left(p_i,  {x}_i^*((p_i, {p}_{-i}),\bm{y}^*(p_i, p_{-i})),\bm{\varphi}) d (\mu_i(p_i) \hspace{-0.6mm}\times\hspace{-0.6mm} {\mu}_{-i}^{*}({p}_{-i}) \right),
	\end{align*}
	for any measure  $\mu_i$.
\end{defi}

Definition \ref{def:mix} states that the expected revenue of supplier $i$  cannot be increased if he  unilaterally deviates from the mixed equilibrium price strategy  $\mu_i^{*}$. Let $F_i^e$ denote the CDF of $\mu_i^*$, i.e., $F_i^e(p_i)=\mu_i^*(\{p\leq p_i\})$. Let $u_i$ and $l_i$ denote the upper support and lower support of the mixed price equilibrium $\mu_i^*$, respectively, i.e., $u_i\hspace{-0.5mm}=\hspace{-0.5mm}\inf\{{p}_i\hspace{-0.5mm}: \hspace{-0.5mm}F_i^e(p_i)=1\}$ and $l_i\hspace{-0.5mm}=\hspace{-0.5mm}\sup\{{p}_i\hspace{-0.5mm}:\hspace{-0.5mm} F_i^e(p_i)=0\}$.    To characterize the mixed price equilibrium, we need to fully characterize the CDF function $F_i^e$ (including $u_i$ and $l_i$) for each $i \in\{1,2\}$. 

Then,  we show that the mixed price equilibrium exists for each type of subgames and characterize some properties of the mixed price equilibrium in Lemma 1.  Lemma 1 can be derived following the same method for the $\text{S}_1\text{S}_1$ case in \cite{capacityprice}. Later, we discuss how to compute the mixed price equilibrium of the $\text{S}_1\text{S}_1$, $\text{S}_1\text{S}_0$, and $\text{S}_0\text{S}_0$ cases, respectively.

\begin{lemma}[characterization of  the mixed price equilibrium]\label{lem:mix}
	For any pair $(\varphi_i,\varphi_{-i} )$, when the demand $D$ falls in the range where  no pure price equilibrium exists as shown in Proposition \ref{prop:pureprice}, the  mixed price equilibrium exists and    has properties as follows.
	
	(i) Both suppliers have the same lower support and the same upper support:
	\vspace{-2mm}
	\begin{align}
	&l_1=l_2=l>0,~u_1=u_2= \bar{p}.\label{eq:4b}
	\end{align}\par\vspace{-2mm}
	
	(ii) The equilibrium electricity-selling revenues $\pi_i^{RE}$ satisfy: 
	\begin{align}
	&\pi_i^{RE}(\bm{\varphi})=\pi_i^R(l,\min(D,y_2^*(l,\varphi_{i})),\bm{\varphi}).\label{eq:5b}
	\end{align}\par\vspace{-1mm}
	
	(iii) For any $i=1,2$, $F_i ^e$ is strictly increasing over $[l,\bar{p}]$, and  has no atoms\footnote{The atom at $p$ means that the left-limit of CDF at $p$ satisfies $F_i^e(p^-)\triangleq \lim_{p'\uparrow p}F_i^e(p')<F_i^e(p)$.} over $[l, \bar{p})$. Also, $F_i ^e$  cannot have  atoms at $\bar{p}$  for both $i=1,2$. 
\end{lemma}

Lemma \ref{lem:mix} shows that both suppliers' mixed-price-equilibrium strategies have the same support and have  continuous CDFs over $[l,\bar{p})$. Based on Lemma \ref{lem:mix}, we next characterize the mixed price equilibrium for the subgames of each type $\text{S}_1\text{S}_1$, $\text{S}_1\text{S}_0$, and $\text{S}_0\text{S}_0$.

\subsubsection{$\text{S}_1\text{S}_1$ subgame (i.e., $\sum \varphi_i=2$)} As shown in Proposition \ref{prop:pureprice}, when the demand satisfies $\min_i y_i^*(\bar{p},\varphi_i)<D< \sum_i y_i^*(\bar{p},\varphi_i)$, there is no pure price equilibrium.  We can characterize a close-form equilibrium revenue for each supplier at the mixed price equilibrium, which has been proved in  \cite{capacityprice}. Furthermore, under the mixed price equilibrium, both suppliers get strictly positive revenues, while they may get zero revenues under the pure price equilibrium  as shown in Proposition \ref{prop:pureprice}.  We show the close-form equilibrium revenue in Appendix.\ref{appendix:s1s1}.

\subsubsection{$\text{S}_1\text{S}_0$ subgame (i.e., $\sum_i\varphi_i=1$)} In the $\text{S}_1\text{S}_0$ subgame,  a mixed price equilibrium arises when $0<D<\sum_i y_i^*(\bar{p},\varphi_i)$. However,   we cannot characterize a close-form equilibrium revenue, as in the $\text{S}_1\text{S}_1$ case due to the penalty cost on the general renewable generations for the without-storage supplier. Instead, 
we can first characterize the CDF of the mixed price equilibrium assuming the lower support $l$ in Theorem \ref{thm:mscdf}, and then show how to compute the lower support $l$ in Proposition \ref{thm:mscomp}. We present the proofs in Appendix.\ref{appendix:proofstage2}.

\vspace{-1mm}
\begin{thm}[$\text{S}_1\text{S}_0$: CDF of  the mixed price equilibrium]\label{thm:mscdf}
	In the $\text{S}_1\text{S}_0$ subgame (i.e., $\sum_i\varphi_1=1$), when $0<D<\sum_i y_i^*(\bar{p},\varphi_i)$, suppose that the common lower support $l_1=l_2=l$ of the mixed price equilibrium is known. Then, the suppliers' mixed equilibrium  price strategies  are characterized by the following CDF $F_i^e$:
	\begin{itemize}
		\item If $\varphi_i=1$, we have
		\begin{align}
		&\hspace{-9mm}F_i^e(p)= \frac{ \pi_{-i}^R\left(p,\min\left(y_{-i}^*(p,\varphi_{-i}),D\right),\bm{\varphi}\right)-\pi_{-i}^{RE}(\bm{\varphi})}{\pi_{-i}^R(p,\min\left(y_{-i}^*(p,\varphi_{-i}),D\right),\bm{\varphi})-\pi_{-i}^R(p, (D-\mathbb{ E }[X_i])^+,\bm{\varphi})}.\label{F1}
		\end{align}
		\item If $\varphi_i=0$, we have
		\begin{align}
		&\hspace{-3mm}F_i^e(p)=\int_l^{\bar{p}} \frac{\pi_{-i}^{RE}(\bm{\varphi})}{p^2\cdot \min\left(y_i^*(p,\varphi_{i}),D\right)-p^2\cdot (D-\mathbb{ E }[X_{-i}])^+}dp.\label{F2}
		\end{align}
		for any $l \leq p< \bar{p}$. 
		
	\end{itemize}	
\end{thm}

As shown in Theorem \ref{thm:mscdf}, supplier $i$'s mixed strategy $F_i^e$ is coupled with the other supplier's equilibrium revenue $\pi_{-i}^{RE}$. Next, we will explain how to compute the lower support $l$.   Toward this end, in \eqref{F1} and \eqref{F2}, we replace the equilibrium lower support  $l$ by a variable $l_i^\dagger$, and replace $F_i^e({p})$  by $F_i^e({p}\mid l_i^\dagger)$ to emphasize that  $F_i^e(p\mid l_i^\dagger)$ is  a function of $l_i^\dagger$.  Lemma \ref{lem:mix} (iii) implies that there exists a solution  $l_i^\dagger$ to the equation $F_i^e(\bar{p}^-\mid l_i^\dagger)=1$ for at least one of the suppliers. Furthermore, we can prove  that $F_i^e(\bar{p}^-\mid l_i^\dagger)$ decreases in $l_i^\dagger$, and hence the  solution (in $l_i^\dagger$) to $F_i^e(\bar{p}^-\mid l_i^\dagger)=1$ is unique. 
Then, we can compute the lower support $l$ in Proposition \ref{thm:mscomp}.

\begin{prop}[{$\text{S}_1\text{S}_0$: computing the lower support $l$}]\label{thm:mscomp}
	Based on the solution $l_i^\dagger$ such that  $F_i^e(\bar{p}^-\mid l_i^\dagger)=1$, $ \forall i=1,2$,    we consider two cases and  compute the lower support $l$ as follows.
	\begin{enumerate}
		\item If  $F_i^e(\bar{p}^-\mid l_i^\dagger)=1$ has a solution  $l_i^\dagger$ for both suppliers, then the equilibrium lower support is $l=\max_i (l_i^\dagger)$.
		\item If  $F_i^e(\bar{p}^-\mid l_i^\dagger)=1$ has a solution  $l_i^\dagger$ for only one supplier $i$, we have this unique solution $l_i^\dagger$ as the equilibrium lower support $l$.
	\end{enumerate}	
\end{prop}

Through Theorem \ref{thm:mscdf} and  Proposition \ref{thm:mscomp}, we can compute the lower support and suppliers' equilibrium revenues. Although we cannot obtain a close-form equilibrium revenue,  in Theorem \ref{prop:comparison}, we can show that in the $\text{S}_1\text{S}_0$ subgame, if two suppliers' random generations have the same mean value, then the with-storage supplier's equilibrium revenue is always strictly  higher than that of the without-storage supplier.

\begin{thm}[$\text{S}_1\text{S}_0$: revenue  comparison]\label{prop:comparison}
	If  $\varphi_i=1$, $\varphi_{-i}=0$ and  $\mathbb{E}[X_i]=\mathbb{E}{[X_{-i}]}$,  then
	$\pi_i^{RE}(\bm{\varphi})>\pi_{-i}^{RE}(\bm{\varphi})$  for both pure and mixed price equilibrium.
	Particularly, if $X_{-i}$  follows a uniform distribution  over $[0,\bar{X}_{-i}]$, we have 	
	\begin{equation}
	\frac{\pi_i^{RE}(\bm{\varphi})}{\pi_{-i}^{RE}(\bm{\varphi})}\geq \left \{
	\begin{aligned}
	&2,~\text{if}~0<D \leq \mathbb{E}[X_i],\\
	&4,~\text{if}~D = \mathbb{E}[X_i],\\
	&\frac{\lambda}{\bar{p}},~\text{if}~D > \mathbb{E}[X_i].\\
	\end{aligned}
	\right.
	\end{equation}	
	
\end{thm}

Theorem \ref{prop:comparison} shows the dominance of the with-storage supplier in the $\text{S}_1\text{S}_0$ subgame, whose electricity-selling revenue can be much higher than that of the without-storage  supplier. The intuition  is that the random generation  makes  the without-storage  supplier at the disadvantage in the market  (due to the penalty cost). This suggests potential economic benefits of storage investment for the supplier.\footnote{Note that Theorem 3 only compares the revenue of the two suppliers. When considering the storage investment cost in Stage \uppercase\expandafter{\romannumeral1} and comparing the suppliers' profit, we will have some surprising results shown in Section \ref{section:stage1} and Section \ref{section:sim}.}  However, investing in storage does not always bring benefits. If both suppliers invest in storage, it may reduce both suppliers' revenues compared with the case that at least one supplier does not invest in storage. We will discuss it later in Proposition \ref{prop:positiverev}.

\subsubsection{$\text{S}_0\text{S}_0$ subgame (i.e., $\sum_i\varphi_i=0$)} In the $\text{S}_0\text{S}_0$ case, both suppliers do not invest in storage and face the  penalty cost.  When $0<D<\sum_i y_i^*(\bar{p},\varphi_i)$, for the mixed price equilibrium, we can neither obtain the close-form equilibrium revenue as in the $\text{S}_1\text{S}_1$ case nor obtain the equilibrium strategy CDF as in Theorem \ref{thm:mscdf} of the $\text{S}_1\text{S}_0$ case. Note  that in the   $\text{S}_1\text{S}_1$ and $\text{S}_1\text{S}_0$ subgames, at least one supplier is not subject to the penalty cost, which makes it possible to characterize the equilibrium strategy CDF or even close-form equilibrium revenue. In this $\text{S}_0\text{S}_0$ subgame, we will characterize a range of the lower support $l$ in Proposition \ref{prop:ns}. 

\begin{prop}[$\text{S}_0\text{S}_0$: lower support]\label{prop:ns}
	In the $\text{S}_0\text{S}_0$ subgame (i.e., $\sum_i\varphi_1=0$), when $0<D<\sum_i y_i^*(\bar{p},\varphi_i)$, the lower support $l$ of the mixed price equilibrium satisfies
	\vspace{-1mm} 
	\begin{align}
	\min_i ~y_i^*(l,\varphi_i)<D\leq \sum_i y_i^*(l,\varphi_i) ~\text{and} ~l <\bar{p}. \label{eq:nslower}
	\end{align}
\end{prop}
The bidding quantity $y_i^*(l,\varphi_i)$ is the minimal bidding quantity of supplier $i$ when he uses the mixed price strategy. Proposition \ref{prop:ns} shows that this minimal bidding quantity cannot be too lower or too higher for both suppliers. 

Note that the   mixed price equilibrium  has a continuous CDF over $[l,\bar{p})$ shown in Lemma \ref{lem:mix}, but we cannot derive it in close form. To have a better understanding of the CDF,   we discretize the price  to approximate the original continuous price set, and  compute the mixed equilibrium for the discrete price set.  The details are shown in Appendix.\ref{appendix:s0s0}.

\subsection{Strictly positive revenue in  the $\text{S}_1\text{S}_0$ and $\text{S}_0\text{S}_0$ subgames}
Analyzing the equilibrium revenues of the three types of subgames, we show in  Proposition \ref{prop:positiverev} that in the $\text{S}_1\text{S}_0$ and $\text{S}_0\text{S}_0$ subgames, both suppliers always get strictly positive revenues. 

\begin{prop}[strictly positive revenue with randomness]\label{prop:positiverev}
	In the $\text{S}_1\text{S}_0$ and $\text{S}_0\text{S}_0$ subgames, 
	each supplier $i$ always gets strictly positive revenue at  (both pure and mixed) equilibrium, i.e., $\pi_i^{RE}>0$.
\end{prop}

This result is counter-intuitive for the following reason. Recall that in the $\text{S}_1\text{S}_1$ subgame, both suppliers can get zero revenue if the demand is below a threshold as shown in Proposition 2. The common wisdom is that when the generation is random, the revenues of suppliers tend to be low due to the penalty cost. In contrast, Proposition \ref{prop:positiverev} shows that the suppliers' revenues are always strictly positive when the generation is random. Thus, the randomness can in fact be beneficial. The underlying reason  should be understood from the point of view  of  market competition.  The randomness  makes  suppliers bid more conservatively  in  their bidding quantities, which leads to less-fierce market competition and thus increases their revenues.

\section{Equilibrium analysis of  Stage \uppercase\expandafter{\romannumeral1}} \label{section:stage1}
In Stage  \uppercase\expandafter{\romannumeral1}, each supplier $i$ has two strategies: (i) investing in storage, i.e., $\varphi_i=1$, and (ii) not investing storage, i.e.,  $\varphi_i=0$, which leads to a  bimatrix game.   For this bimatrix game, we can analyze the equilibrium strategy by simply comparing the  profits for each strategy pair of the two suppliers.  Note that while the electricity-selling revenue is given in the results of Section \ref{section:stage2}, the profit also depends on the storage cost.  To calculate the storage investment cost, we also propose a probability-based method using real data to characterize the storage capacity for the with-storage supplier in Section \ref{section:capacity}.

Each supplier's profit can be calculated by taking the  expectation of the equilibrium revenue in the local energy market at each hour, and  subtracting storage investment cost over the investment horizon (scaled into one hour). Note that suppliers' storage-investment strategy pairs $\bm{\varphi}=(\varphi_1,\varphi_2)$  lead to four possible subgames: $\text{S}_1\text{S}_1$ subgame (i.e.  $\sum_i\varphi_i=2$), $\text{S}_1\text{S}_0$ subgame (i.e., $\sum_i\varphi_i=1$, including two cases: $(\varphi_1,\varphi_2)=(1,0)$ and $(\varphi_1,\varphi_2)=(0,1)$), and $\text{S}_0\text{S}_0$ subgame (i.e.  $\sum_i\varphi_i=0$).  
Taking the expectation of equilibrium revenue over all the hours in the investment horizon,  we denote supplier $i$'s equilibrium revenue in the $\text{S}_1\text{S}_1$ and $\text{S}_0\text{S}_0$ subgames as $\pi_i^{\text{S}_1\text{S}_1}$ and $\pi_i^{\text{S}_0\text{S}_0}$, respectively. For the $\text{S}_1\text{S}_0$ subgame, we denote the with-storage and without-storage supplier $i$'s equilibrium revenue as $\pi_i^{\text{S}_1\text{S}_0|\text{Y}}$  and $\pi_i^{\text{S}_1\text{S}_0|\text{N}}$, respectively.   For illustration,  we list the profit table with all four strategy pairs in Table \ref{tab:profit}. 

\begin{table*}[ht] \normalsize
	\centering
	\renewcommand\arraystretch{1.1}
	
	\begin{tabular}{|p{4cm}<{\centering}|p{4cm}<{\centering}|p{4cm}<{\centering}|}	
		\hline
		& Supplier 2: invest &
		Supplier 2: not invest\\
		\hline
		Supplier 1: invest&	$(\pi_1^\text{$\text{S}_1\text{S}_1$}-C_1, \pi_2^{\text{S}_1\text{S}_1}-C_2)$&$(\pi_1^{\text{S}_1\text{S}_0|\text{Y}}-C_1, \pi_2^{\text{S}_1\text{S}_0|\text{N}})$\\
		\hline
		Supplier 1: not invest&		$(\pi_1^{\text{S}_1\text{S}_0|\text{N}}, \pi_2^{\text{S}_1\text{S}_0|\text{Y}}-C_2)$&$(\pi_1^{\text{S}_0\text{S}_0}, \pi_2^{\text{S}_0\text{S}_0})$\\
		\hline
	\end{tabular}
	\vspace{-1mm}
	\caption{Supplier's profits under different  $\bm{\varphi}$.}
	\label{tab:profit}
\end{table*}

Next, we will first derive the conditions for each storage-investment strategy pair to be an equilibrium, respectively. Then, we analyze the equilibrium with respect to the parameters of storage cost and demand. Finally, we show that both suppliers can get strictly positive profits in this storage-investment game.

\subsection{Conditions of pure storage-investment equilibrium} We will characterize the conditions on the storage cost and the subgame equilibrium revenue for each strategy pair to become  an equilibrium, respectively. 

First, we define the pure storage-investment equilibrium in Definition \ref{defi:stoeq}, which states that neither supplier has an incentive to  deviate from his storage-investment decision  at the equilibrium.

\begin{defi}[pure storage-investment equilibrium]\label{defi:stoeq}
	A storage-investment vector $\bm{\varphi}^*$ is a pure storage-investment equilibrium if the profit satisfies 
	$\Pi_i\left(\varphi_i^*,\varphi_{-i}^* \right)\geq   ~\Pi_i\left(\varphi_i,\varphi_{-i}^* \right),$
	\text{for any} ~$\varphi_{i}\neq \varphi_i^*$, and any $i=1,2$.
\end{defi}

Based on Definition \ref{defi:stoeq}, we characterize the conditions on the  storage cost and the subgame equilibrium revenue  for the storage-investment pure equilibrium  in Theorem \ref{thm:stoeq}, the proof of which is presented in Appendix.\ref{appendix:proofstage1}.

\begin{thm}[conditions of pure storage-investment equilibrium]\mbox{}\label{thm:stoeq}
	\begin{itemize}
		
		\item $\text{S}_0\text{S}_0$ case is an equilibrium if  $C_i\in [\pi_i^{\text{S}_1\text{S}_0|\text{Y}}-\pi_i^{\text{S}_0\text{S}_0},+\infty)$, for both $i=1,2$.
		\item $\text{S}_1\text{S}_0$ case is an equilibrium (where $\varphi_i=1$ and $\varphi_{-i}=0$ ) if  $C_i\in [0, \pi_i^{\text{S}_1\text{S}_0|\text{Y}}-\pi_i^{\text{S}_0\text{S}_0}]$ and $C_{-i}\in [\pi_{-i}^{\text{S}_1\text{S}_1}-\pi_{-i}^{\text{S}_1\text{S}_0|\text{N}},+\infty)$.
		\item $\text{S}_1\text{S}_1$ case is an  equilibrium if  $C_i\in [0, \pi_i^{\text{S}_1\text{S}_1}-\pi_i^{\text{S}_1\text{S}_0|\text{N}}]$, for both $i=1,2$.
	\end{itemize}
	
	If $C_i$ satisfies none of  the conditions above, there exists no pure storage-investment equilibrium.\footnote{Note that if  $\pi_i^{\text{S}_1\text{S}_0|\text{Y}}-\pi_i^{\text{S}_0\text{S}_0}<0$ or $\pi_i^{\text{S}_1\text{S}_1}-\pi_i^{\text{S}_1\text{S}_0|\text{N}}<0$, then the set $ [0, \pi_i^{\text{S}_1\text{S}_0|\text{Y}}-\pi_i^{\text{S}_0\text{S}_0}]=\emptyset$ or  $[0, \pi_i^{\text{S}_1\text{S}_1}-\pi_i^{\text{S}_1\text{S}_0|\text{N}}]=\emptyset$. This means that the condition $C_i\in [0, \pi_i^{\text{S}_1\text{S}_0|\text{Y}}-\pi_i^{\text{S}_0\text{S}_0}]$ or $C_i\in [0, \pi_i^{\text{S}_1\text{S}_1}-\pi_i^{\text{S}_1\text{S}_0|\text{N}}]$ cannot be satisfied.  }
\end{thm}

Theorem \ref{thm:stoeq} shows that  the storage-investment equilibrium depends on the comparison between the storage cost and the revenue difference between the cases $\text{S}_1\text{S}_0$ and $\text{S}_1\text{S}_0$, or the cases $\text{S}_1\text{S}_0$ and $\text{S}_1\text{S}_1$. Also, Theorem \ref{thm:stoeq}  implies that a lower storage cost will incentivize  the supplier to invest in storage. 

According to Theorem \ref{thm:stoeq}, given the storage cost and  the expected equilibrium revenue of each subgame, we can characterize the pure equilibrium for nearly all values of $C_i$. However, if storage cost $C_i$  satisfies none of the conditions in  Theorem \ref{thm:stoeq}, there will be no pure price equilibrium. Note that when there is no pure storage-investment equilibrium, we can always characterize the mixed equilibrium as the game in Stage \uppercase\expandafter{\romannumeral1} is a finite game\cite{gamex}. We show how to compute the mixed equilibrium  in Appendix.\ref{appendix:proofstage1}.

Since we cannot characterize close-form equilibrium revenues   for the $\text{S}_1\text{S}_0$ and $\text{S}_0\text{S}_0$ subgames, it remains challenging to characterize  the  storage-investment equilibrium with respect to the system parameters, e.g., the storage cost and  demand. In the next subsection, we will focus on deriving insights of  the storage-investment equilibrium  in some special and practically interesting cases. 

\begin{figure}[t]
	\centering	
	{\includegraphics[width=2.4in]{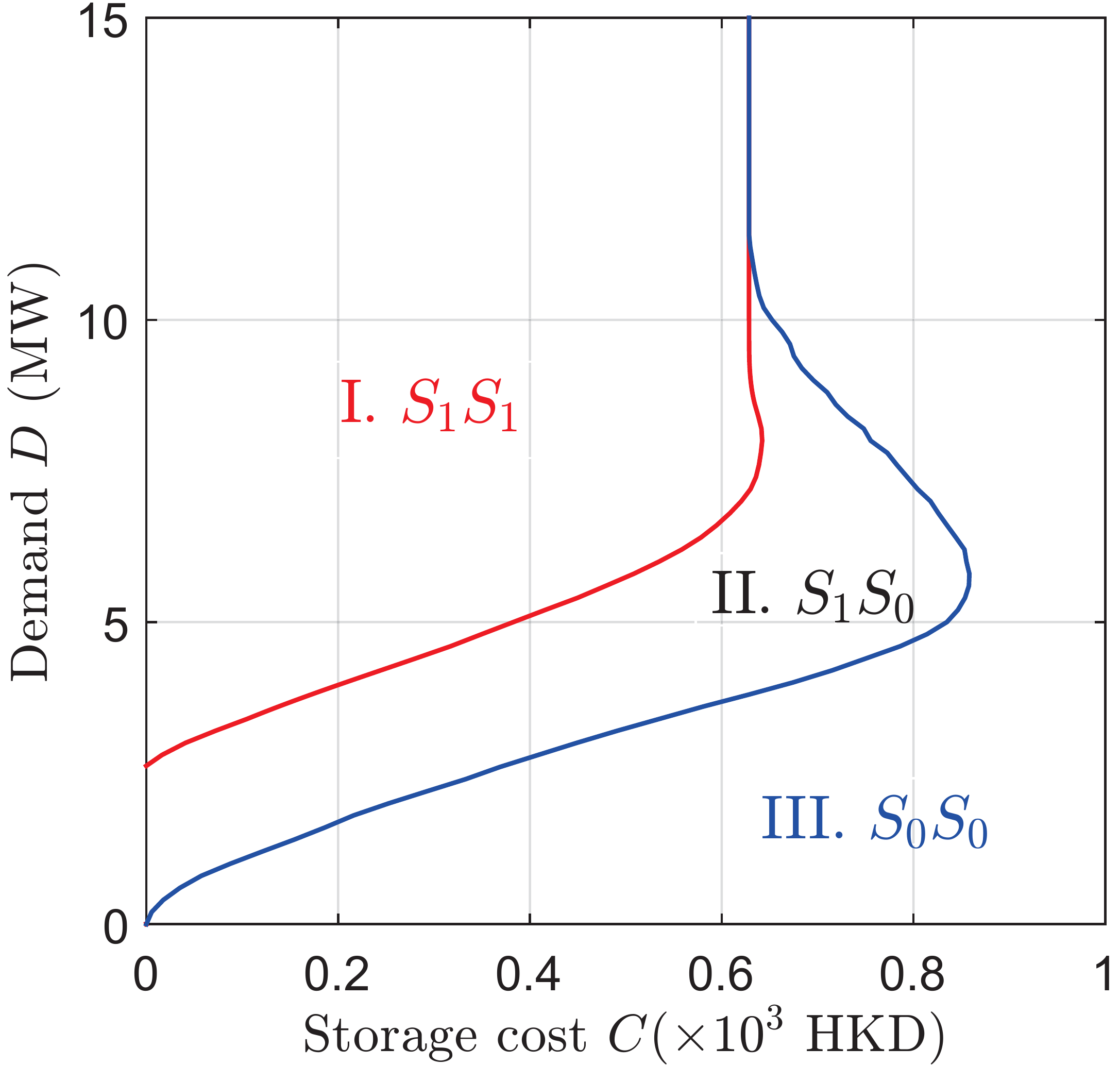}}
	\vspace{-1mm}
	\caption{\small  Equilibrium split with storage cost and demand at $\lambda=1.5$ HKD/kWh.}
	\vspace{-3mm}
	\label{fig:subfig:lam1} 
\end{figure}

\vspace{-1mm}
\subsection{Impact of storage cost and  demand on storage-investment equilibrium} We analyze the impact of storage cost and demand on the storage-investment equilibrium and have the analytical results for the cases when: (i) the storage cost $C_i$ is sufficiently large; (ii) the demand $D^{m,t}$ is sufficiently large or small. We present all the proofs in Appendix.\ref{appendix:proofstage1}.

To better illustrate the storage-investment equilibrium, we  show one simulation result of the equilibrium split (i.e., the storage-investment equilibrium with respect to parameters such as the demand and the storage cost)  in Figure \ref{fig:subfig:lam1}, and the details of the simulation setup are presented in Section \ref{section:sim}. In this simulation, for the illustration purpose, we consider the same demand $D$ for any hour $t$ of any month $m$. We also consider two homogeneous suppliers (with the same storage cost, the same renewable energy capacity and the same renewable energy distribution) to reveal the impact on storage-investment decision.\footnote{We can prove that a  pure Nash equilibrium of storage investment always exists in this  homogeneous  case. However, for the heterogeneous  case,  we cannot theoretically prove that the  pure Nash equilibrium always exists. In Appendix, we simulate an example with two heterogeneous  suppliers (with different  capacities of renewables) and show the storage-investment equilibrium in such a heterogeneous case.}    In Figure \ref{fig:subfig:lam1} (where  the penalty price is $\lambda=1.5$ Hong Kong dollars (HKD) per kWh), with respect to the demand and storage cost, the storage-investment equilibrium is divided into three regions:  Region  \uppercase\expandafter{\romannumeral1} of $\text{S}_1\text{S}_1$ (the left side of the red curve), Region \uppercase\expandafter{\romannumeral2} of $\text{S}_1\text{S}_0$ (between the red curve and the blue curve), and Region  \uppercase\expandafter{\romannumeral3} of $\text{S}_0\text{S}_0$ (the right side of the blue curve). 

First,  for the impact of  the storage cost, a higher storage cost will  discourage suppliers from investing in storage as implied in Theorem \ref{thm:stoeq}.   We will further show that when the storage cost is higher than a threshold, no suppliers will invest in storage no matter what the demand or penalty. However, counter-intuitively, we also find that in the case of a zero storage cost, not both suppliers will invest in storage once the demand is lower than a certain threshold.

As shown in Figure \ref{fig:subfig:lam1},   when the storage cost is larger than a threshold, i.e., $C>0.86\times 10^3$ HKD, the $\text{S}_0\text{S}_0$ case will be the only equilibrium (independent of the demand $D$)  and no suppliers invest in storage. We show this property in Proposition \ref{prop:stocost}. The reason is that the benefit from investing in storage  is bounded. When the storage cost is greater than a threshold corresponding to the bounded benefit, no suppliers will choose to invest in storage. 
\begin{prop}\label{prop:stocost}
	There exists a threshold $C_i^{\text{S}_0\text{S}_0}$ such that if the storage cost satisfies $C_i>C_i^{\text{S}_0\text{S}_0}$ for both $i=1,2$, the $\text{S}_0\text{S}_0$ case will be the unique pure storage-investment equilibrium.
\end{prop}

However, as shown in Figure \ref{fig:subfig:lam1}, when the demand is smaller than a certain threshold, i.e., $D<2.8$ MW, the $\text{S}_1\text{S}_1$ case cannot be a pure equilibrium even when the storage cost $C=0$. We show this property in Proposition \ref{prop:stodemandl}. The reason is that when the demand is smaller than a certain threshold, in the $\text{S}_1\text{S}_1$ case, both suppliers can only  get zero revenues (as shown in Proposition \ref{prop:pureprice}) due to the competition. Thus, if the $\text{S}_1\text{S}_1$ case is the storage-investment state where both suppliers invest in storage,   one supplier can always deviate to not investing in storage, which can bring him a strictly positive profit as implied in Proposition \ref{prop:positiverev}.

\begin{prop}\label{prop:stodemandl}
	If the demand satisfies $0<D^{m,t}\leq \min_i \mathbb{ E }[X_i^{m,t}]$ for any $t$ and $m$, the $\text{S}_1\text{S}_1$ case cannot be the equilibrium.
\end{prop}
\vspace{-1mm}

Second, for the impact of demand, we already show that at a sufficiently low demand, the $\text{S}_1\text{S}_1$ case cannot be the equilibrium  in Proposition \ref{prop:stodemandl}. We will further show that if the demand is higher than a certain threshold, each supplier has a dominant strategy of whether to invest in storage based on his storage cost, which does not depend on the other supplier's decision.  For example, at $D>11$ MW  in Figure \ref{fig:subfig:lam1}, for these two homogeneous suppliers,  if the storage cost is higher than a threshold, i.e., $C>0.63\times 10^3$ HKD, each supplier will not invest in storage (i.e., $\text{S}_0\text{S}_0$); otherwise, each supplier will invest (i.e., $\text{S}_1\text{S}_1$). We show this property in Proposition \ref{prop:stodemandh}. The reason is that if the demand is large enough, both suppliers can bid the highest price and sell out the maximum bidding quantity. Thus, there is no competition between suppliers, and they will make storage-investment decisions based on their own storage costs.	
\vspace{-1mm}
\begin{prop}\label{prop:stodemandh}
	There exists $D^{m,t,th}>0$ and $C_i^\text{th}>0$, such that when the demand satisfies $D^{m,t}\geq D^{m,t,th}$ for any $t$ and $m$, supplier $i$ has the dominant strategy $\varphi_{i}^*$ as follows.\footnote{ We characterize the close-form threshold $D^{m,t,th}>0$ and $C_i^\text{th}>0$ in Appendix.\ref{appendix:proofstage1}.} 
	\begin{equation}
	\varphi_{i}^*= \left \{
	\begin{aligned}
	&1,~\text{if}~\text{the~storage~cost}~ C_i\leq C_i^\text{th},\\
	&0, ~\text{if}~\text{the~storage~cost}~ C_i> C_i^\text{th}.
	\end{aligned}
	\right.
	\end{equation}	
\end{prop}

\subsection{Strictly positive profits of suppliers} We show that in  suppliers' competition  facing the cost of storage investment, both suppliers can get strictly positive profits.
\vspace{-2mm}
\begin{prop}[strictly positive profit]\label{prop:stoprofit}
	Both suppliers will get strictly positive profits at the storage-investment equilibrium.
\end{prop}

This proposition also shows the benefit of the uncertainty of renewable generation, which is similar to Proposition \ref{prop:positiverev}. Recall that if both suppliers have stable outputs,  they may get zero revenue (shown in Proposition \ref{prop:pureprice}) and thus get negative profit considering the storage cost. However, with the random generation, both suppliers will get strictly positive profits at the storage-investment equilibrium even facing the storage cost. We will explain it as follows.  Note that in the $\text{S}_0\text{S}_0$  case or the $\text{S}_1\text{S}_0$ case, the without-storage supplier always gets a strictly positive revenue (shown in Proposition \ref{prop:positiverev})  with a zero storage cost. 
In the $\text{S}_1\text{S}_0$ case or the $\text{S}_1\text{S}_1$ case, if the with-storage supplier gets a non-positive profit, he can always deviate to not investing in storage. This deviation provides him a strictly positive profit, which implies that the supplier will always get strictly positive profit.

\section{Characterization of storage capacity}\label{section:capacity}
We propose a probability-based  method using  historical data of renewable generations to compute  the storage capacity. Note that suppliers charge and discharge the storage to maintain his output at the mean value of the random renewable generations as shown in \eqref{eq:chdis}.\footnote{It is interesting to size the variable storage capacity considering the possibility of not completely smoothing out the renewable output. However,  it is quite challenging to characterize such an equilibrium storage capacity in closed-form,  which we will study as future work.} Therefore, the charge and discharge amounts are also random variables, and  we characterize the storage capacity such that its energy level will not exceed the storage capacity with a targeted probability. In this part, we focus on the storage with 100\% charge and discharge efficiency and no degradation cost. In Appendix.\ref{appendix:stotage},  we  show that a lower charge/discharge efficiency and the consideration of degradation cost will increase the total storage cost of a supplier, which further affects the storage-investment equilibrium.


To begin with, we set a probability target $\alpha$, and we aim to find a storage capacity $S_i$ such that the energy level in the storage  exceeds the  capacity with a probability no greater than $\alpha$.   Specifically, the with-storage supplier $i$  will charge and discharge storage with value $CD_i^{m,t}$ at hour $t$ of month $m$ as shown in \eqref{eq:chdis}.  We assume that the initial energy level of storage is fixed for all the months and denote it as $S^l_i$. Note that the energy level of storage is the sum of the charge and discharge over the time, and is constrained by the storage capacity. Starting from the initial energy level $S^l_i$, the probability  that energy level exceeds the   minimum capacity (i.e., zero) and  the  maximum capacity (i.e., $S_i$) of the storage in a day of month $m$ is $\max_{t'\in\mathcal{T}}\text{Pr}(\sum_{t=1}^{t'} CD_i^{m,t}+S^l_i<0)$ and $\max_{t'\in\mathcal{T}} \text{Pr}(\sum_{t=1}^{t'} CD_i^{m,t}+S^l_i> S_i)$, respectively. Considering all months $m$, we aim to choose the storage capacity $S_i$ so that the following hold: 
\begin{align}
&\mathbb{ E }_m\big[\max_{t'\in\mathcal{T}}\text{Pr}(\sum_{t=1}^{t'} CD_i^{m,t}+S^l_i<0)\big]\leq \alpha,\label{eq:pl}\\
&\mathbb{ E }_m\big[\max_{t'\in\mathcal{T}} \text{Pr}(\sum_{t=1}^{t'} CD_i^{m,t}+S^l_i> S_i)\big]\leq \alpha.\label{eq:pu}
\end{align}

Then,  we  describe how to use historical data \cite{hkob} to  compute the storage capacity that satisfies the probability threshold as in \eqref{eq:pl} and \eqref{eq:pu}.  we will first characterize an upper bound for the probability that energy level exceeds the given storage capacity in terms of the random variable $CD_i^{m,t}$, and then we propose Algorithm 1  to compute the required storage capacity to satisfy \eqref{eq:pl} and \eqref{eq:pu}.

First,  given the underflow capacity $S_i^l>0$ and  overflow capacity $S_i^u\triangleq S_i-S_i^l>0$, we characterize   an upper bound 	$Pr^{l,m}(S_i^l)$ for 	$ \max_{t'}\text{Pr}(\sum_{t=1}^{t'} CD_i^{m,t}+S_i^l<0)$ and an upper bound 	$Pr^{u,m}(S_i^u)$ for $\max_{t'}\text{Pr}(\sum_{t=1}^{t'} CD_i^{m,t}+S_i^l> S_i)$, respectively. We characterize these upper bounds based on Markov inequality \cite{concentration}, which is shown in Proposition \ref{prop:bound}.

\begin{prop}[Markov-inequality-based upper bound] \label{prop:bound}
	Given $S_i^l>0$ and $S_i^u>0$, the Markov-inequality-based upper bounds are shown as follows.
	
	\begin{itemize}
		\item For the upper bound	$Pr^{l,m}(S_i^l)$:
		\begin{align}
		Pr^{l,m}(S_i^l)=\max_{t'} \min_{s>0}  B^l(s), \label{eq:sl}
		\end{align} 
		where	 $B^l(s)\triangleq e ^ { - s S_i^l} \cdot \mathbb { E } \left[ e ^ { s \sum_{t=1}^{t'} -CD_i^{m,t} }\right] $.
		
		\item For the upper bound	$Pr^{u,m}(S_i^u)$:
		\begin{align}
		Pr^{u,m}(S_i^u)\triangleq~\max_{t'}\min_{s>0} B^u(s),\label{eq:su}
		\end{align}
		where	 $B^u(s)\triangleq e ^ { - s S_i^u} \cdot \mathbb { E } \left[ e ^ { s \sum_{t=1}^{t'} CD_i^{m,t} }\right] $.
	\end{itemize}
\end{prop}

Note that  $Pr^{l,m}(S_i^l)$ and $Pr^{u,m}(S_i^u)$ are  decreasing in $S_i^l$ and $S_i^u$, respectively. Also, $Pr^{l,m}(S_i^l)\rightarrow 0$  as  $S_i^l\rightarrow +\infty$, and   $Pr^{u,m}(S_i^u)\rightarrow 0$  as  $S_i^u\rightarrow +\infty$. These show  that a larger capacity will decrease the probability that the charge/discharge exceeds the capacity. Also, for any  probability threshold $\alpha>0$,  we can always find a  capacity, such that the probability  that energy level exceeds the capacity is below  $\alpha$.

Second, we propose Algorithm \ref{algorithm:sapacity}  to characterize the storage capacity $S_i$ based on the   historical data of $CD_i^{m,t}$ (derived from the renewable generation data of $X_i^{m,t}$). We use the underflow capacity $S_i^l$ for supplier $i$ as an example for illustration, and the overflow capacity $S_i^u$  follows the same procedure. Specifically,  for the underflow capacity $S_i^l$, we search it in an increasing order from zero as in Step 4. Given $S_i^l$,  for each month $m$, we calculate the exceeding probability $Pr^{l,m}(S_i^l)$ according to \eqref{eq:sl} as in Steps 5-7. Note that based on the data samples of $\sum_{t=1}^{t'} -CD_i^{m,t}$,  $B^l(s)$ is strictly convex in $s$.  Thus, for any $S_i^l>0$, the value of $ \min_{s>0}  B^l(s)$ can be efficiently computed using Newton's method\cite{newtonmethod}. Further,  we conduct an exhaustive search for ${t'\in\mathcal{T}} $ to obtain $Pr^{l,m}(S_i^l)$. We calculate the expected exceeding probability $\mathbb{ E }_m[Pr^{l,m}(S_i^l)]$ over months as in Step 8. We obtain the minimal underflow capacity $S_i^l$ if the exceeding probability  satisfies $\mathbb{ E }_m[Pr^{l,m}(S_i^l)]\leq \alpha$ as  in Step 9. Similarly, we can get the minimal overflow capacity $S_i^u$. The required  storage capacity is calculated as in Step 11.

\begin{algorithm} 
	\caption{Storage capacity $S_i$}  
	\label{alg:B}  
	\begin{algorithmic}[1]  \label{algorithm:sapacity}
		\STATE {\textbf{initialization}: set iteration index $S_i^l=S_i^u=0$, step size $\Delta S$;}
		\FOR {each $k\in\{l,u\}$}
		\REPEAT
		\STATE $S_i^k:=S_i^k+\Delta S;$
		\FOR {each $m\in\mathcal{M}$}
		\STATE Supplier $i$ calculates $Pr^{k,m}(S_i^k)$ according to \eqref{eq:sl} or \eqref{eq:su};
		\ENDFOR
		\STATE Supplier $i$ calculates $\mathbb{ E }_m[Pr^{k,m}(S_i^k)]$;		
		\UNTIL{	$$\mathbb{ E }_m[Pr^{k,m}(S_i^k)]\leq \alpha;$$ }
		\ENDFOR
		\STATE Each supplier $i$ computes $$S_i=S_i^l+S_i^u;$$
		\STATE {\textbf{output}}: $S_i$.
	\end{algorithmic}  
\end{algorithm} 

\begin{figure}[ht]
	\centering
	{\includegraphics[width=2.7in]{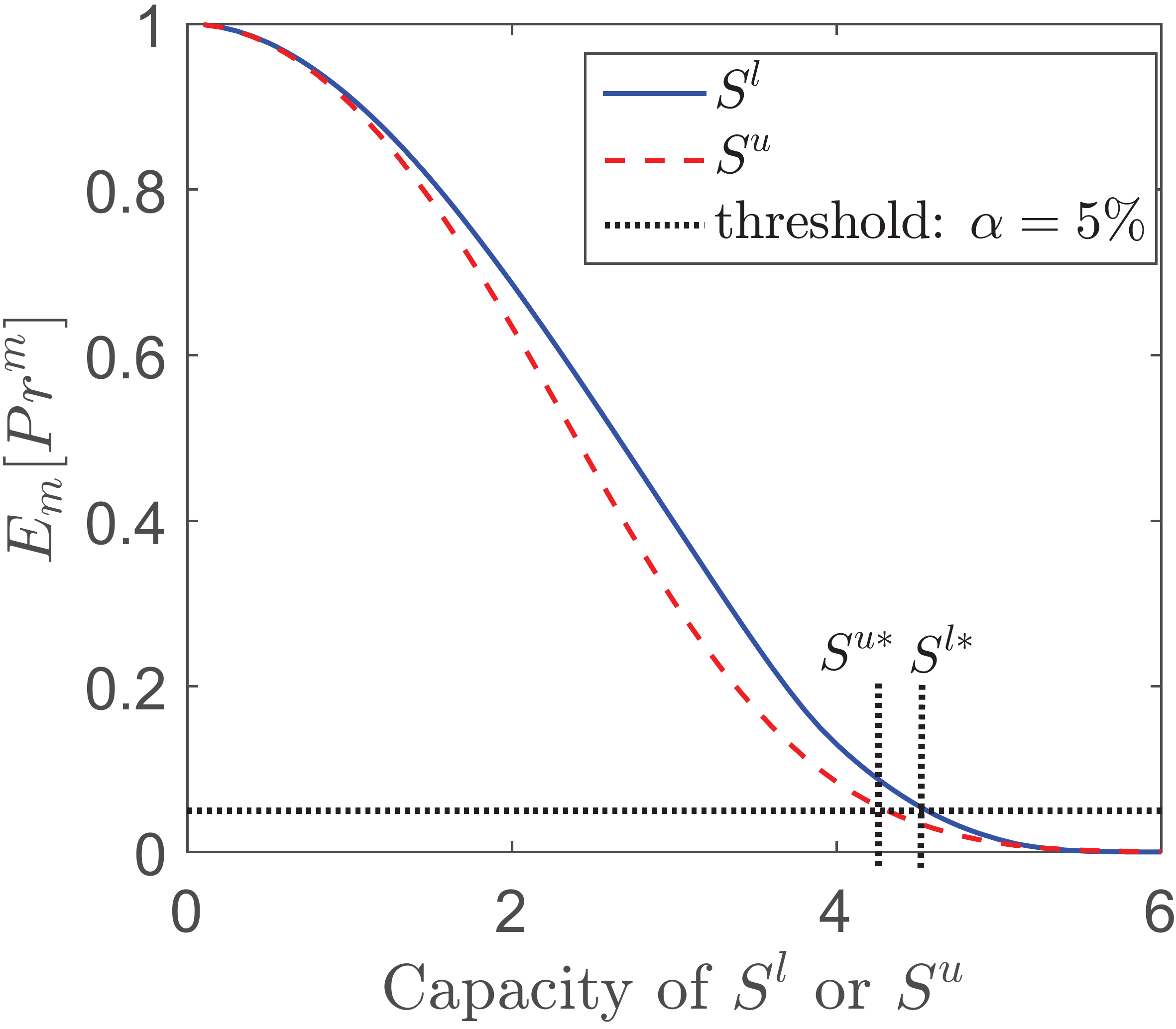}}
	\caption{ Characterization of storage capacity. }
	\label{fig:subfig:capacity2} 
\end{figure}
As an illustration, we calculate and show the underflow probability  $\mathbb{ E }_m[Pr^{l,m}(S_i^l)]$ and overflow probability $\mathbb{ E }_m[Pr^{u,m}(S_i^u)]$   in the blue solid curve and red dashed curve respectively in Figure 	\ref{fig:subfig:capacity2}. The probability of $\mathbb{ E }_m[Pr^{l,m}(S_i^l)]$ ($\mathbb{ E }_m[Pr^{u,m}(S_i^u)]$,  respectively)  decreases with respect to the capacity $S_i^l$ ($S_i^u$, respectively). If the capacity $S_i^l$ ($S_i^u$, respectively) is small and close to zero, the exceeding probability  $\mathbb{ E }_m[Pr^{l,m}(S_i^l)]$ ($\mathbb{ E }_m[Pr^{u,m}(S_i^u)]$, respectively) will  approach one. However, when the capacity is large and close to a certain value (e.g., 6 in Figure 	\ref{fig:subfig:capacity2}), the corresponding exceeding probability will be close to zero. We choose the probability threshold $\alpha=5\%$ and obtain the corresponding minimal capacity $S_i^{l*}$ and $S_i^{u*}$ as marked in Figure \ref{fig:subfig:capacity2}.

\section{Simulation}\label{section:sim}
In simulations, in addition to some analytical  properties of storage-investment equilibrium shown in Section \ref{section:stage1},  we will further investigate the impact of the penalty, storage cost, and demand on suppliers' profits. We will show some counter-intuitive results due to the competition between suppliers. For example, a higher penalty, a higher storage cost,  and a lower demand can even increase a supplier's  profit at the storage-investment equilibrium. Furthermore, the first supplier who invests in storage may benefit  less than the competitor who does not invest in storage.   We will illustrate the detailed results in the following.

\subsection{Simulation setup}
In simulations, we consider two homogeneous suppliers (with the same renewable capacity,  generation distribution, and  storage cost) to show the storage-investment equilibrium. We also consider a fixed demand $D$ for all the hours and months for illustration. Next, we explain the  empirical distribution of renewable generation as well as parameter configurations of the penalty price $\lambda$,  demand $D$, and  storage cost $C$.

\subsubsection{Empirical distribution of renewable generation} We use the historical data of solar energy generation in Hong Kong from the year 1993 to year 2012 \cite{hkob} to approximate the continuous CDF of suppliers' renewable generations. Specifically, we cluster the renewable generations at hour $t$ of all  days  into $M=12$ types (months) considering the seasonal effect.  We use daily data (from the year 1993 to year 2012) of renewable energy in month $m$ at hour $t$ to approximate the distribution of renewable generation at hour $t$  of month $m$. Based on the discrete data, we characterize a continuous empirical CDF to model the distribution of  renewable power.  We present the details of the characterization of empirical CDF  in Appendix.\ref{appendix:sim}.

Furthermore,  to check the reliability of the empirical distribution, we consider two sample data sets: one set consists of  all the data samples from the year  1993 to  2012, and the other consists of the data samples from another specific year (e.g., 2013). We conduct Kolmogorov-Smirnov test \cite{massey1951kolmogorov} using the Matlab function \textit{kstest2} to test whether these two data sets are  from the same continuous distribution \cite{kstest2}. The result shows that most of the hours of a month can pass the test. Also, our model is general for any continuous distribution of renewable generations. Interested readers can also use other data or other distributions of renewable energy to test the results.  

\begin{figure}[t]
	\centering
	\includegraphics[width=2.4in]{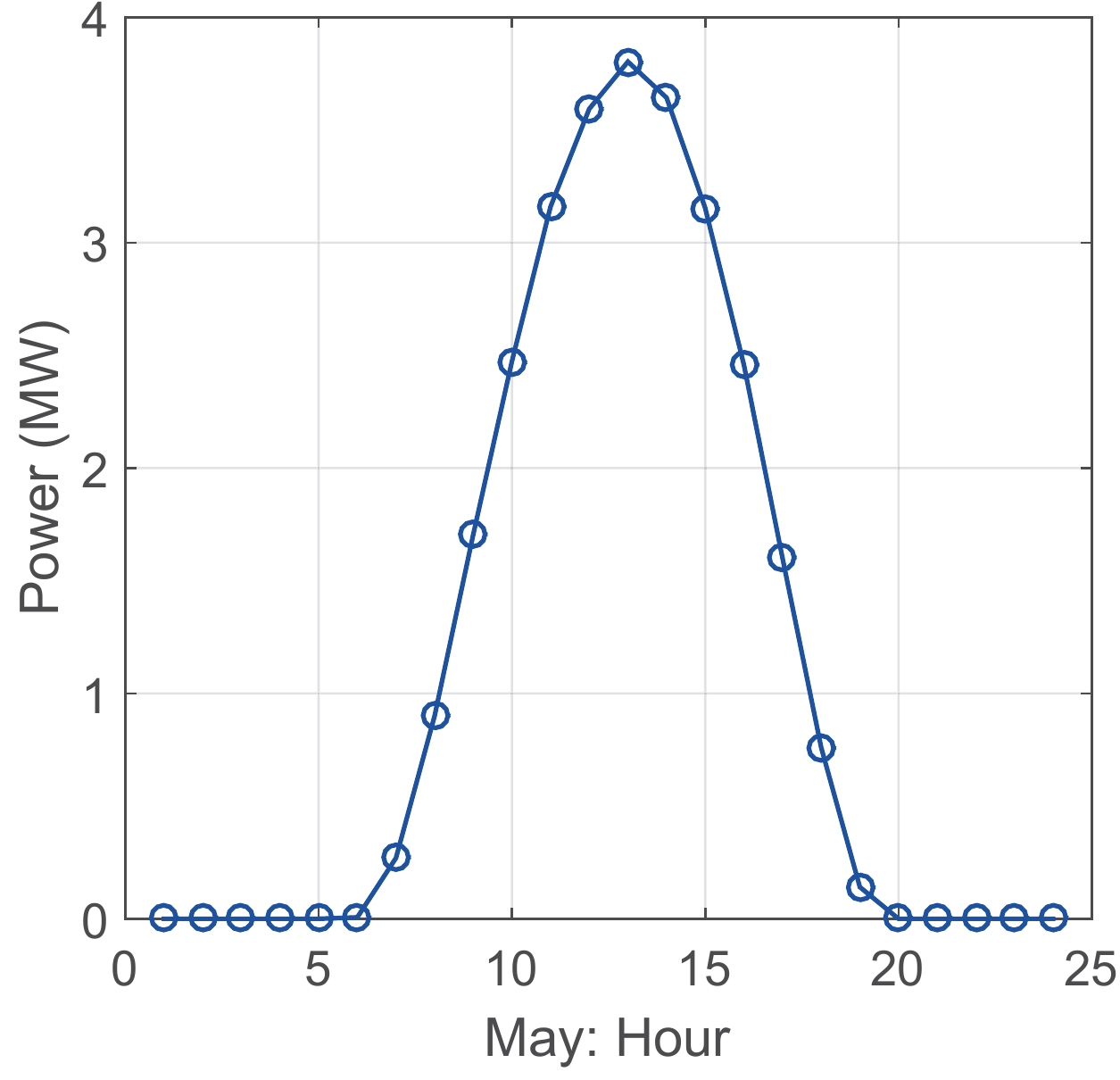}
	\vspace{-2mm}
	\caption{\small  Average solar energy of different hours in May.}
	\label{fig:solar}
\end{figure}

\subsubsection{Parameters configuration }
We explain the configuration of the parameters of the penalty price $\lambda$,  demand $D$, and  storage cost $C$, respectively. We set  the parameters to reflect the real-world practice, and study the impact of the  parameters on the market equilibrium. 

\begin{figure*}[t]
	\centering
	\subfigure[]{
		\label{fig:subfig:payoff1} 
		\raisebox{-2mm}{\includegraphics[width=2.32in]{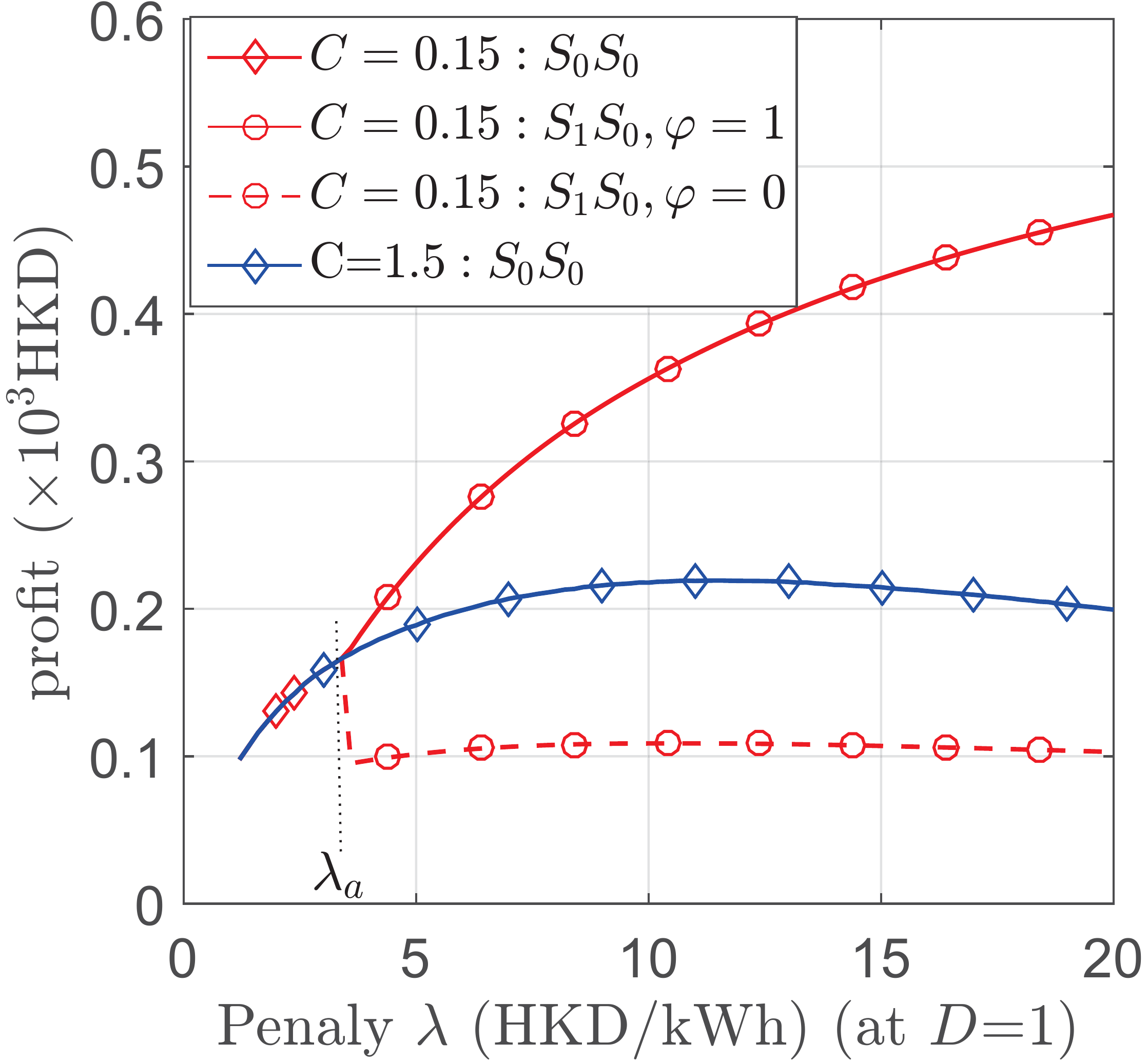}}}
	\hspace{-2mm}
	\subfigure[]{
		\label{fig:subfig:price} 
		\raisebox{-2mm}{\includegraphics[width=2.32in]{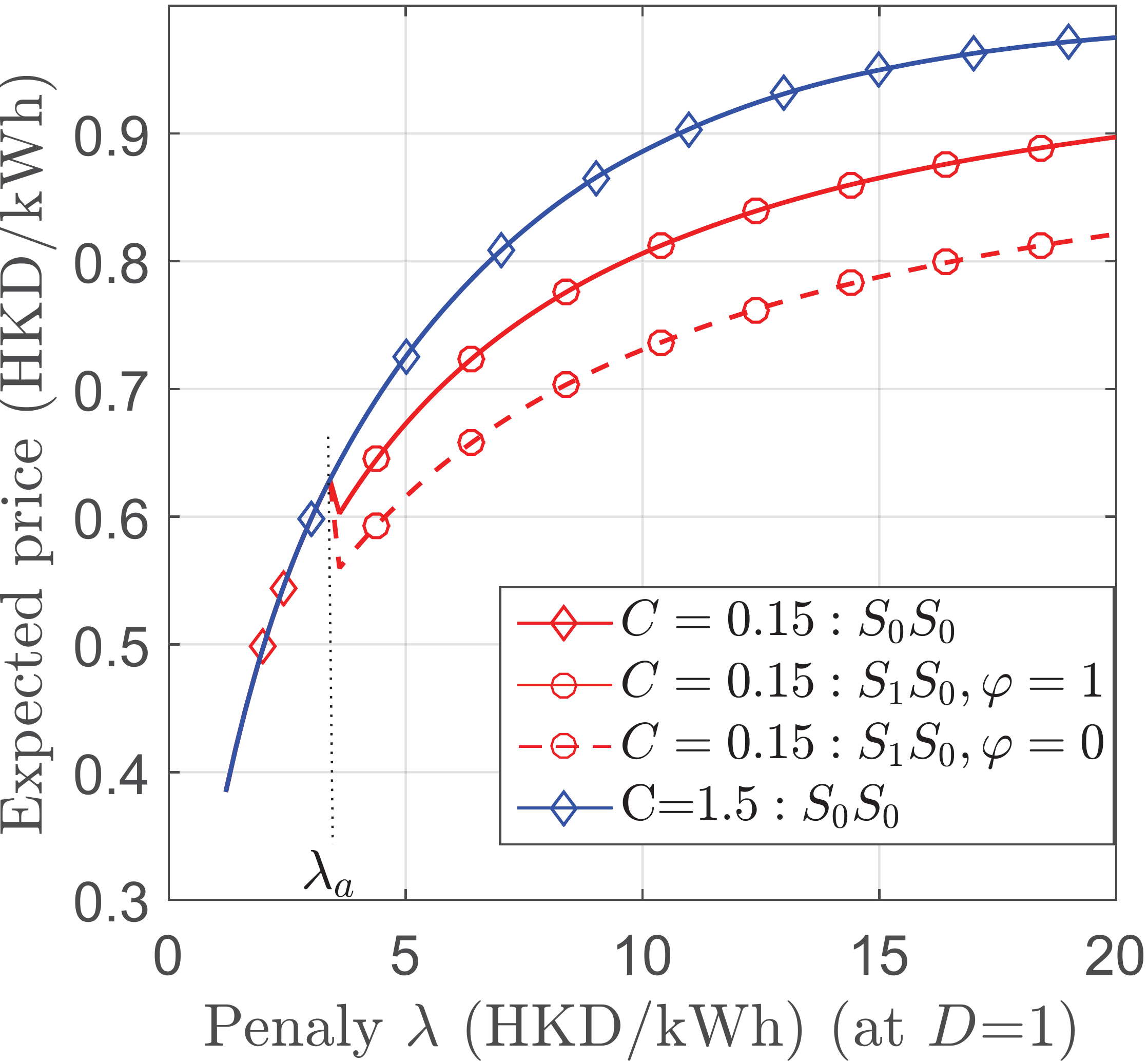}}}
	\hspace{-2mm}
	\subfigure[]{
		\label{fig:subfig:payoff2} 
		\raisebox{-2mm}{\includegraphics[width=2.352in]{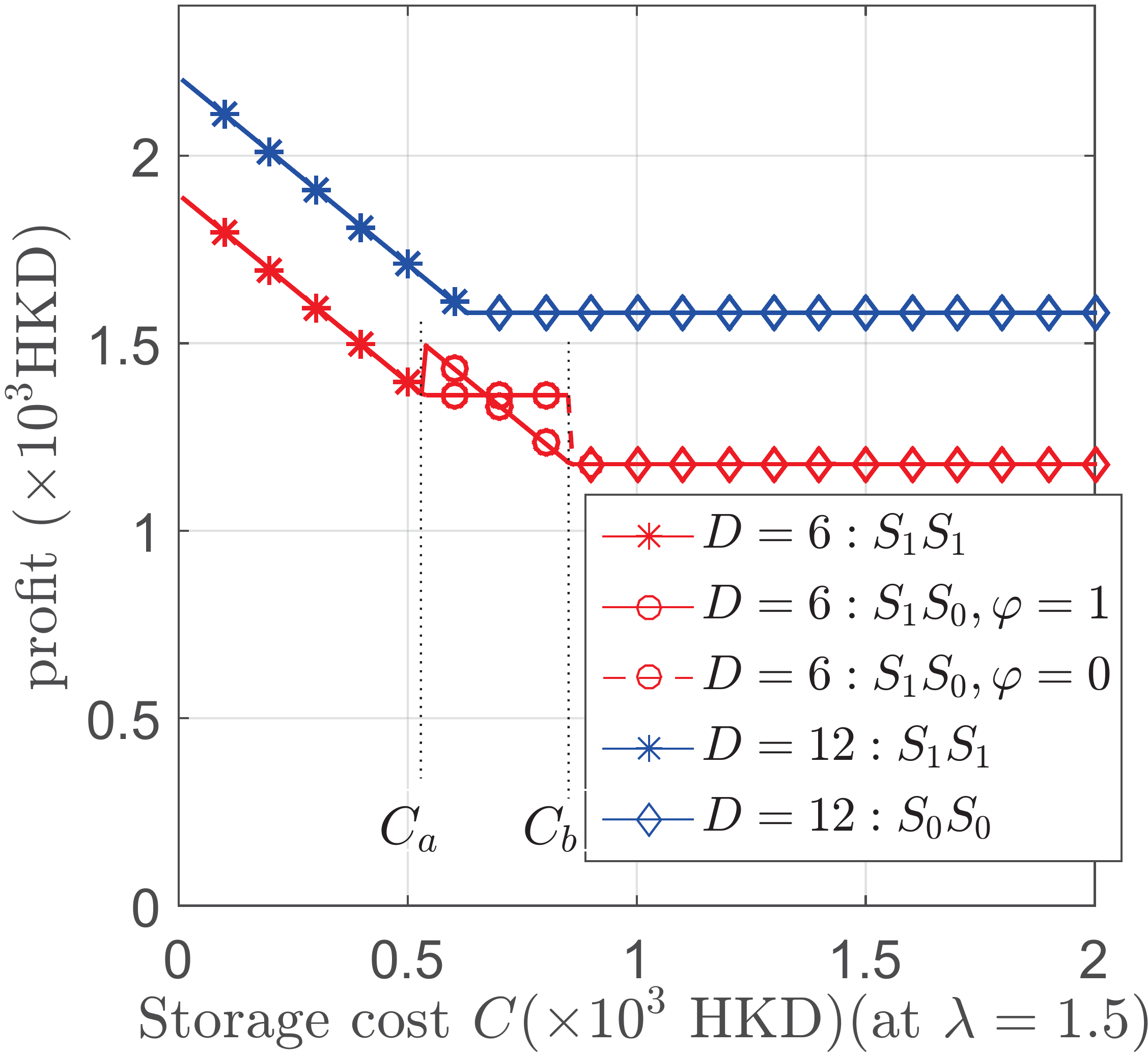}}}
	\hspace{-2mm}
	\subfigure[]{
		\label{fig:subfig:payoff3} 
		\raisebox{-2mm}{\includegraphics[width=2.36in]{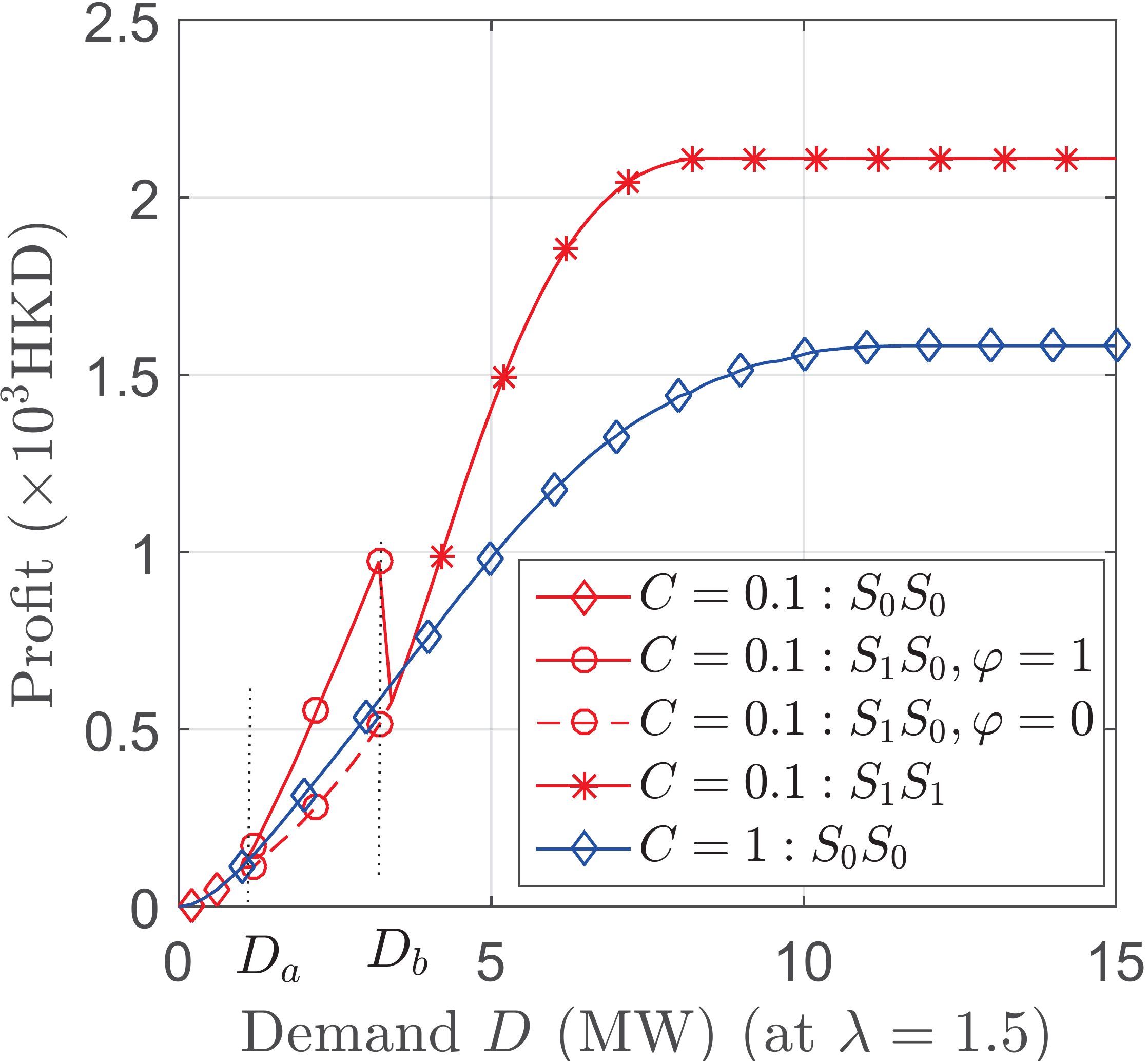}}}
	\vspace{-2mm}
	\caption{(a) Profit of suppliers with penalty ($D=1$ MW); (b) Expected bidding price of suppliers with penalty ($D=1$ MW); (c) Profit of suppliers with storage cost ($\lambda=1.5$ HKD/kWh); (d) Profit of suppliers with demand ($\lambda=1.5$ HKD/kWh). } 
	\vspace{-2mm}
	\label{fig:payoff}
\end{figure*}

\begin{itemize}
	\item The penalty $\lambda$: We choose the price cap $\bar{p}=1$ HKD/kWh, since the electricity price for residential users in Hong Kong is around  1 HKD/kWh \cite{hkhome}. Note that a penalty price satisfies $\lambda>\bar{p}$. In Figure \ref{fig:payoff}(a), we will consider a wide range of the ratio $\frac{\lambda}{\bar{p}}\in [1.2,20]$ to demonstrate the impact of the penalty. In Figures \ref{fig:payoff}(b)(c)(d), we fix the penalty price  $\lambda=1.5$ HKD/kWh  and focus on illustrating the impact of other parameters.
	\item The demand $D$: In Figure \ref{fig:payoff}(d), we will discuss a wide range of demand from 0 MW to 15MW to show the  impact of the demand. As a  comparison, in Figure \ref{fig:solar}, we show the average renewable power across hours in May.    In  Figure \ref{fig:payoff}(a) and (b), we fix the demand at $D=1$ MW to show the impact of other parameters ($\lambda$ and $C$). In  Figure \ref{fig:payoff}(c), we choose a larger demand $D=12$ MW and a smaller demand $D=6$ MW to show the impact of demand on the equilibrium profit.	
	\item The Storage cost $C_i$: Recall that the storage investment cost is  $C_i=c_i \kappa_i S_i$. There are different types of storage technologies with diverse  capital costs and lifespans.  For example, the pumped hydroelectric storage is usually cheap, and can last for 30 years with the capital cost $c_i=40\sim 800$ HKD/kWh, while the Li-ion battery can last 15 years with the capital cost about $c_i=1600\sim 9000$ HKD/kWh \cite{storagecost2017}.
	We choose the annual interest rate  $ r_i=5\%$, and the storage capacity for the with-storage supplier is characterized  as 43 MWh by Algorithm 1. We capture the impact of  parameters $c_i$ and $\kappa_i$ through the storage  cost $C_i$.   According to the calculation of storage investment cost $C_i=c_i \kappa_i S_i$, we can calculate that  the (hourly) investment  cost $C_i$ of the  pumped hydroelectric storage is $0.012\times 10^3-0.255\times 10^3$ HKD and the   cost  of the  Li-ion battery  is $0.76\times 10^3-4.36\times 10^3$  HKD. This shows that the storage cost can have a wide range.\footnote{Note that we only consider the investment cost in the storage cost. In practice, there are also other costs that need to be included, such as maintenance cost.} Then, in Figures \ref{fig:payoff}(c), we will consider a wide range of  storage costs from 0 to $2\times 10^3$  HKD. Although zero storage cost is not very practical, we use it to show a low storage cost and capture the  entire range of the impact of the storage costs. In Figure \ref{fig:payoff}(a)(b)(d), we choose lower storage costs ($0.1\times 10^3$ and $0.15 \times 10^3$  HKD) and  higher storage costs ($1\times 10^3$ and $1.5\times 10^3$ HKD) to show the different results under different storage costs. 	
\end{itemize}

\subsection{Simulation results} 

We will discuss the impact of penalty, storage cost, and demand on suppliers' profits, and show some counter-intuitive results due to the competition between suppliers.

\subsubsection{The impact of penalty on suppliers' profits}
\textit{Although a higher penalty $\lambda$ can increase the penalty cost on the without-storage supplier, surprisingly, we find that a higher penalty can also increase this supplier's profit,  due to the reduced market competition in the energy market.}

We show how suppliers' profits and expected bidding prices at the storage-investment equilibrium change with the penalty (at demand $D=1$ MW) in	Figure \ref{fig:subfig:payoff1} and \ref{fig:subfig:price}, respectively. Different colors represent different storage costs. The diamond marker shows that $\text{S}_0\text{S}_0$ is the storage-investment equilibrium, and the circle marker shows that  $\text{S}_1\text{S}_0$ is the equilibrium. Also, when  $\text{S}_1\text{S}_0$  is the equilibrium,  the solid lines and dashed lines distinguish the with-storage supplier and without-storage supplier,  respectively.

First, we show that at the equilibrium where both suppliers do not invest in storage (i.e., $\text{S}_0\text{S}_0$), a higher penalty $\lambda$ can increase both suppliers' profits. As shown in Figure \ref{fig:subfig:payoff1}, when the storage cost is high at $C=1.5\times 10^3$ HKD, both suppliers will not invest in storage for any value of  the penalty $\lambda$ from $1.2$ HKD/kWh to $20$ HKD/kWh (in blue curve with diamond marker). In this case, both suppliers' profits can first increase (at $\lambda<11$ HKD/kWh) and then decrease (at $\lambda>11$ HKD/kWh) with $\lambda$  (in blue curve).   The intuition for the increase of profit at $\lambda<11$ HKD/kWh is that a higher penalty  decreases both suppliers' bidding quantity if the bidding price remains the same. This reduces the market competition and enables both suppliers to bid a higher price in the local energy market as shown in Figure \ref{fig:subfig:price} (in blue curve).  However, the increased penalty also increases the penalty cost on suppliers, so the suppliers' profits will also decrease if the penalty is too high (at $\lambda>11$ HKD/kWh).

Second, we show that   at the equilibrium where one supplier  invests in storage and one does not (i.e., $\text{S}_1\text{S}_0$), a higher penalty $\lambda$ can also increase both suppliers' profits. We consider a low storage cost $C=0.15 \times 10^3$ HKD as in red curves in	Figure \ref{fig:subfig:payoff1} and Figure \ref{fig:subfig:price}. 
We see that if $\lambda$ is low (at $\lambda<\lambda_a$), both suppliers will not invest in storage (i.e., $\text{S}_0\text{S}_0$), and their profits increase with penalty shown in	Figure \ref{fig:subfig:payoff1} (at $\lambda<\lambda_a$ in red curve with diamond marker that  overlaps with blue curve).    As the penalty increases (at $\lambda>\lambda_a$), the equilibrium will change from $\text{S}_0\text{S}_0$ to $\text{S}_1\text{S}_0$, since a higher penalty and a lower storage cost can enable a supplier to enjoy more benefits by investing in storage. We discuss the profit of the with-storage supplier and without-storage supplier respectively as follows. 

\begin{itemize}
	\item For the with-storage supplier, as shown in Figure \ref{fig:subfig:payoff1}, when $\lambda>\lambda_a$, his profit increases  as penalty increases (in red solid curve), which can be much higher than the without-storage supplier (in red dashed curve). The reason is that in the $\text{S}_1\text{S}_0$ case,  the penalty cost  makes the with-storage supplier dominate over the without-storage one. The with-storage supplier can  bid higher prices than the without-storage supplier as shown in Figure \ref{fig:subfig:price} (in red solid curve and red dashed curve), and he also does not need to pay the penalty cost.
	\item However, for the without-storage supplier, as shown in Figure \ref{fig:subfig:payoff1}, his profit also slightly increases  as the penalty increases around $\lambda_a<\lambda<10$ HKD/kWh (in red dashed curve). The intuition is that a higher penalty gives the advantage to the with-storage supplier, which  reduces the market competition and increases both suppliers' bidding price as shown in Figure \ref{fig:subfig:price} (in red curves). Thus, it can also benefit the without-storage supplier. However, as shown in Figure \ref{fig:subfig:payoff1}, if the penalty further increases to $\lambda>10$ HKD/kWh (in red dashed curve), the without-storage supplier's profit will also decrease due to the increased  penalty cost.
\end{itemize}

\subsubsection{The impact of storage cost on suppliers' profits} \textit{Intuitively, a higher storage cost will discourage a supplier  from investing in storage, which  generally decreases a supplier's  profit. However, we find that it may also increase a supplier's profit if the other supplier changes his strategy due to the increased storage cost.}

We show how suppliers' profits at the storage-investment equilibrium change with the storage cost in	Figure \ref{fig:subfig:payoff2}. Different colors represent different demands. The diamond marker, circle marker, and star marker correspond to different  storage-investment  equilibria  of  $\text{S}_0\text{S}_0$,   $\text{S}_1\text{S}_0$, and $\text{S}_1\text{S}_1$, respectively. For the $\text{S}_1\text{S}_0$ case,  the solid lines and dashed lines distinguish the with-storage supplier and without-storage supplier,  respectively.

As shown in Figure 	\ref{fig:subfig:payoff2} (in both red curve and blue curve), generally the higher storage cost decreases suppliers' profits. However, we show that the opposite may be true using the example of $D=6$ MW (in red curve). 
When the demand is  at $D=6$ MW (in red curve),  as the storage cost increases, the equilibrium changes from $\text{S}_1\text{S}_1$ (when $C<C_a$), to  $\text{S}_1\text{S}_0$ (when $C_a<C<C_b$), and finally to   $\text{S}_0\text{S}_0$ (when $C>C_b$). When the equilibrium changes from  $\text{S}_1\text{S}_1$  to $\text{S}_1\text{S}_0$ at the threshold $C=C_a$, one  with-storage supplier in the original $\text{S}_1\text{S}_1$ case has a higher (upward jumping) profit,  after the other supplier chooses not to invest in storage due to the high storage cost. This changes the equilibrium from  $\text{S}_1\text{S}_1$ to $\text{S}_1\text{S}_0$, which reduces the competition and  gives more advantages to the with-storage supplier.

\subsubsection{The impact of demand on suppliers' profits}	\textit{Intuitively,  a higher demand will increase a supplier's  profit. However, we show that a higher demand  may also decrease a supplier's profit if the other supplier changes his strategy due to the increased demand.}

We show how suppliers' profits at the storage-investment equilibrium change  with the demand in	Figure \ref{fig:subfig:payoff3}. Different colors represent different storage costs. The diamond marker, circle marker, and star marker correspond to different  storage-investment  equilibria  of  $\text{S}_0\text{S}_0$,   $\text{S}_1\text{S}_0$, and $\text{S}_1\text{S}_1$, respectively. For the $\text{S}_1\text{S}_0$ case,  the solid lines and dashed lines distinguish the with-storage supplier and without-storage supplier respectively.

As shown in Figure 	\ref{fig:subfig:payoff3} (in both red curve and blue curve), generally a higher demand increases a supplier's profit. However, we  show that the opposite may be true using the example of $C=0.1\times 10^3$ HKD (in red curve). When the storage cost is low at $C=0.1\times 10^3$ HKD (in red curve), as the demand increases, the equilibrium changes from $\text{S}_0\text{S}_0$  (when $D<D_a$),  to  $\text{S}_1\text{S}_0$  (when  $D_a<D<D_b$), and finally to  $\text{S}_1\text{S}_1$ (when $D>D_b$). When the equilibrium changes from $\text{S}_1\text{S}_0$ to $\text{S}_1\text{S}_1$ at the threshold $D=D_b$, the with-storage supplier in the original $\text{S}_1\text{S}_0$ case has a smaller (downward jumping) profit, after the other supplier also chooses to invest in storage due to the high demand. This changes the equilibrium  from  $\text{S}_1\text{S}_0$ to $\text{S}_1\text{S}_1$, which increases the market competition and weakens the advantage of the with-storage supplier in the original $\text{S}_1\text{S}_0$ case. Furthermore, when the storage cost is high at $C=1.5\times 10^3$ HKD  (in blue curve with diamond marker),  both suppliers will not invest in storage  independent of the demand.

\subsubsection{First-mover disadvantage and advantage}Intuitively, the first supplier who invests in storage can benefit  more than the without-storage competitor. \textit{However, we find that if the storage cost is high, the first-mover supplier in investing storage can also benefit  less than the free-rider competitor who does not invest in storage. }

As shown in  Figure \ref{fig:subfig:payoff2} at  $D=6$ MW (in red curve), the $\text{S}_1\text{S}_0$ case is the equilibrium when the storage cost is in the range $C_a<C<C_b$.  If the storage cost is low at $C_a<C<0.7\times 10^3 \text{ HKD}$,  the  with-storage supplier's profit  is higher than the without-storage supplier's profit.  However, if the storage cost is high at  $0.7\times 10^3 \text{ HKD}<C<C_b$,  the  with-storage supplier's profit  is lower than the without-storage supplier. This shows both advantage and disadvantage of the first-mover. Although in some situations investing storage will increase the supplier's profit,  he can get more profits if he waits for the other to invest first when the storage cost is high. However, if the storage cost is low, he should be the first to invest storage in order to get a higher profit.

\section{Extensions: A more general oligopoly model}\label{section:extenstion}

We build a more general oligopoly model and  extend some of the theoretical results and insights from the duopoly case  to the  oligopoly case. Compared with the duopoly model, the only difference of the oligopoly model is that the number of suppliers can be more than two, i.e., $|\mathcal{ I}|\geq 2$. Following the analysis of the duopoly model, we also analyze the equilibrium  in Stage II and Stage I in the oligopoly case and derive some insights.  Specifically,  in Stage II, we extend the theoretical results of the price-quantity competition equilibrium. In Stage I, we  generalize analytical results of  the impact of storage cost and demand on the storage-investment equilibrium. Furthermore,  we show that some of the key insights from the duopoly case, e.g.,  the uncertainty of renewable generation can be beneficial to suppliers,   still hold in the oligopoly case. Next, we will discuss the extensions of Stage II and Stage I in detail, respectively. We include all the proofs of the propositions in Appendix.\ref{appendix:proofoligopoly}.

\subsection{ Stage II Analysis} For Stage II,  the weakly dominant strategy of bidding quantities still hold for the case of more than two suppliers. We generalize the conditions on the existence of the pure price equilibrium and show that the mixed price equilibrium also exists in the oligopoly case. Furthermore, we show that suppliers get  positive revenues  at the mixed price equilibrium. We show the extended analysis in detail as follows.

\subsubsection{Weakly dominant strategy for bidding quantities} The weakly dominant strategies for bidding quantities still hold as in Theorem \ref{thm:quantity}.
\subsubsection{Existence of the pure price equilibrium}  We derive the conditions for the existence of the pure  price equilibrium among suppliers, and generalize Proposition \ref{prop:pureprice}. Specifically, we consider a general subgame in Stage II  denoted as  ${S}^{{\mathcal{U}|\mathcal{V}}}$, where suppliers in the set  $\mathcal{U}$ invest in storage and suppliers in the set $\mathcal{V}$ do not invest. Recall  we denote the set of all the suppliers as $\mathcal{ I}$, and  we have $\mathcal{U}\bigcup \mathcal{V}=\mathcal{I}$.   The case $\mathcal{U}=\mathcal{I}$ means that all the suppliers invest in storage, and the case $\mathcal{V}=\mathcal{I}$ means that no supplier invests in storage. We show the existence of the pure  price equilibrium in Proposition \ref{prop:purepricennn}.

\begin{prop}[existence of the pure price equilibrium  in the oligopoly case]\mbox{}\label{prop:purepricennn}	
	Considering a subgame 	$\text{S}^{{\mathcal{U}|\mathcal{V}}}$ of storage investment among suppliers in Stage II, the existence of the pure	price equilibrium depends on the demand $D$ as follows:
	\begin{itemize}
		\item If  $D \geq \sum_{i\in \mathcal{I}} y_i^*(\bar{p},\varphi_i)$, there exists a pure  price equilibrium $p_i^*=\bar{p}$, with an equilibrium revenue $\pi_i^{RE}=\lambda \int_{0}^{F_i^{-1}(\bar{p}/\lambda)}xf_i(x)dx$ for any $i\in \mathcal{V}$ and $\pi_i^{RE}=\bar{p} \mathbb{E}[X_i]$ for any $i\in \mathcal{U}$.
		\item If  $D\leq \sum_{i\in\mathcal{U}} y_i^*(\bar{p},\varphi_i)-y_j^*(\bar{p},\varphi_j)$ for any  $j\in\mathcal{ U}$, there exists a pure  price equilibrium $p_i^*=0$,  with an equilibrium revenue $\pi_i^{RE}=0$, for any $i\in \mathcal{ I}$.
		\item If there exists $j\in\mathcal{ U}$ such that  $\sum_{i\in\mathcal{U}} y_i^*(\bar{p},\varphi_i)-y_j^*(\bar{p},\varphi_j)<D < \sum_{i\in \mathcal{I}} y_i^*(\bar{p},\varphi_i)$,
		there is no  pure price  equilibrium.
	\end{itemize}
\end{prop}

Similar to the duopoly case, the result of this proposition can be interpreted as follows. If the demand is higher than the threshold $\sum_{i\in \mathcal{I}} y_i^*(\bar{p},\varphi_i)$, all the suppliers  can bid  the price cap to sell the maximum quantities.   If the demand is very low such that $D\leq \sum_{i\in\mathcal{U}} y_i^*(\bar{p},\varphi_i)-y_j^*(\bar{p},\varphi_j)$ for any  $j\in\mathcal{ U}$, the competition is fierce and all the suppliers  bid zero price.   However, if the demand is in the middle,  there will be no pure price equilibrium. 

Note that if the number of with-storage suppliers is no greater than one, i.e., $\mid  \mathcal{ U}  \mid\leq 1$, the condition  that   there exists $j\in\mathcal{ U}$ such that $D\leq \sum_{i\in\mathcal{U}} y_i^*(\bar{p},\varphi_i)-y_j^*(\bar{p},\varphi_j)$ cannot be satisfied. It means that there will be no pure equilibrium of  $p_i^*=0$ for any  demand $D>0$.

\subsubsection{Existence of the mixed price equilibrium} For the case in Proposition \ref{prop:purepricennn} that there exists no pure price equilibrium, we show that there exists a mixed price equilibrium. However, the characterization of mixed strategy is highly non-trivial for the  oligopoly case and it is difficult to completely generalize Lemma \ref{lem:mix}. We generalize it partially as Proposition \ref{lem:mixnnn}  to show the existence of the mixed price equilibrium and show that all the suppliers get positive revenues at the mixed price equilibrium.

\begin{prop}[mixed price equilibrium  in the oligopoly case]\label{lem:mixnnn}
	For any $\bm{\varphi}$, when there is no pure price equilibrium, a  mixed price equilibrium exists and  		
	the equilibrium electricity-selling revenues $\pi_i^{RE}$ satisfies $\pi_i^{RE}(\bm{\varphi})>0,  \text{~for any}~ i\in \mathcal{ I} $.

\end{prop}

The equilibrium revenue for the case where all the suppliers invest storage (i.e., $\mathcal{U}=\mathcal{I}$) has been characterized in \cite{capacityprice}. When there are  two suppliers, we can also characterize the cumulative distribution function (CDF) of the mixed price strategy for the case of one investing storage and one not investing in storage as in Theorem \ref{thm:mscdf}.  However, when $I>2$, for any case where $\mid \mathcal{U} \mid<I$, it is highly non-trivial to characterize the corresponding CDF  analytically.

\subsection{ Stage I Analysis} For Stage I, for the general oligopoly case, we show that a mixed storage-investment equilibrium always exists. We can also  generalize the analytical results of  the impact of storage cost and demand on the storage-investment equilibrium for those settings where (i) the storage cost is sufficiently large; and (ii) the demand is sufficiently large or small. Furthermore,  some of the key insights, e.g., the uncertainty of renewable generation can be beneficial to suppliers, will still hold for the oligopoly case. We discuss the extensions in details in the following.

\subsubsection{Existence of the storage-investment equilibrium}  A mixed equilibrium of storage investment always exists. Note that each supplier has two strategies: investing in storage  and not investing in storage. Numerically, we can check the pure storage-investment equilibrium by the Nash equilibrium definition. Also, a mixed equilibrium of storage investment always exists due to the finite numbers of storage-investment strategies \cite{gamex}.

\subsubsection{Impacts of the storage cost and  demand on storage-investment equilibrium} Some analysis of the impact of the storage cost and  demand on storage-investment equilibrium in the duopoly case  can also be extended.  Specifically, we can extend  Propositions \ref{prop:stocost},  \ref{prop:stodemandl} and \ref{prop:stodemandh} and  to the oligopoly case, which  generalizes the analytical results for the settings where (i) the storage cost is sufficiently large; and (ii) the demand is sufficiently large or small.

First, since the benefit from investing in storage  is bounded, we can show that when the storage cost is greater than a threshold, no suppliers will choose to invest in storage. 

\begin{prop}\label{prop:stocostnnn}
	There exists a threshold $C_i^{\text{no}}$ such that if the storage cost satisfies $C_i>C_i^{\text{no}}$ for any $i\in \mathcal{ I}$, the $S^{\emptyset| \mathcal{ I}}$ case (i.e., no suppliers investing in storage) will be the unique pure storage-investment equilibrium.
\end{prop}

Second, in the subgame $S^{\mathcal{ U}|\mathcal{ V}}$ where $\mid \mathcal{ U}\mid \geq 2$,  if the demand is too low, all the suppliers may get zero revenue in the energy market as implied in Proposition \ref{prop:purepricennn}. This will make the with-storage suppliers deviate to not investing in storage. Thus, we have the proposition as follows.

\begin{prop}\label{prop:stodemandlnnn}
	In the subgame $S^{\mathcal{ U}|\mathcal{ V}}$, 	if the demand satisfies $0<D^{m,t}\leq \min_{j\in \mathcal{ U}} (\sum_{i\in\mathcal{U}} y_i^*(\bar{p},\varphi_i)-y_j^*(\bar{p},\varphi_j))$  for any $t$ and $m$,   the case  $S^{\mathcal{ U}|\mathcal{ V}}$ (i.e., suppliers in set $\mathcal{ U}$  invest in storage and suppliers in set $\mathcal{V}$  do not invest in storage) cannot be a pure storage-investment equilibrium.
\end{prop}

Third,  as in Proposition \ref{prop:purepricennn}, when the demand is higher than certain threshold, all the suppliers can bid the price cap to sell all his bidding quantity. In this case,  there is no competition between suppliers, and they will make storage-investment decisions independently  based on their own storage costs. We show this proposition as follows.	

\begin{prop}\label{prop:stodemandhnnn}
	There exist $D^{m,t,th}>0$ and $C_i^\text{th}>0$, such that when the demand satisfies $D^{m,t}\geq D^{m,t}_{\text{th}}$ for any $t$ and $m$, supplier $i$ has the dominant strategy $\varphi_{i}^*$ as follows.
	\begin{equation}
	\varphi_{i}^*= \left \{
	\begin{aligned}
	&1,~\text{if}~\text{the~storage~cost}~ C_i\leq C_i^\text{th},\\
	&0, ~\text{if}~\text{the~storage~cost}~ C_i> C_i^\text{th}.
	\end{aligned}
	\right.
	\end{equation}	
\end{prop}

\subsubsection{Positive profits at the storage-investment equilibrium}  We can further extend Proposition \ref{prop:stoprofit} to show the benefit of the uncertainty to the equilibrium profit. We show that  in  suppliers' competition  (even with the potential cost of the storage investment), all the  suppliers can get strictly positive profits at the equilibrium. 
\begin{prop}[strictly positive profit]\label{prop:stoprofitnnn}
	All the suppliers will get strictly positive profits at the storage-investment equilibrium.
\end{prop}	
This proposition  shows the benefit of the renewable generation randomness. If all the suppliers have stable outputs,  they may get zero revenue as implied in Proposition \ref{prop:purepricennn}	 and thus get negative profit under possible storage cost. However, with the random generation, all the  suppliers will get strictly positive profit at the storage-investment equilibrium even considering the storage cost. The intuition is that if one supplier invests in storage and gets non-positive profit, he can always choose not to invest in storage. This at least saves him the cost of storage investment, which increases his profit.  Also, note that when no supplier invests in storage, all the suppliers can get  positive profits.  Therefore,  only the state where all the suppliers get positive profits can be an equilibrium.

In summary, we can extend some of our major theoretical results and insights to the oligopoly case of more than two suppliers. Some of the key insights from the duopoly case, e.g.,  the uncertainty of renewable generation can be beneficial to suppliers,   still hold in the oligopoly case. However, we are not able to analytically extend all insights to the oligopoly case due to the complexity of analysis.   We would like to explore it in our future work.

\section{Conclusion}\label{section:con}
We study a duopoly two-settlement local energy market where renewable energy suppliers compete to sell electricity to consumers with or without energy storage. We formulate the interactions between suppliers and consumers as a three-stage game-theoretic model.  We characterize a price-quantity competition equilibrium in the local energy market, and further  characterize a storage-investment equilibrium at the beginning of  the investment horizon between suppliers. Surprisingly, we find the uncertainty of renewable generation can increase suppliers' profits compared with the case where both suppliers invest in storage and stabilize the outputs. In simulations, we show more counterintuitive results due to the market competition. For example, a higher penalty, a higher storage cost, and a   lower demand may increase a supplier's  profit. We also show that the first-mover in investing in storage  may  benefit less than the  free-rider competitor who does not invest in storage. In the future work, we will  size the variable storage capacity considering the possibility of not completely smoothing out the renewable output.

	\newpage

\section*{Appendix }\label{appendix:a}
This appendix is organized as follows:

\begin{itemize}
	\item Section \ref{appendix:s1s1}: We show the equilibrium revenue of suppliers in the  $\text{S}_1\text{S}_1$ case, when the demand satisfies $\min_i y_i^*(\bar{p},\varphi_i)<D< \sum_i y_i^*(\bar{p},\varphi_i)$ and there is no pure price equilibrium but the mixed price equilibrium. 
	\item Section \ref{appendix:s0s0}: We show how we discretize the continuous price set to approximate the mixed price equilibrium in the  $\text{S}_0\text{S}_0$ case.
	\item Section \ref{appendix:stotage}: For the storage capacity characterization, we first show the proof of the propositions in Section \ref{section:capacity}, and then we present the model of the imperfect storage.
	\item Section \ref{appendix:sim}: For the simulations,	we first show the characterization of the continuous CDF for the renewable-generation distribution using historical data, and then we simulate an example of two heterogeneous suppliers.
	\item Section \ref{appendix:proofstage3}:  We prove the theorems and propositions of Stage III.
	\item Section \ref{appendix:proofstage2}: We prove the theorems and propositions of Stage II.
	\item Section \ref{appendix:proofstage1}: We prove the theorems and propositions of Stage I.
		\item Section \ref{appendix:proofoligopoly}: We prove the propositions in the oligopoly model.
\end{itemize}

\vspace{5mm}

\section{Appendix: Mixed price equilibrium of $\text{S}_1\text{S}_1$ subgame}\label{appendix:s1s1}
 As shown in Proposition \ref{prop:pureprice}, when the demand satisfies $\min_i y_i^*(\bar{p},\varphi_i)<D< \sum_i y_i^*(\bar{p},\varphi_i)$, there is no pure price equilibrium.  We can characterize a close-form equilibrium revenue for each supplier at the mixed price equilibrium in Proposition \ref{thm:bothpricen}, which has been proved in  \cite{capacityprice}. 

\begin{prop}[$\text{S}_1\text{S}_1$: mixed-equilibrium revenue]\label{thm:bothpricen}
 In the $\text{S}_1\text{S}_1$ case (i.e., $\sum_{ i = 1 }\varphi_i=2$), if  $\min_i y_i^*<D< \sum_i y_i^*$,    there exists no pure price equilibrium but exists the mixed price equilibrium, with the equilibrium revenue as follows.
\begin{equation}
\pi_i^{RE}(\bm{\varphi})=\left \{
\begin{aligned}
&\bar{p}(D-y_{-i}^*),~{if}~y_i^*> y_{-i}^*,\\
&\frac{\bar{p}(D-y_i^*) y_i^*}{\min(y_{-i}^*,D)},~otherwise,\notag\\
\end{aligned}
\right.
\end{equation}
where $y_i^*=\mathbb{ E }[X_i]$ and $y_{-i}^*=\mathbb{ E }[X_{-i}]$ as characterized in Theorem \ref{thm:quantity}.
\end{prop}

According to Proposition \ref{thm:bothpricen}, one supplier's equilibrium revenue is related to the other supplier's bidding quantity (i.e., mean value of generations). Specifically, one supplier's equilibrium revenue decreases if the other supplier's bidding quantity increases. Furthermore, under the mixed price equilibrium, both suppliers get strictly positive revenues while they may get zero revenues when the demand is below the threshold $\min_i y_i^*$  as shown in Proposition \ref{prop:pureprice} under the pure price equilibrium.  
\vspace{5mm}
\section{Appendix: Mixed price equilibrium of $\text{S}_0\text{S}_0$ subgame}\label{appendix:s0s0}

 In the $\text{S}_0\text{S}_0$ case, both suppliers do not invest in storage and face the general penalty cost.  When $0<D<\sum_i y_i^*(\bar{p},\varphi_i)$, the  mixed price equilibrium  has a continuous CDF over $[l,\bar{p})$ shown in Lemma \ref{lem:mix}, but we cannot derive it in close form. To have a better understanding of the CDF,   we discretize the price  to approximate the original continuous price set, and  compute the mixed equilibrium for the discrete price set. 

Specifically, we discretize the price between $(0,\bar{p}]$ into $\{\Delta p,2\Delta p,3\Delta p,..., \bar{p}-\Delta p,\bar{p}\}$ with a small $\Delta p>0$. We search for the lower support in the range given in \eqref{eq:nslower}  in the following way. Given a lower support $l'$, the mixed strategy of each supplier has the support $\{l',l'+\Delta p, l'+\Delta p,..., \bar{p}\}$ that approximates the original  continuous support  $[l,\bar{p}]$. For each supplier,  each of price strategies in the support yields the same expected revenue, which can be used to construct a set of linear equations and calculate the mixed equilibrium. If the probability of each price for each supplier is between $(0,1)$, 
then the lower support $l'$ is feasible; otherwise, there exists the price that should be excluded from the support $\{l',l'+\Delta p, l'+\Delta p,..., \bar{p}\}$ and  the lower support $l'$ is not feasible. We calculate the equilibrium revenue according to Lemma \ref{lem:mix} (ii).

\section{Appendix: Characterization of storage capacity}\label{appendix:stotage}
We will first prove Proposition  \ref{prop:bound} and show some  properties of the  upper bound $Pr^{l,m}(S_i^l)$ and  $Pr^{u,m}(S_i^u)$. Then, we discuss the imperfect storage model and show how it affects the storage cost.

\subsection{Proof of Proposition \ref{prop:bound}}

\textbf{Proof}:  Below, we illustrate the upper bound $Pr^{l,m}(S_i^l)$. The upper bound  $Pr^{u,m}(S_i^u)$ can be derived analogously.

Given $t'\in \mathcal{T}$, we have 	
\vspace{-2mm}
\begin{align}
\text{Pr}(\sum_{t=1}^{t'} -CD_i^{m,t}> S_i^l)&=\text{Pr} \left( e ^ { s  \sum_{t=1}^{t'} -CD_i^{m,t} \geq e ^ { s S_i^l} }\right)\leq e ^ { -s S_i^l } \cdot \mathbb { E } \left[ e ^ { s \sum_{t=1}^{t'} -CD_i^{m,t} } \right]\triangleq B^l(s),\label{eq:capacity}
\end{align}
for any $s>0$. The inequality in \eqref{eq:capacity} is due to the Markov inequality.\footnote{This inequality is also known as Chernoff bound, which can achieve a tight probability bound\cite{mousavi2010tight}. }  Given $S_i^l>0$,  we can find a tight upper bound  for the probability  $\text{Pr}(\sum_{t=1}^{t'} -CD_i^{m,t}> S_i^l)$ by minimizing the RHS in \eqref{eq:capacity} over $s$. Therefore,	$Pr^{l,m}(S_i^l)=\max_{t'} \min_{s>0}  B^l(s)$. \qed

\subsection{Properties of some  properties of the  upper bound $Pr^{l,m}(S_i^l)$ and  $Pr^{u,m}(S_i^u)$. }
We have properties for $Pr^{l,m}(S_i^l)$ and 	$Pr^{u,m}(S_i^u)$ as follows.

\begin{prop}[properties of the upper bounds] \label{prop:boundpro}
	Given $S_i^l>0$ and $S^u>0$, the Markov-inequality-based upper bounds have properties  as follows.
	
	\begin{enumerate}
		\item $Pr^{l,m}(S_i^l)\leq 1$ and $Pr^{u,m}(S_i^u)\leq 1$.
		\item  $Pr^{l,m}(S_i^l)$ and $Pr^{u,m}(S_i^u)$ are  decreasing in $S_i^l$ and $S_i^u$, respectively.
		\item  $Pr^{l,m}(S_i^l)\rightarrow 0$  as  $S_i^l\rightarrow +\infty$, and   $Pr^{u,m}(S_i^u)\rightarrow 0$  as  $S_i^u\rightarrow +\infty$.
	\end{enumerate}
\end{prop}

\textbf{Proof}: The first property is because $\min_{s>0}  B^l(s)\leq B^l(0^-)=1$ and $\min_{s>0}  B^u(s)\leq B^u(0^-)=1$. The second property is straightforward from the function $B^l(s)$ and $B^u(s)$. The third property is because $CD_i^{m,t}$ is bounded. Thus,  $B^l(s)\rightarrow 0$  as  $S_i^l\rightarrow +\infty$, and  $B^u(s)\rightarrow 0$  as  $S_i^u\rightarrow +\infty$. \qed

Proposition \ref{prop:boundpro} shows that a larger capacity decreases the charge/discharge exceeding probability. Also, for any positive probability threshold $\alpha$,  we can always find a sufficiently large capacity to let the exceeding probability below  $\alpha$. This lays the foundation for Algorithm 1.

\subsection{Generalization of imperfect storage model}

We consider the  imperfect energy storage in two aspects:  (i) less-than-100\% charge and discharge efficiency and (ii) the degradation cost incurred by the charge and discharge.  Next, we will explain how the storage charge and discharge are determined in our work, and then further discuss how the imperfect storage impacts the total storage cost and investment equilibrium.

To begin with, we explain the model of the  storage charge and discharge as well as the energy level of the storage in our work. Specifically, the with-storage supplier charges and discharges the energy storage to stabilize his renewable output at the mean value.  Thus, the charge and discharge power is only dependent on the random variable of renewable  generations. At hour $t$ of renewable-generation-type (month) $m$, we denote the charge amount as $CD_i^{m,t+}\geq 0$ and the discharge amount as $CD_i^{m,t-}\geq 0$.  These values are characterized  based on the random generation $X_i^{m,t}$ as follows:
\begin{align}
&CD_i^{m,t+}=(X_i^{m,t}-\mathbb{ E }[X_i^{m,t}])^+,\label{eq:newc1n}\\
&CD_i^{m,t-}=(X_i^{m,t}-\mathbb{ E }[X_i^{m,t}])^-,\label{eq:newc2n}
\end{align}
where $g^+\triangleq \max(g,0)$ and  $g^-\triangleq\max(-g,0)$.
Furthermore, we denote the charge efficiency as $\eta_i^c$ and the discharge efficiency as $\eta_i^d$. The energy level in the storage can be calculated by adding the charge and discharge over time at month  $m$ as follows.
\begin{align}
e_i^{m,t}=e_i^{m,t-1}+\eta_i^c CD_i^{m,t+}- CD_i^{m,t-}/\eta_i^d. \label{eq:cd}
\end{align}
Next, we discuss how the degradation cost and the less-than-100\% charge and discharge efficiency   impact the total storage cost.
\subsubsection{Degradation cost}

We show that the degradation cost will increase the total cost of deploying the  storage for the with-storage supplier.  The degradation cost is caused by the charge and discharge of the storage. In the ideal case, we do not include the degradation cost as part of the  storage cost. With the degradation,  the total  cost of deploying the storage will be higher. One widely used model in the literature for the  degradation cost is a linear model \cite{hao2}\cite{datasto}.  We denote the unit cost of charge and discharge as $c_i^o$. Thus, the expected degradation cost $C_i^o$ (in each hour) is 
\begin{align}	
&C_i^o=\mathbb{ E }_{m,t}[c_i^o CD_i^{m,t+}+c_i^o CD_i^{m,t-}], \label{eq:degradation}
\end{align}
which can be calculated based on the historical data of $X_i^{m,t}$.

Therefore,   We can simply add  \eqref{eq:degradation} to the original storage cost. We calculate the total storage cost as $ C_i'=C_i+C_i^o$, which includes  both investment cost and the degradation cost.

\subsubsection{Charge and discharge efficiency}  
 The lower charge and discharge efficiency will increase the storage capacity and thus increase the total storage cost. Our goal is to characterize a minimum storage capacity such that the energy level $e_i^{m,t}$ will  exceed the storage capacity with a probability no greater than $\alpha$. As shown in \eqref{eq:cd}, the charge and discharge efficiency ($\eta_i^c,~ \eta_i^d$) will affect the energy level $e_i^{m,t}$. Compared with the perfect storage model with $\eta_i^c=\eta_i^d=100\%$, the difference in the imperfect storage model is that $\eta_i^c<100\%$ and  $\eta_i^d<100\%$. With the charge and discharge efficiency,  we modify \eqref{eq:pl} and  \eqref{eq:pu} in Section \ref{appendix:stotage}.A  into the following.
 \begin{align}
 &\mathbb{ E }_m\big[\max_{t'\in\mathcal{T}}\text{Pr}(\sum_{t=1}^{t'} \eta_i^c CD_i^{m,t+}- CD_i^{m,t-}/\eta_i^d+S^l_i<0)\big]\leq \alpha,\label{eq:pla}\\
 &\mathbb{ E }_m\big[\max_{t'\in\mathcal{T}} \text{Pr}(\sum_{t=1}^{t'} \eta_i^c CD_i^{m,t+}- CD_i^{m,t-}/\eta_i^d+S^l_i> S_i)\big]\leq \alpha.\label{eq:pua}
 \end{align}
Similarly, we can follow Algorithm \ref{algorithm:sapacity} in Section \ref{appendix:stotage}.A  to compute $S_i$ given the probability threshold $\alpha$.

\begin{figure}[ht]
		\centering
		\includegraphics[width=3.4in]{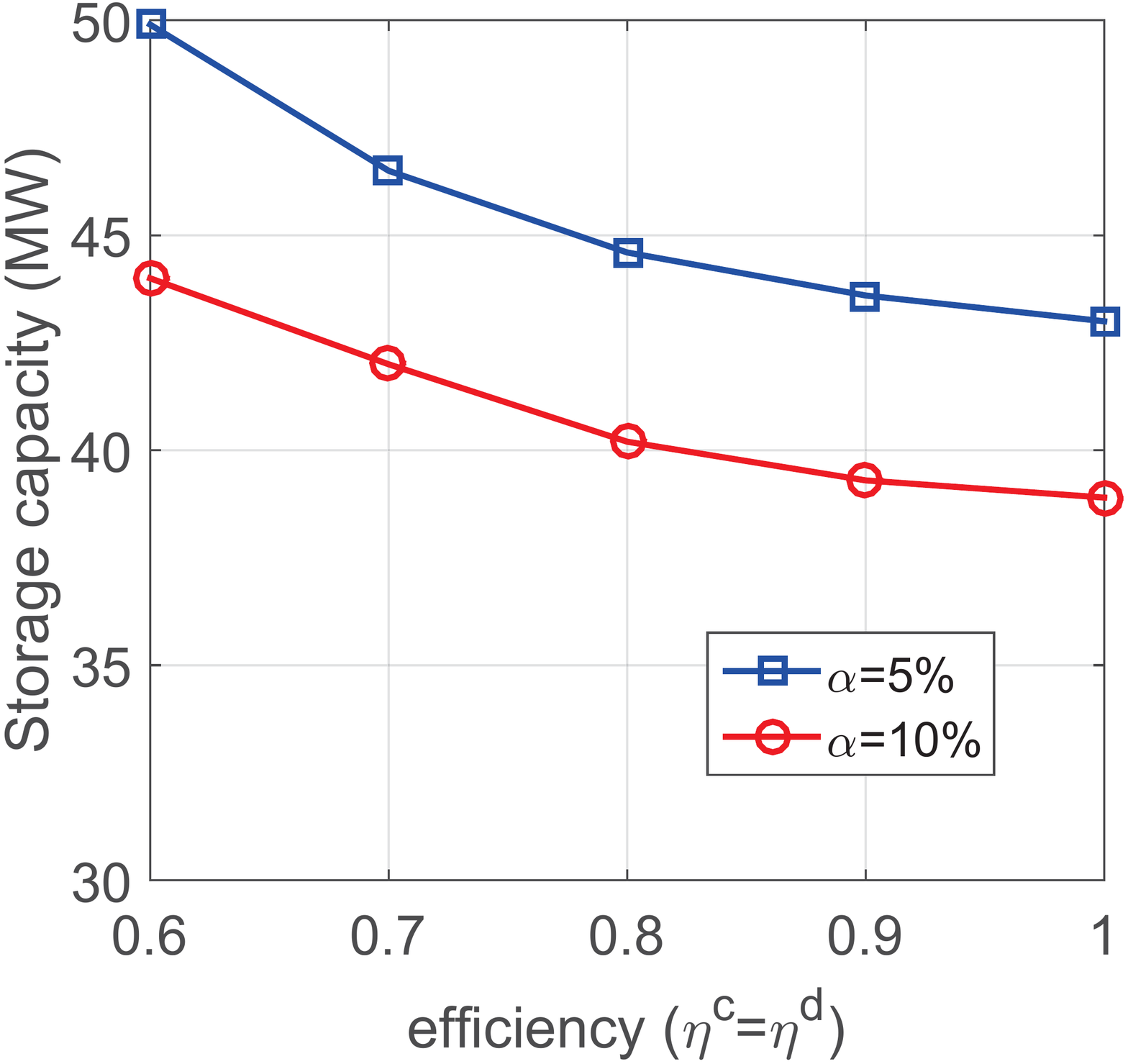}
		\vspace{-2mm}
		\caption{\small  Storage capacity with charge/discharge efficiency.}
		\label{fig:eff}
	\end{figure}

According to Algorithm 1 that computes the storage capacity, we show how charge/discharge efficiency impacts the storage capacity in Figure \ref{fig:eff}. The blue curve shows the case where the probability   that the energy level exceeds the capacity  is smaller than 5\% and the red curve shows the case where the probability   that the energy level exceeds the capacity  is smaller than 10\%. We see that as the efficiency decreases, the required storage capacity  increases (which further increases the storage investment cost).
	
In summary, compared with the case of perfect storage, a lower charge/discharge efficiency with the degradation cost will increase the total storage cost of a supplier. In Section VI, we present some analytical results of  the storage cost' s impact on the  storage-investment equilibrium. In Section VIII, we also show the simulation results of the impact of the  storage cost on the suppliers' profits.  These discussions can capture the impact of the imperfect storage.

\vspace{5mm}
\section{Appendix: Simulations}\label{appendix:sim}

We will first show the details of how we approximate the continuous CDF for the renewable-generation distribution using historical data. Then, we show a simulation result for two heterogeneous suppliers.

\subsection{Empirical distribution of renewable generations}
\begin{figure}[ht]
	\centering
	\subfigure[]{
		\label{fig:subfig:ecdf_ill} 
		\raisebox{-4mm}{\includegraphics[width=2.2in]{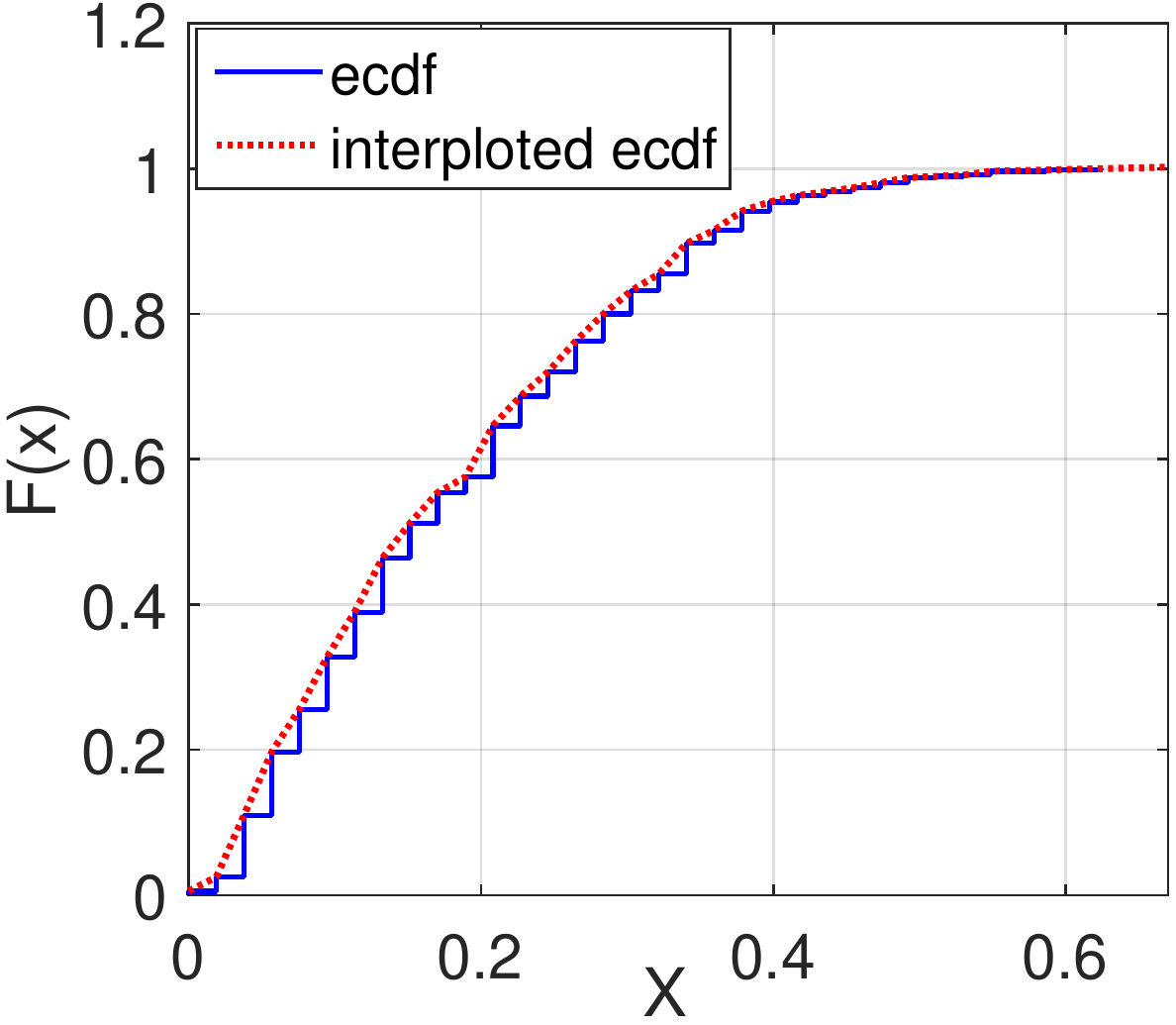}}}
	\hspace{-3mm}
	\subfigure[]{
		\label{fig:subfig:ecdf} 
		\raisebox{-4mm}{\includegraphics[width=2.2in]{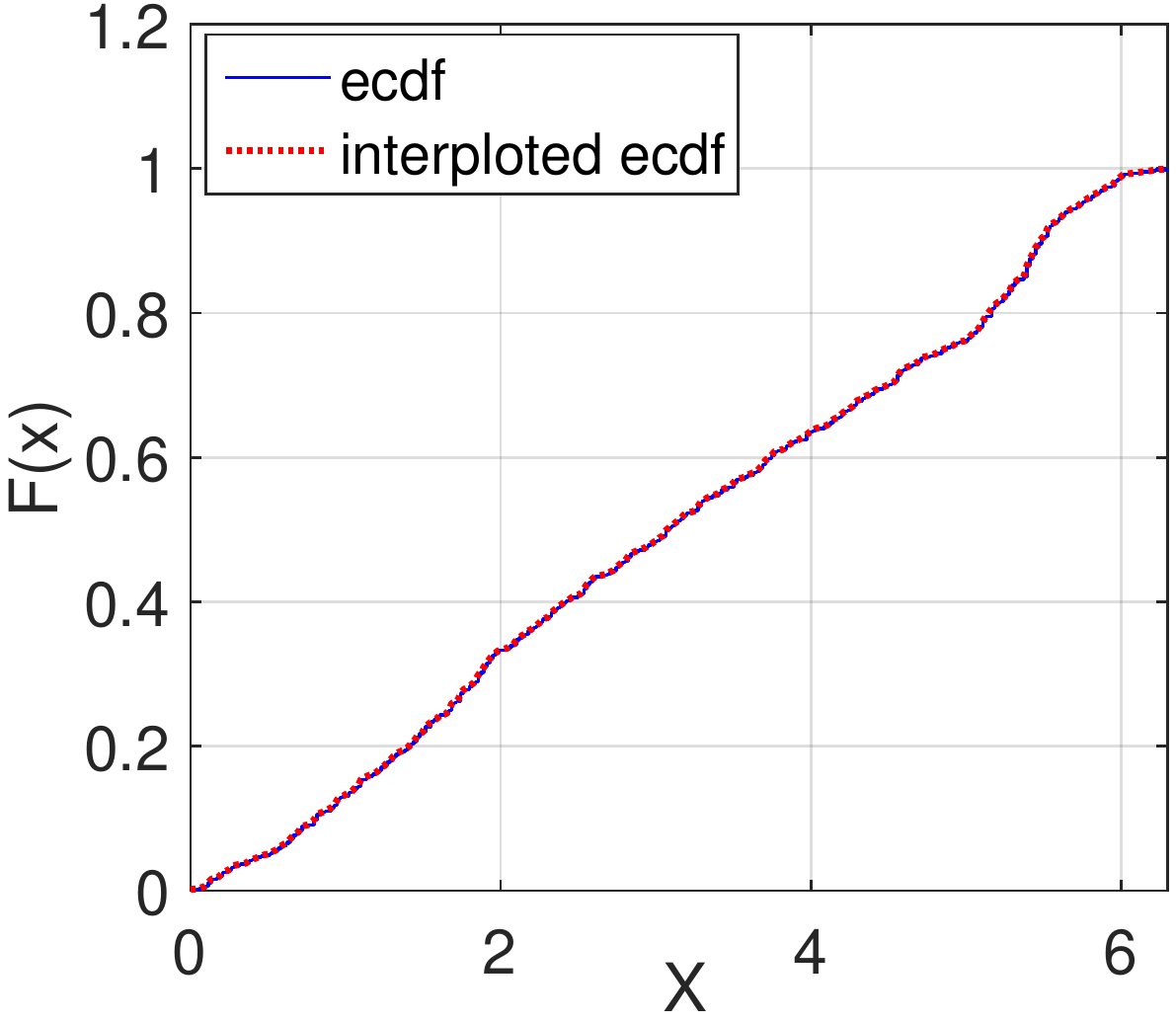}}}
	
	\vspace{-4mm}
	\caption{(a) Illustration of ECDF  and linearly-interpolated ECDF; (b) ECDF and linearly-interpolated ECDF at hour 9 of May. }
	\vspace{-3mm}
	\label{fig:ecdf}
	\vspace{-2mm}
\end{figure}

We use the historical data of solar energy in Hong Kong from the year 1993 to year 2012 \cite{hkob} to approximate the continuous CDF of suppliers' renewable generations. Specifically, we cluster the renewable generations at hour $t$ of all  days  into $M=12$ types (months) considering the seasonal effect.  We use daily data (from the year 1993 to year 2012) of renewable energy in month $m$ at hour $t$ to approximate the distribution of renewable generation at hour $t$  of month $m$. Based on the discrete data, we first use an \emph{empirical cumulative distribution function} (ECDF) to model the renewable power distribution.\footnote{Given a sample of real-world data $X_1,X_2,\ldots,X_m$, the standard ECDF $\widehat { F } ( x ):~R \rightarrow [0,1]$ is defined as $\widehat { F } ( x ) = \frac { 1 } { m } \sum _ { i = 1 } ^ { m } I \left( X _ { i } \leq x \right)$, where $I(\cdot)$ is the  indicator function\cite{ecdf}.} Note that our model is built on the continuous CDF of suppliers' renewable generations. Thus, we further use linear  interpolation to set up the continuous ECDF from the ECDF\cite{simulation}. We illustrate the ECDF and linearly-interpolated ECDF in Figure \ref{fig:subfig:ecdf_ill}, where the stepwise blue solid curve represents the ECDF and the red dotted curve represents the  linearly-interpolated ECDF. For the illustration of renewable generation distribution,  we show the ECDF and linearly-interpolated ECDF of hour $t=9$ of month  $m=5$ (May) in Figure \ref{fig:subfig:ecdf}. Through the linearly-interpolated ECDF $F_i$, we can also compute the value $F_i^{-1}(\cdot)$ efficiently.

\subsection{Simulations of two heterogeneous suppliers}

 We simulate an example with two heterogeneous suppliers. Note that  we can prove that a  pure Nash equilibrium of storage investment will always exist in the homogeneous case (with the same storage cost, the same renewable energy capacity and the same renewable energy distribution). However, for the general heterogeneous case,  we cannot theoretically prove that the  pure Nash equilibrium always exists. In our following example of heterogeneous suppliers, the pure Nash equilibrium of storage investment still exists. 
 
  Specifically,   we consider that supplier 2's renewable generation capacity is twice as much as the capacity of supplier 1, where both suppliers have the same distribution of renewable energy.  For comparison, we consider the homogeneous case as in the simulation of the main text where each supplier's renewable generation  capacity is equal to supplier 1's capacity of the heterogeneous case. In the following, we first assume that the storage investment  cost is the same across the two suppliers, and study the storage-investment equilibrium with respect to the storage cost  and demand  in the homogeneous (capacity) case and heterogeneous (capacity) case, respectively. Then, we allow the storage investment cost  to also differ across the two suppliers in the heterogeneous case, and study the storage-investment equilibrium with respect to the  two suppliers' different storage costs.

\begin{figure}[t]
	\centering
	\subfigure[]{  
		\label{fig:subfig:cd1} 
		\raisebox{-2mm}{\includegraphics[width=2.52in]{./figure/split_lamb2p}}}
	\hspace{-1mm}
	\subfigure[]{
		\label{fig:subfig:cd2} 
		\raisebox{-2mm}{\includegraphics[width=2.52in]{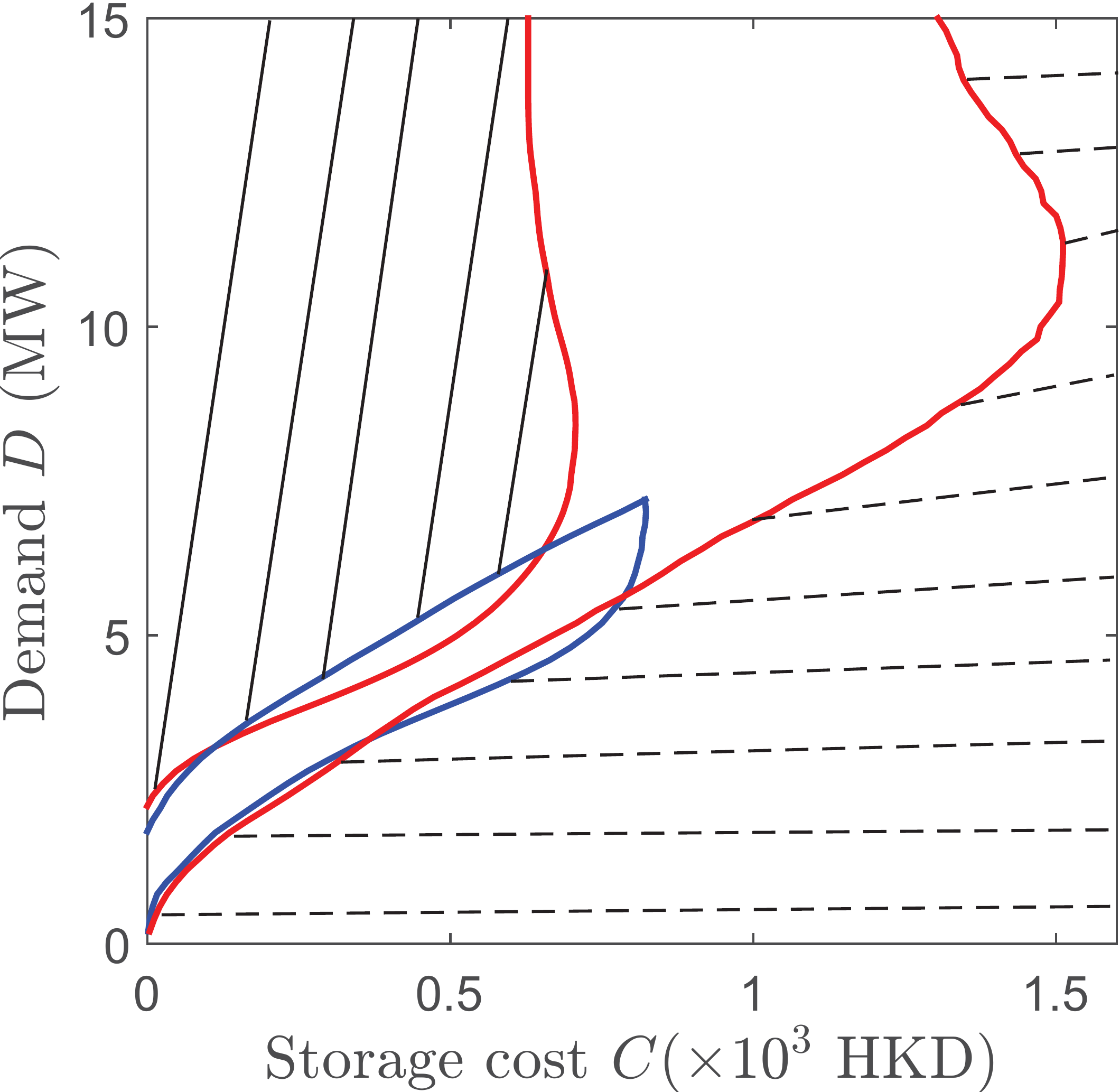}}}
	
	\vspace{-5mm}
	\caption{(a)  Equilibrium split in the homogeneous case; (b) Equilibrium split in the heterogeneous case.} 
	\vspace{-2mm}
	\label{fig:heter}
\end{figure}

We first consider the case that two suppliers' bear the same investment cost of storage, so as to focus on showing the impact of different capacities of renewables.\footnote{Note that the two suppliers have different storage capacities due to the different capacities of renewables. We choose  different unit costs of storage capacity and let two suppliers have the same storage investment cost.}
Figure  \ref{fig:subfig:cd1}  shows the equilibrium split in terms of demand and storage cost under the homogeneous case. Note that this figure has been shown as Figure \ref	{fig:subfig:lam1}  of the main text.  Figure  \ref{fig:subfig:cd2}   shows the equilibrium split in terms of demand and storage cost under the heterogeneous case.

\begin{itemize}
	\item    In Figure \ref{fig:subfig:cd1}, in Region I, both-investing-storage is one equilibrium;  in Region III, neither-investing-storage is one equilibrium; in Region II, one investing in storage and one not investing in storage will be one equilibrium. 
	\item  
	In Figure \ref{fig:subfig:cd2}, in the solid-grid region, both-investing-storage is one equilibrium; in  the dash-grid region,   neither-investing-storage is one equilibrium; in the region bounded by the red curve,   supplier 1 does not invest in storage while supplier 2 should invest in storage; and in the region bounded by the blue curve, supplier 1  invests in storage while supplier 2 does not invest in storage. 
\end{itemize}

\begin{figure}[t]
	\centering
	\includegraphics[width=2.7in]{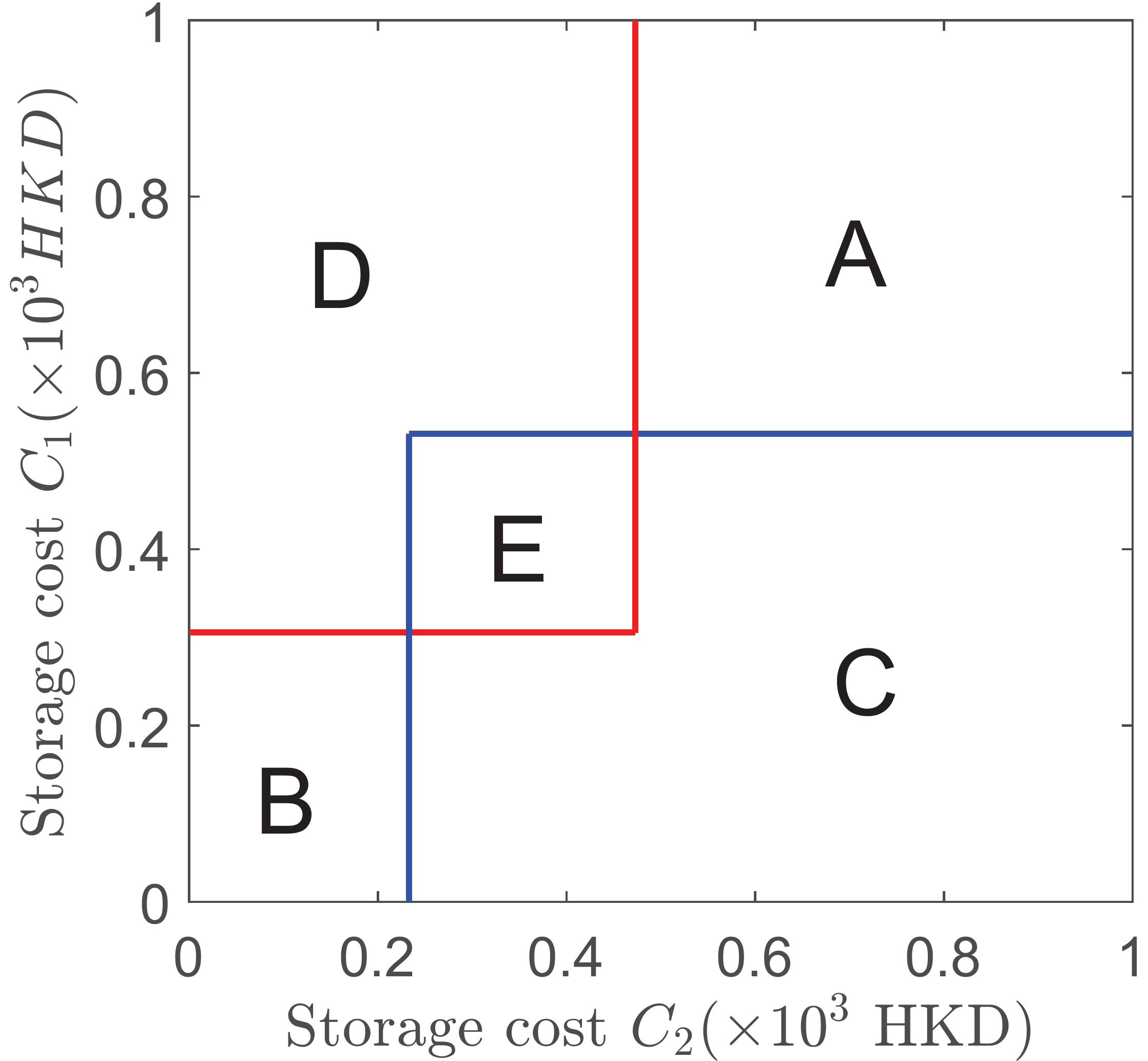}
	\vspace{-2mm}
	\caption{\small  Equilibrium split with storage cost.}
	\label{fig:cost}
\end{figure}
Generally, in the heterogeneous case, we see that the region where supplier 2 should invest in storage is larger than the region of supplier 1. The intuition is that supplier 2 has a larger capacity of renewables,  which gives her advantage in the competition. When both suppliers face the same high storage cost greater than  1000 HKD as in Figure \ref{fig:subfig:cd2}, supplier 1 will not invest in storage at the equilibrium for any demand $D$ but supplier 2 may still invest in storage when the demand is high. Also, the region of only supplier 1 investing in storage and only supplier 2 investing in storage can overlap under the heterogeneous case, which means that only supplier 1 investing in storage and only supplier 2 investing in storage are both equilibria.

Next we consider  the case that  two heterogeneous suppliers bear different storage investment costs.  We choose a certain demand ($D=4$ MW) and show the equilibrium split with respect to the storage cost of the two suppliers in  Figure \ref{fig:cost}. In Figure \ref{fig:cost}, if the storage costs of  supplier 1 and supplier 2 lie in Region A, neither supplier will invest in storage. In Region B, both suppliers will invest in storage.  In Region D,  only supplier 2 invests in storage and supplier 1 will not invest in storage. In Region C,  only supplier 1 invests in storage and supplier 2 will not invest in storage. However, in Region E, only supplier 1 investing in storage and only supplier 2 investing in storage, are both equilibria.

\vspace{5mm}
\section{Appendix: Proofs of Stage \uppercase\expandafter{\romannumeral3}}\label{appendix:proofstage3}

To prove Proposition \ref{prop:stage3}, we will discuss the following two cases and  analyze the objective function of Problem \eqref{eq:consumer} based on linear functions. For notation simplicity, we omit the superscript $m,t$ in the corresponding variables and  parameters.

\begin{itemize}
	\item If $p_1=p_2=p$, we rewrite the objective function \eqref{sg2:ob} as
	\begin{align}
	(P_g-p)(x_1+x_2). \label{sg2:obn}
	\end{align}
	
	Since $P_g-p>0$, the optimal value  is achieved at the maximum value of  $x_1+x_2$, i.e., $\min(D, y_1+y_2)$ according to the constraints \eqref{sg2:c1} and \eqref{sg2:c2}. 
	
	
	\item If $p_1\neq p_2$, we assume $p_1>p_2$ without loss of generality. We rewrite the objective function \eqref{sg2:ob} as
	\begin{align}
	(P_g-p_2)(x_1+x_2)+(p_1-p_2)x_1. \label{sg2:obn2}
	\end{align}
	Since $P_g-p_2>0$ and $p_1-p_2>0$, the optimal value  is achieved at the maximum value of  $x_1+x_2$ and  the maximum value of $x_1$ as follows:
	\begin{align}
	&x_1^*+x_2^*=\min(D, y_1+y_2),\\
	&x_1^*=\min(y_1,D).
	\end{align}
	Then, we obtain the optimal solution  $x_2^*=\min(D, y_1+y_2)-\min(y_1,D)$, which is equivalent to  $x_2^*=\min(D-\min(y_1,D),y_2)$. 
\end{itemize}

Combining the above two cases, we have Proposition 1 proved. \qed

\textit{Remark 1}: Proposition \ref{prop:stage3} can be easily extended to the oligopoly case with more than 2 suppliers.

\textit{Remark 2}: Given the other supplier $-i$'s bidding price $p_{-i}$ and bidding quantity $y_{-i}$, the supplier $i$'s payoff function generally is not continuous in price $p_i$ at $p_i=p_{-i}$ due to the discontinuous change of the optimal capacity $x_i^*$. This shows that given the other supplier $-i$'s decisions, supplier $i$'s payoff function generally is discontinuous.

\section{Appendix: Proofs of Stage \uppercase\expandafter{\romannumeral2}}\label{appendix:proofstage2}

\subsection{Proof of Theorem \ref{thm:quantity}}

To prove Theorem \ref{thm:quantity}, the key step is to  show that given  price $p_i$, the revenue  function $\pi_i^R\hspace{-1mm}\left(p_i, x_i^*(\bm{p},\bm{y}),\bm{\varphi}\right)$ of supplier $i$ with respect to  $x_i^*(\bm{p},\bm{y})$ is increasing on the interval $(0,y_i^*)$  and decreasing  on the interval $ (y_i^*,+\infty)$. Then, combined with Proposition 1, we can prove that $y_i^*$ will be the weakly dominant strategy for the bidding quantity. We discuss the weakly dominant strategy for supplier $i$ with $\varphi_{i}=1$ and $\varphi_{i}=0$, respectively.

\subsubsection{Case of $\varphi_{i}=1$}We will prove that the weakly dominant strategy of bidding quantity  for the with-storage supplier $i$ (i.e., $\varphi_i=1$) is  $y_i^*\left(p_i, \varphi_i\right)=\mathbb{ E }[X_i]$.   Given any price $p_i\leq \bar{p}<\lambda$, the function $\pi_i^R\hspace{-1mm}\left(p_i, x_i^*(\bm{p},\bm{y}),\bm{\varphi}\right)$ with respect to $x_i^*(\bm{p},\bm{y})$ is linearly increasing on the interval $ (0,\mathbb{ E }[X_i])$  and linearly decreasing  on the interval $ (\mathbb{ E }[X_i],+\infty)$. Thus, given any price $p_i$, we always have
\begin{align} 
 \pi_i^R\hspace{-1mm}\left(p_i, x_i^*(\bm{p},\bm{y}),\bm{\varphi}\right)\leq \pi_i^R\hspace{-1mm}\left(p_i, \mathbb{ E }[X_i],\bm{\varphi} \right)\label{eq:proof_quantity 1}
\end{align}
Then, we discuss a total of three cases to show  that with-storage supplier's revenue cannot be better off if he chooses strategy $y_i$ other than $y_i^*\left(p_i, \varphi_i\right)=\mathbb{ E }[X_i]$. For notation simplicity, we use $y_i^*$ to represent  $y_i^*\left(p_i, \varphi_i\right)$ in the later discussion.

(a) If $y_i<y_i^*=\mathbb{ E }[X_i]$, according to Proposition 1, we have 
\begin{align} 
x_i^*(\bm{p},({y}_i,y_{-i})) \leq  x_i^*(\bm{p},({y}_i^*,y_{-i}))\leq  \mathbb{ E }[X_i],~ \text{for~any~} y_{-i},
\end{align}
which  (according to \eqref{eq:proof_quantity 1}) implies
\begin{align} 
\pi_i^R\left(p_i, x_i(\bm{p},({y}_i,y_{-i})),\bm{\varphi})\leq \pi_i(p_i, x_i(\bm{p},({y}_i^*,y_{-i})),\bm{\varphi}\right).
\end{align}

(b) If  $y_i >y_i^*=\mathbb{ E }[X_i]$ and $x_i^*(\bm{p},({y}_i,y_{-i}))>\mathbb{ E }[X_i]$, according to Proposition 1,  we have $$x_i^*(\bm{p},({y}_i^*,y_{-i}))=\mathbb{ E }[X_i],$$ which (according to \eqref{eq:proof_quantity 1}) implies
$$\pi_i^R(p_i, x_i^*(\bm{p},({y}_i,y_{-i})),\bm{\varphi}) \leq \pi_i^R(p_i,\mathbb{ E }[X_i],\bm{\varphi})=\pi_i^R(p_i, x_i^*(\bm{p},({y}_i^*,y_{-i})),\bm{\varphi}).$$ 

(c) If  $y_i>y_i^*=\mathbb{ E }[X_i]$ and $x_i^*(\bm{p},({y}_i,y_{-i}))\leq \mathbb{ E }[X_i]$, according to Proposition 1,  we have
$$x_i^*(\bm{p},({y}_i,y_{-i}))= x_i^*(\bm{p},({y}_i^*,y_{-i})),$$ which implies
$$\pi_i^R(p_i, x_i^*(\bm{p},({y}_i,y_{-i})),\bm{\varphi})= \pi_i^R(p_i, x_i^*(\bm{p},({y}_i^*,y_{-i})),\bm{\varphi}).$$

Combining the above three conditions (a)-(c), we complete the proof that $y_i^*=\mathbb{ E }[X_i]$ if $\varphi_i=1$.

\subsubsection{Case of $\varphi_i=0$} We prove  the weakly dominant strategy of bidding quantity for the without-storage supplier $i$ (i.e., $\varphi_i=0$) is $y_i^*=F_i^{-1}(\frac{p_i}{\lambda})$.  We take the derivative of $\pi_i^R\hspace{-1mm}\left(p_i, x_i^*(\bm{p},\bm{y}),\bm{\varphi}\right)$ with respect to $x_i^*(\bm{p},\bm{y})$ and 
 give any $p_i>0$, it is easy to show that the function $\pi_i^R\left(p_i, x_i^*(\bm{p},\bm{y}),\bm{\varphi}\right)$ is increasing on the interval  interval $ (0,F_i^{-1}(\frac{p_i}{\lambda}))$  and decreasing  on the interval $ (F_i^{-1}(\frac{p_i}{\lambda}),+\infty)$. Thus, given any price $p_i$, we always have
\begin{align} 
\pi_i^R\hspace{-1mm}\left(p_i, x_i^*(\bm{p},\bm{y}),\bm{\varphi}\right)\leq \pi_i^R\hspace{-1mm}\left(p_i, F_i^{-1}(\frac{p_i}{\lambda}),\bm{\varphi} \right).\label{eq:proof_quantity 0}
\end{align}

Then, we can follow the proof step for $y_i^*$ in the case of $\varphi_{i}=1$ and prove that  $y_i^*=F_i^{-1}(\frac{p_i}{\lambda})$   for supplier $i$ with $\varphi_i=0$. \qed

\subsection{Proof of Proposition \ref{prop:pureprice}}

We verify the pure price equilibrium according to Definition \ref{def:pureprice} that the supplier cannot be better off if he deviates unilaterally. Towards this end, note that for supplier $i$ with or without storage, the revenue function $\pi_i^R\left(p_i, x_i^*(\bm{p},\bm{y}),\bm{\varphi}\right)$ is strictly increasing   with respect to both the price $p_i$  and the selling quantity $x_i^*(\bm{p},\bm{y})$ that is in the range $[0, y_i^*(p_i,\varphi_i)]$ (without considering the other supplier's coupled decisions).  We will discuss the three types of subgames respectively.
\subsubsection{The type $\text{S}_0\text{S}_0$ (i.e., $\sum_i\varphi_i=0$)} We first prove that when
$D \geq \sum_i y_i^*(\bar{p},{\varphi}_i)$,  $p_1=p_2=\bar{p}$ is a pure price equilibrium and show that this pure price equilibrium is unique. Then, we show that when
$D < \sum_i y_i^*(\bar{p},{\varphi}_i)$, there exists no pure price equilibrium.

(a) The case of 	$D \geq \sum_i y_i^*(\bar{p},{\varphi}_i)$.

We first prove that when
$D \geq \sum_i y_i^*(\bar{p},{\varphi}_i)$,  $p_1=p_2=\bar{p}$ is a pure price equilibrium. When $p_1=p_2=\bar{p}$, according to Proposition 1, the total selling energy quantities of supplier 1 and supplier 2 satisfy
\begin{align}
\sum_i x_i^*\left((\bar{p},\bar{p}),({y}_1^*(\bar{p},{\varphi}_1), {y}_2^*(\bar{p},{\varphi}_2)) \right)&=
\min(D, {y}_1^*(\bar{p},{\varphi}_1)+ {y}_2^*(\bar{p},{\varphi}_2))\\&={y}_1^*(\bar{p},{\varphi}_1)+ {y}_2^*(\bar{p},{\varphi}_2).\label{eq:proof_pureprice_1}
\end{align}
Since $x_i^*(\bm{p},\bm{y})\leq y_i^*(p_i,\varphi_i)$ always holds for any $i=1,2$, based on  \eqref{eq:proof_pureprice_1}, we have 
\begin{align}
&x_i\triangleq  x_i^*((\bar{p},\bar{p}),({y}_1^*(\bar{p},{\varphi}_1), {y}_2^*(\bar{p},{\varphi}_2)) )=
{y}_i^*(\bar{p},{\varphi}_i).
\end{align}
We will show that both suppliers cannot be better off if they deviate from such a bidding strategy. Without loss of generality, if supplier $1$ bids a price $p_1'< \bar{p}$ unilaterally, according to Proposition 1, we have
\begin{align}
x_1'\triangleq{x}_1^*(({p}_1',\bar{p}),({y}_1^*(p_1',{\varphi}_1), {y}_2^*(\bar{p},{\varphi}_2))&= \min \left\{D, {y}_1^*(p_1',{\varphi}_1)\right\}\\&={y}_1^*(p_1',{\varphi}_1)\\&<x_1.
\end{align}
Since the revenue function $\pi_i^R\left(p_i, x_i^*(\bm{p},\bm{y}),\bm{\varphi}\right)$ is strictly increasing   with respect to the price $p_i$ and the selling quantity $x_i^*(\bm{p},\bm{y})\leq y^*$, we have 
\begin{align}
\pi_1^R\left(p_1', x_1',\bm{\varphi}\right)<\pi_1^R\left(\bar{p}, x_1,\bm{\varphi}\right),
\end{align}
which shows that supplier 1's revenue decreases if he deviates from the price $\bar{p}$. This proves that $p_1=p_2=\bar{p}$ is a pure price equilibrium.

Next, we show that this equilibrium is unique. Without loss of generality, suppose that
supplier 1 bids a price $p_1'<\bar{p}$ while the other supplier  bids a price $p_{2}'\leq \bar{p}$. Since 	$D \geq \sum_i y_i^*(\bar{p},{\varphi}_i)$, according to Proposition 1, each supplier's maximum bidding quantity will be sold out and  we  have
\begin{align}
x_1^*(\bm{p}',\bm{y}^*(\bm{p}',\bm{\varphi}))= y_1^*(p_1',\varphi_1)\leq y_1^*(\bar{p},\varphi_1).
\end{align}
Therefore, supplier 1 can always increase his price $p_i'$ to $\bar{p}$, which will increase his revenue due to the increased price and non-decreasing selling quantity. Thus,  any price pair $(p_1,p_2) \neq (\bar{p},\bar{p})$ can't be an equilibrium.

(b) {Case of $0<D<  \sum_i y_i^*(\bar{p},{\varphi}_i)$ }.

{We will assume that both suppliers bid the pure prices and will discuss a total of three cases in the following to show that no pure price strategy can be an equilibrium.  
	
	First, suppliers' bidding prices are not equal, and we assume $p_i<p_{-i}$ without loss of generality. the lower-price supplier can always increase the price by a small $\varepsilon>0$  such that $p_i'=p_i+\varepsilon<p_{-i}$. Then,  the bidding price  $p_i'>p_i$  and the selling quantity at $p_i'$ denoted as ${x}_i'$ satisfies	${x}_i'= \min \left\{D, y_i^*(p_i+\varepsilon,\varphi_i)\right\}\geq x_i=\min \left(D, y_i^*(p_i,\varphi_i)\right)$.  In this case, we denote  the revenue at the original price   $p_i$  as $\pi_i$, and the revenue at the price   $p_i'$  as $\pi_i'$.  We have $\pi_i'>\pi_i$ since  $p_i'>p_i$  and ${x}_i'\geq x_i$. Thus, the unequal bidding price cannot be an equilibrium. 
	
	Second, two suppliers bid the same positive price, i.e., $p_1=p_1=p>0$. Based on Proposition 1,  the selling quantities of two suppliers satisfy the following condition
	\begin{align}
	\sum_i	x_i^*(\bm{p},\bm{y}^*(\bm{p},\bm{\varphi}) )=
	\min(D, y_1^*({p},\varphi_1)+y_2^*({p},\varphi_2)).
	\end{align} 
	For simplicity,	we denote the original selling quantity of supplier 1 and supplier 2 as $x_1$ and $x_2$, respectively when $p_1=p_1=p>0$.  Then we discuss two cases (i) and (ii).
	
	\begin{itemize}

		\item (i) When $D<  y_1^*({p},\varphi_1)+y_2^*({p},\varphi_2)$, we have
		\begin{align}
		x_1+x_2=
		D.\label{eq:proof_ndev}
		\end{align}

		In this case, if  supplier 1  reduces the price by a small $\varepsilon_1>0$ to a price $p_1'=p-\varepsilon_1$ unilaterally, we have 
		\begin{align}
		&x_1'\triangleq {x}_1^*(({p}-\varepsilon_1,{p}),(y_1^*({p-\varepsilon_1},\varphi_1),y_2^*({p},\varphi_2))= \min \left\{D, y_1^*({p-\varepsilon_1},\varphi_1)\right\}.
		\end{align}
		
		If  supplier $2$  reduces the price by a small $\varepsilon_2>0$ to a price $p_2'=p-\varepsilon_2$ unilaterally , we have 
		\begin{align}
		x_2'\triangleq{x}_2^*(({p},{p}-\varepsilon_2),(y_1^*({p},\varphi_1),y_2^*({p-\varepsilon_2},\varphi_2))=\min \left\{D, y_2^*(p-\varepsilon_2,\varphi_2)\right\}.
		\end{align} 
		We choose  small $\varepsilon_1$ and $\varepsilon_2$ such that $D<  y_1^*({p}-\varepsilon_1,\varphi_1)+y_2^*({p}-\varepsilon_2,\varphi_2)$  holds.
		Then, we have
		\begin{align}	
		x_1'+x_2'=\min \left\{D, y_1^*({p-\varepsilon_1},\varphi_1)\right\}+\min \left\{D, y_2^*(p-\varepsilon_2,\varphi_2)\right\}.\label{eq:proof_dev1}
		\end{align}
		Combining \eqref{eq:proof_ndev} and \eqref{eq:proof_dev1}, we see that at least one supplier $i$ can always reduce the price by a small $\varepsilon_i>0$ unilaterally such that the selling quantity increases by  $$x_i'-x_i>\frac{1}{2}\min(D, y_1^*({p-\varepsilon_1},\varphi_1),y_2^*({p-\varepsilon_2},\varphi_2) ).$$ Since we can choose a sufficiently small $\varepsilon_i, \forall i=1,2$, the revenue $\pi_i$ will increase  due to the  increased selling quantity $x_i'-x_i$ (with an upward jumping).

		\item (ii) When $  y_1^*({p},\varphi_1)+y_2^*({p},\varphi_2)\leq D< y_1^*(\bar{p},\varphi_1)+y_2^*(\bar{p},\varphi_2)$, we have 
		\begin{align}
		x_1+x_2=y_1^*({p},\varphi_1)+y_2^*({p},\varphi_2).\label{eq:ndev2}
		\end{align} 
		Both suppliers can sell out the bidding  quantities completely as follows.
		\begin{align}
		x_1=y_1^*({p},\varphi_1),~ x_2=y_2^*({p},\varphi_2).\label{eq:ndev3}
		\end{align} 
		
		Note that    $D-y_2^*({p},\varphi_2)\geq y_1^*({p},\varphi_1)=x_1$. Supplier $1$ can always increase his price $p$ to $p'=\bar{p}>p$ unilaterally, and $x_1'=\min(y_1^*({p'},\varphi_1), D-y_2^*({p},\varphi_2))$. Since we also have $y_1^*({p'},\varphi_1)\geq y_1^*({p},\varphi_1)=x_1$, supplier $1$'s obtained demand $x_1'$ at $p'$ will not decrease, i.e.,  $x_i'\geq x_i$. Thus, the revenue of supplier  $i$  after increasing the price will also increase.
	\end{itemize}
	
	In summary, when two suppliers bid the same positive price, one supplier can always deviate so as to obtain a higher revenue, which shows that the equal positive bidding prices cannot be pure price equilibrium.
	
	Third, both suppliers bid the price at zero: $p_1=p_2=0$. In this case, both suppliers have zero revenues: $\pi_1^R=\pi_2^R=0$. Note that both without-storage suppliers   will also bid the zero quantity $y_i^*(p_i,\varphi_i)=0$  as shown in Theorem \ref{thm:quantity}. Thus, any supplier $i$  can always set a positive price $p_i'>0$  to obtain the positive demand since the other supplier bid zero quantity. This makes his revenue $\pi_1^{R'}>0$ after increasing the price. There, the pure price strategy $p_1=p_2=0$ cannot be the equilibrium

	So far, for the {case of $0<D<  \sum_i y_i^*(\bar{p},{\varphi}_i)$ }, we have discussed all the three cases  of the pure price strategies  but none of them is an equilibrium. Thus, there exists no pure price equilibrium when $0<D<  \sum_i y_i^*(\bar{p},{\varphi}_i)$.

	\subsubsection{The type $\text{S}_1\text{S}_0$ (i.e., $\sum_i\varphi_i=1$)} Following the same arguments as in the type $\text{S}_0\text{S}_0$, we can first prove that when
	$D \geq \sum_i y_i^*(\bar{p},{\varphi}_i)$,  $p_1=p_2=\bar{p}$ is a pure price equilibrium and show that this pure price equilibrium is unique. Furthermore, we can show that when
	$D < \sum_i y_i^*(\bar{p},{\varphi}_i)$, there exists no pure price equilibrium.
	
	\subsubsection{The type $\text{S}_1\text{S}_1$ (i.e., $\sum_i\varphi_i=2$)}
	The results have been proved in the paper \cite{capacityprice}.
	
	In conclusion, we have Proposition \ref{prop:pureprice} proved.
	
	\subsection{Proof of Theorem \ref{thm:mscdf}}
	We prove Theorem \ref{thm:mscdf} based on Lemma \ref{lem:mix}  that has been shown in \cite{capacityprice}.  However, based on Lemma \ref{lem:mix}, deriving the mixed price equilibrium  in our  model is still not straightforward compared with \cite{capacityprice}. That is because,  in \cite{capacityprice}, supplier's bidding quantity  is upper-bounded by his deterministic production quantity, while in our model, without-storage supplier's bidding quantity is upper-bounded by a function of price. The difference significantly increases the complexity of the analysis in our work. 
	
	%
	%
	
	{ To prove Theorem \ref{thm:mscdf}, we will utilize a basic property of mix strategy equilibrium as shown in Lemma 2\cite{mic}. In Lemma 2, we  use  $\pi_i^{RM}(\mu_i,\mu_{-i},\bm{\varphi})$ to   denote the expected revenue of supplier $i$ at any arbitrary mixed price strategy $(\mu_1,\mu_2)$, which is defined as follows.}
	\begin{align*}
	\pi_i^{RM}(\mu_i,\mu_{-i},\bm{\varphi})\hspace{-0.9mm}=\int_{{[0,\bar{p}]}^{2}}\hspace{-0.6mm}\pi_i^R (p_i,  {x}_i^*((p_i, {p}_{-i}),\bm{y}^*(p_i, p_{-i})),\bm{\varphi}) d (\mu_i(p_i) \hspace{-0.6mm}\times\hspace{-0.6mm} {\mu}_{-i}({p}_{-i}) )
	\end{align*}
	
	\begin{lemma}\label{lemma:mix2}
	 $\pi_i^{RM}(p_i,\mu_{-i}^*,\boldsymbol{\varphi}) = \pi_i^{RE}(\boldsymbol{\varphi})$, for all $p_i\in[l,\bar{p}]$, where $\pi_i^{RE}$ is the equilibrium revenue \cite{mic}.
	\end{lemma}

	{Lemma \ref{lemma:mix2} shows that the equilibrium revenue $\pi_i^{RE}$ of supplier $i$ is equal to  the expected revenue when he plays any pure strategy $p_i$ in the support, i.e., $p_i\in[l,\bar{p}]$,  against the mixed  strategy $\mu_{-i}^*$ of the other supplier  at the equilibrium.}

	Based on Lemma \ref{lem:mix} and Lemma \ref{lemma:mix2}, we will characterize the equilibrium revenue $\pi_i^{RE}$  as well as the CDF of the mixed price equilibrium  $F_i^e(p)$  using the lower support $l$ over $p\in[l,\bar{p})$.\footnote{Note that $F_2^e(p)$ may not be continuous at $p=\bar{p}$ as indicated in Lemma \ref{lem:mix}. } We make the analysis of the with-storage supplier (i.e., $\varphi_i=1$) and without-storage supplier (i.e., $\varphi_i=0$) as follows.
	
	\subsubsection{With-storage supplier $i$ (i.e.,  $\varphi_i=1$)} For supplier $i$,  based on Lemma 2, the equilibrium revenue $\pi_i^{RE}$  can be characterized  by  the expected revenue when he plays any pure strategy $p_i \in [l,\bar{p})$  against the mixed  strategy of supplier $-i$ (with CDF $F_{-i}^e$ and PDF $f_{-i}^e$) at the equilibrium as follows
	\begin{align}
	\pi_i^{RE}(\bm{\varphi})&=\pi_i^{RM}(p_i,\mu_{-i}^*,\bm{\varphi})\notag \\&=\underbrace{p_i\min(D,\mathbb{ E }[X_i])\cdot(1-F_{-i}^e(p_i))}_{p_i\leq p_{-i}} \notag\\&~~~~+
	\underbrace{p_i\int_{l}^{p_i}\min \left(D-\min(y_{-i}^*(p_{-i},\varphi_{-i}),D),\mathbb{ E }[X_i]\right)\cdot f_{-i}^e(p_{-i})dp_{-i}}_{p_i> p_{-i}}\label{eq:par1o}.
	\end{align}
		Note that in \eqref{eq:par1o}, $D-\min(y_{-i}^*(p_{-i},\varphi_{-i}),D)\leq \mathbb{ E }[X_i]$ will always hold for any $ p_{-i}\in [l,\bar{p}]$, i.e., $D-\min(y_{-i}^*(l,\varphi_{-i}),D)\leq \mathbb{ E }[X_i]$, or
	\begin{align}
	D\leq \mathbb{ E }[X_i]+y_{-i}^*(l,\varphi_{-i}).
	\end{align}
	
	\noindent This helps us simplify the second part ``$p_i> p_{-i}$" in \eqref{eq:par1o}. We can prove this by contradiction  as follows. If $D-\min(y_{-i}^*(l,\varphi_{-i}),D)> \mathbb{ E }[X_i]$, there exists a small  $\varepsilon>0$ such that $D-\min(y_{-i}^*(l+\varepsilon,\varphi_{-i}),D)> \mathbb{ E }[X_i]$ still holds. Based on \eqref{eq:par1o}, we have 
	\begin{align}
	&\pi_i^{RE}=\pi_i^{RM}(l,\mu_{-i}^*,\bm{\varphi})={l\cdot \min(D,\mathbb{ E }[X_i])},\label {eq:pif2}
	\end{align} 
	and for any $\varepsilon>0$,
	\begin{align}
	\pi_i^{RE}(\bm{\varphi})&=\pi_i^{RM}(l+\varepsilon,\mu_{-i}^*,\boldsymbol{\varphi})\notag \\&={(l+\varepsilon)\cdot \min(D,\mathbb{ E }[X_i]) (1-F_{-i}^e(l+\varepsilon))}+{(l+\varepsilon)\int_{l}^{l+\varepsilon} \mathbb{ E }[X_i]\cdot f_{-i}^e(p_{-i})dp_{-i}}\notag \\
	&={(l+\varepsilon)\cdot \min(D,\mathbb{ E }[X_i]) (1-F_{-i}^e(l+\varepsilon))}+{(l+\varepsilon)\cdot \mathbb{ E }[X_i]\cdot F_{-i}^e(l+\varepsilon)}\notag \\
	&\geq (l+\varepsilon)\cdot \min(D,\mathbb{ E }[X_i])\label{eq:pif1}.
	\end{align}
	Then, we can see that  $\eqref {eq:pif2}$ and $\eqref {eq:pif1}$ contradict with each other, and thus $D-\min(y_{-i}^*(p_{-i},\varphi_{-i}),D)\leq \mathbb{ E }[X_i]$ will always hold for $p_2\in [l,\bar{p}]$, which enables us to simplify \eqref{eq:par1o}.
	
	Since $\pi_i^{RE}(\bm{\varphi})=\pi_i^{RM}(p_i,\mu_{-i}^*,\bm{\varphi})$ is constant over $p_i\in [l,\bar{p})$, the derivative of $\pi_i^{RM}(p_i,\mu_{-i}^*,\bm{\varphi})$  with respect to $p_i$ is zero over $p_i\in [l,\bar{p})$, i.e.,
	\begin{align}
	\frac{\partial \pi_i^{RM}(p_i,\mu_{-i}^*,\bm{\varphi})}{\partial p_i}=&{\min(D,\mathbb{ E }[X_i])(1-F_{-i}^e(p_i))}+p_i\min(D,\mathbb{ E }[X_i])(-f_{-i}^e(p_i))\notag\\
	+&{\int_{l}^{p_i} \left(D-\min(y_{-i}^*(p_{-i}),D)\right)f_{-i}^e(p_{-i})dp_{-i}}
	+{{p_i} \left(D-\min(y_{-i}^*(p_i,\varphi_{-i}),D)\right)f_{-i}^e(p_i)} \label{eq:par1} \notag\\
	=&0.
	\end{align}
	
	Combining \eqref{eq:par1} with \eqref{eq:par1o}, we have the PDF of mixed  price strategy at the equilibrium for  without-storage supplier ${-i}$'s  as follows.
	\begin{align}
	&f_{-i}^e(p)= \frac{\pi_i^{RE}(\bm{\varphi})}{p^2\cdot \min\{y_{-i}^*(p,\varphi_{-i}),D\}-p^2\cdot [D-\mathbb{E}{[X_i]}]^+},\label{eq: mf2}
	\end{align}
	which is  characterized by the equilibrium revenue $\pi_i^{RE}$ of supplier $i$.
	
	\subsubsection{Without-storage supplier $i$ (i.e.,  $\varphi_i=0$)} For supplier $i$ without storage, similarly,  based on Lemma 2,  the  equilibrium revenue $\pi_i^{RE}(\bm{\varphi})$ can be characterized by the expected revenue when he plays any pure strategy $p_i \in [l,\bar{p})$   against the mixed  strategy of supplier ${-i}$ (with CDF $F_{-i}^e$) at the equilibrium as follows
	\begin{align}
	\pi_i^{RE}(\bm{\varphi})\hspace{-0.5mm}=&\hspace{-0.5mm}\pi_i^{RM}(p_i,\mu_{-i}^*,\bm{\varphi})\hspace{-0.5mm}\\=&\hspace{-0.5mm}\underbrace{\pi_i^R(p_i,\min \left(D, y_i^*(p_i,\varphi_i),\bm{\varphi}\right) )\cdot (1-F_{-i}^e(p_i))}_{p_i\leq p_{-i}} \notag\\&\hspace{-4mm}+
	\underbrace{\pi_i^R\left (p_i, \min(D-\min(\mathbb{ E }[X_{-i}],D),y_i^*(p_i,\varphi_i)),\bm{\varphi}\right) \cdot F_{-i}^e(p_i)}_{p_i>p_{-i}}. \label{eq: mpi2}
	\end{align}
	Similarly,  we have that $D-\min(\mathbb{ E }[X_{-i}],D)\leq y_i^*(p_i,\varphi_i)$ always holds for any $ p_i\in [l,\bar{p}]$.
	Then, according to \eqref{eq: mpi2},  we have the PDF of the mixed price strategy at the equilibrium for the  with-storage supplier $-i$ as follows.
	\begin{align}
	&F_{-i}^e(p)= \frac{ \pi_i^R\left(p,\min\{y_i^*(p,\varphi_i),D\},\bm{\varphi}\right)-\pi_i^{RE}(\bm{\varphi})}{\pi_i^R\left(p,\min\{y_i^*(p,\varphi_i),D\},\bm{\varphi}\right)-\pi_i^R\left(p, [D-\mathbb{ E }[X_{-i}]]^+,\bm{\varphi}\right)}\label{eq: mf1},
	\end{align}
	which is  characterized by the equilibrium revenue $\pi_{i}^{RE}$ of supplier $i$.
	
	In conclusion, if $\varphi_i=1$, we have
	\begin{align}
	&F_i^e(p)= \frac{ \pi_{-i}^R\left(p,\min\{y_{-i}^*(p,\varphi_{-i}),D\},\bm{\varphi}\right)-\pi_{-i}^{RE}(\bm{\varphi})}{\pi_{-i}^R(p,\min\{y_{-i}^*(p,\varphi_{-i}),D\},\bm{\varphi})-\pi_{-i}^R(p, (D-\mathbb{ E }[X_i])^+,\bm{\varphi})}.
	\end{align}
	
	If $\varphi_i=0$, we have
	\begin{align}
	&F_i^e(p)=\int_l^{\bar{p}} \frac{\pi_{-i}^{RE}(\bm{\varphi})}{p^2\cdot \min\{y_i^*(p,\varphi_{i}),D\}-p^2\cdot (D-\mathbb{ E }[X_{-i}])^+}dp.
	\end{align}
	for any $l \leq p< \bar{p}$. 
	
	\qed
	
	\subsection{Proof of Proposition \ref{thm:mscomp}} 
	
	To prove Proposition \ref{thm:mscomp}, we first show that $F_i^e(\bar{p}^-\mid l_i^\dagger)$ is always decreasing in $l_i^\dagger,\forall i$, based on which we can prove Proposition  \ref{thm:mscomp} (1) by contradiction. Then, we can have Proposition \ref{thm:mscomp} (2) proved directly from Lemma \ref{lem:mix} (iii).
	
	We now prove that $F_i^e(\bar{p}^-\mid l_i^\dagger)$ is always decreasing with $l_i^\dagger$, for both $\varphi_i=1$ and $\varphi_i=0$. 
	\subsubsection{With-storage supplier $i$ (i.e, $\varphi_i=1$)}
	For the without-storage supplier $i$, according to \eqref{F1}, we have
	\begin{align}
	&F_i^e(\bar{p}^-\mid l_i^\dagger)= \frac{ \pi_{-i}^R\left(\bar{p},\min\{y_{-i}^*(\bar{p},\varphi_{-i}),D\},\bm{\varphi}\right)-\pi_{-i}^{RE}(\bm{\varphi})}{\pi_{-i}^R(\bar{p},\min\{y_{-i}^*(p,\varphi_{-i}),D\},\bm{\varphi})-\pi_{-i}^R(\bar{p}, (D-\mathbb{ E }[X_i])^+,\bm{\varphi})}.\label{FF1}
	\end{align}
	
	Note that the equilibrium revenue function $\pi_{-i}^{RE}(\bm{\varphi})$ (shown in Lemma \ref{lem:mix} (iii)) is increasing in the lower support  $l_i^\dagger$, and thus $F_i^e(\bar{p}^-\mid l_1^\dagger)$ is decreasing in $l_i^\dagger$. 
	
	\subsubsection{Without-storage supplier $i$ (i.e, $\varphi_i=0$)} For the without-storage supplier $i$, according to \eqref{F2}, we have
	\begin{align}
	&F_i^e(\bar{p}^-\mid l_i^\dagger)=\int_{l_i^\dagger}^{\bar{p}} \frac{{l_i^\dagger}\cdot\min(D,\mathbb{ E }[X_i])}{p^2\cdot \min\{y_i^*(p,\varphi_i),D\}-p^2\cdot [D-\mathbb{ E }[X_i]]^+}dp.
	\end{align}
	We take the first-order derivative of $F_i^e(\bar{p}^-\mid l_i^\dagger)$ with respect to $l_i^\dagger$ and obtain
	\begin{align}
	\frac{\partial  F_i^e(\bar{p}^-\mid l_i^\dagger)}{\partial l_i^\dagger}&= \int_{l_i^\dagger}^{\bar{p}} \frac{\min(D,\mathbb{ E }[X_i])}{p^2\cdot \min\{y_i^*(p,\varphi_i),D\}-p^2\cdot (D-\mathbb{ E }[X_i])^+}dp\notag \\ &~~~~~~~-\frac{\min(D,\mathbb{ E }[X_i])}{{l_i^\dagger}\cdot \min\{y_i^*(l_i^\dagger,\varphi_i),D\}-{l_i^\dagger}\cdot (D-\mathbb{ E }[X_i])^+}.\label{ss}
	\end{align}
	Further, we take the derivative of \eqref{ss} with respect to $l_i^\dagger$ again and have
	\begin{align}
	\frac{\partial^2  F_i^e(\bar{p}^-\mid l_i^\dagger)}{\partial {l_i^\dagger}^2}=- \frac{1}{{l_i^\dagger}}\cdot \frac{\partial\frac{\min(D,\mathbb{ E }[X_i])}{\min\{y_i^*(l_i^\dagger,\varphi_i),D\}- (D-\mathbb{ E }[X_i])^+}}{\partial l_i^\dagger}.
	\end{align}
	Note that $\frac{\min(D,\mathbb{ E }[X_i])}{\min\{y_i^*(l_i^\dagger,\varphi_i),D\}- (D-\mathbb{ E }[X_i])^+}$ decreases in $l_i^\dagger$ because $y_i^*(l_i^\dagger,\varphi_i)$ increases in $l_i^\dagger$. Thus, we always have
	\begin{align}
	\frac{\partial^2 F_i^e(\bar{p}^-\mid l_i^\dagger)}{\partial {l_i^\dagger}^2}\geq 0,
	\end{align}
	which shows that $\frac{\partial  F_i^e(\bar{p}^-\mid l_i^\dagger)}{\partial l_i^\dagger}$ is non-decreasing with $l_i^\dagger$. Then, we choose $l_i^\dagger=\bar{p}$ and  have
	\begin{align}
	\frac{\partial  F_i^e(\bar{p}^-\mid l_i^\dagger)}{\partial l_i^\dagger}&=-\frac{\min(D,\mathbb{ E }[X_i])}{{\bar{p}}\cdot \min\{y_i^*(\bar{p},\varphi_i),D\}-{\bar{p}}\cdot (D-\mathbb{ E }[X_i])^+}\notag \\&<0,\notag 
	\end{align}
	which holds for all $l_i^\dagger\leq \bar{p}$. The reason is that $D<\mathbb{ E }[X_i]+y_i^*(\bar{p},\varphi_i)$ in the subgame $\text{S}_1\text{S}_0$ without the pure price equilibrium. Therefore,  we have that ${ F_i^e(\bar{p}^-\mid l_i^\dagger)}$ decreases with $l_i^\dagger$.
	
	Till now, we have shown that $F_i^e(\bar{p}^-\mid l_i^\dagger)$ is always decreasing in $l_i^\dagger$ for both $\varphi_i=1$ and $\varphi_i=0$. Then, we can prove Proposition  \ref{thm:mscomp} (1) by contradiction. According to Lemma \ref{lem:mix} (iii), if  $F_i^e(\bar{p}^-\mid l_i^\dagger)=1$ has a  solution $l_i^{\dagger*}$ for both suppliers $i=1,2$, either  $l=\max (l_1^{\dagger*},l_2^{\dagger*})$ or  $l=\min (l_1^{\dagger*},l_2^{\dagger*})$ will hold. If $l=\min (l_1^{\dagger*},l_2^{\dagger*})$, without loss of generality,  we assume $l_1^{\dagger*}< l_2^{\dagger*}$ and $l=l_1^{\dagger*}$. Note that $F_1^e(\bar{p}^-\mid l_1^{\dagger*})=1$ and hence $F_2^e(\bar{p}^-\mid l_2^{\dagger*})=1$. Since $F_2^e(\bar{p}^-\mid l_2^\dagger)$ is decreasing with $l_2^\dagger$, then $F_2^e(\bar{p}^-\mid l_1^{\dagger*})>1$, which is a contradiction of the CDF. Therefore, we can only choose $l=\max (l_1^{\dagger*},l_2^{\dagger*})$ and we have Proposition  \ref{thm:mscomp} (1) proved. Furthermore, according to  Lemma \ref{lem:mix} (iii), we have that $F_i^e(\bar{p}^-)=1$ ~\text{is true for at least one of the suppliers. } Thus, if we have only one  solution of $l_i^{\dagger}$ among $i=1$ and  $i=2$, it  must be the equilibrium lower support, which has Proposition  \ref{thm:mscomp} (2) proved. \qed

	\subsection{Proof of Theorem \ref{prop:comparison}} 
	
	We first prove that	
	$\pi_i^{RE}>\pi_{-i}^{RE}$ always holds for a general distribution for the renewable generation $X_i$ if $\varphi_i=1$, $\varphi_{-i}=0$ and  $\mathbb{E}[X_i]=\mathbb{E}{[X_{-i}]}$. Then, we consider the case  that $X_{-i}$  follows a uniform distribution.
	
	\subsubsection{A general distribution for $X_i$} We consider the cases of pure price equilibrium and  mixed price equilibrium respectively, and   characterize suppliers' revenue as follows.

	(a) The case with pure price equilibrium: According to Proposition \ref{prop:pureprice}	and Lemma \ref{lem:mix} (ii), we have	
	\begin{equation}
	\pi_i^{RE}(\bm{\varphi})= \left \{
	\begin{aligned}
	&\bar{p}\min(\mathbb{ E }[X_i],D),~\text{if}~\varphi_i=1,\\
	&\pi_i^R(\bar{p},\min(D,y_i^*(\bar{p},\varphi_{i})),\bm{\varphi}),~\text{if}~\varphi_i=0.
	\end{aligned}
	\right.
	\end{equation}	 
	Note that $D\geq \mathbb{ E }[X_i]+y_i^*(\bar{p},\varphi_{i})$ when there is the pure price  equilibrium.	Therefore, if $\varphi_i=1$ and $\varphi_{-i}=0$, we have 
	\begin{align}
	\pi_i^{RE}(\bm{\varphi})&=\bar{p}\mathbb{ E }[X_i]. \label{eq:proof_c1}\\
	\pi_{-i}^{RE}(\bm{\varphi})&= \pi_{-i}^R(\bar{p},y_{-i}^*(\bar{p},\varphi_{-i}),\bm{\varphi})\\
	&=\lambda \int_0^{F_{-i}^{-1}(\frac{\bar{p}}{\lambda})}xf_{-i}(x)dx.\\
	&=\lambda \int_0^{F_{-i}^{-1}(\frac{\bar{p}}{\lambda})}xdF_{-i}(x)\\
	&=\bar{p}F_{-i}^{-1}(\frac{\bar{p}}{\lambda})-\lambda \int_0^{F_{-i}^{-1}(\frac{\bar{p}}{\lambda})}F_{-i}(x)dx\label{eq:proof_mcom1}\\
	&< \bar{p}F_{-i}^{-1}(\frac{\bar{p}}{\lambda})-\bar{p} \int_0^{F_{-i}^{-1}(\frac{\bar{p}}{\lambda})}F_{-i}(x)dx. \label{eq:proof_mcom}
	\end{align}
	
	Based on \eqref{eq:proof_mcom},  we consider the following function $h(x)$ for any $p>0$ and $0\leq x<\bar{X}_{-i}$. Note that $F_{-i}^{-1}(\frac{\bar{p}}{\lambda})<\bar{X}_{-i}$ since $\bar{p}<\lambda$.
	\begin{align}
	h(x)={p}x-p \int_0^{x}F_{-i}(x)dx. \label{eq:proof_pro3}
	\end{align}
	The, we have 
	\begin{align}
	h'(x)={p}-p F_{-i}(x)>0,
	\end{align}
	which shows that $h(x)$ increases in $x$. Since $F_{-i}^{-1}(\frac{\bar{p}}{\lambda})< \bar{X}_{-i}$, according to \eqref{eq:proof_mcom}, we have
	\begin{align}
	\pi_{-i}^{RE}(\bm{\varphi})
	&<\bar{p}\bar{X}_{-i}-\bar{p} \int_0^{\bar{X}_{-i}}F_{-i}(x)dx\\
	&=\bar{p}\mathbb{ E }[X_{-i}]\leq	\pi_i^{RE}(\bm{\varphi})  \label{eq:proof_c2}.
	\end{align}
	Based on \eqref{eq:proof_c1} and \eqref{eq:proof_c2}, if $\mathbb{ E }[X_{-i}] \leq \mathbb{ E }[X_{i}]$, then we always have 
	\begin{align}
	\pi_{-i}^{RE}(\bm{\varphi})<	\pi_i^{RE}(\bm{\varphi}). \label{eq:proof_rev}
	\end{align}
	
	(b) The case without pure price equilibrium: The proof procedure is the similar to the case (a)  with pure price equilibrium. The difference is to replace $\bar{p}$ into the lower support $l$, i.e., 
	
	\begin{equation}
	\pi_i^{RE}(\bm{\varphi})= \left \{
	\begin{aligned}
	&l\cdot \min(\mathbb{ E }[X_i],D),~\text{if}~\varphi_i=1,\\
	&\pi_i^R(l,\min(D,y_i^*(l,\varphi_{i})),\bm{\varphi}),~\text{if}~\varphi_i=0.
	\end{aligned}
	\right.
	\end{equation}
	
	We will discuss the following two cases.
	\begin{itemize}
		\item $\mathbb{ E }[X_i]\leq D$: If $\varphi_i=1$ and $\varphi_{-i}=0$, we have
		\begin{align}
			\pi_i^{RE}(\bm{\varphi})&= l\cdot \mathbb{ E }[X_i].\\
		\pi_{-i}^{RE}(\bm{\varphi})&=\pi_{-i}^R(l,\min(D,y_{-i}^*(l,\varphi_{-i})),\bm{\varphi})\\&\leq \pi_{-i}^R(l,y_{-i}^*(l,\varphi_{{-i}}),\bm{\varphi}).
		\end{align} 
		
		We can follow the same argument as in (a) with the pure price equilibrium to show that $\pi_{i}^{RE}> \pi_{-i}^{RE}$ if $\mathbb{ E }[X_i]\geq \mathbb{ E }[X_{-i}]$. The only difference is to replace $\bar{p}$ by $l$.
		
		\item $\mathbb{ E }[X_i]> D$: If $\varphi_i=1$ and $\varphi_{-i}=0$, we have
		\begin{align}
			\pi_i^{RE}(\bm{\varphi})&= l\cdot D.\\
		\pi_{-i}^{RE}(\bm{\varphi})&=\pi_{-i}^R(l,\min(D,y_{-i}^*(l,\varphi_{-i})),\bm{\varphi}).
		\end{align} 
		
		\begin{itemize}
			\item If $y_{-i}^*(l,\varphi_{{-i}})\leq D$, we have
			\begin{align}
			\pi_{-i}^{RE}(\bm{\varphi})&=\pi_{-i}^R(l,y_{-i}^*(l,\varphi_{-i}),\bm{\varphi})\\
			&\leq ly_{-i}^*(l,\varphi_{{-i}})-l \int_0^{y_{-i}^*(l,\varphi_{{-i}})}F_{-i}(x)dx~(\text{as in}~ \eqref{eq:proof_mcom})\\
			&<lD\\
			&=	\pi_i^{RE}(\bm{\varphi}).
			\end{align} 
			\item If $y_{-i}^*(l,\varphi_{{-i}})> D$, we have
			\begin{align}
			\pi_{-i}^{RE}(\bm{\varphi})&=\pi_{-i}^R(l,D,\bm{\varphi})\\
			&= lD-\lambda \int_0^{D}(D-x)f_{-i}(x)dx\\
			&< lD\\
			&=	\pi_i^{RE}(\bm{\varphi}).
			\end{align} 
			
		\end{itemize}
	\end{itemize}
	
	Therefore, for the case (b) without pure price equilibrium, we also have that  $\pi_{i}^{RE}> \pi_{-i}^{RE}$ if $\mathbb{ E }[X_i]\geq \mathbb{ E }[X_{-i}]$. Combining case (a) with the pure price equilibrium, for a general distribution of $X_i$,  we prove that  $\pi_{i}^{RE}> \pi_{-i}^{RE}$ if $\mathbb{ E }[X_i]\geq \mathbb{ E }[X_{-i}]$.
	
	\subsubsection{Uniform distribution of $X_{-i}$}
We will derive the revenues (at both pure and mixed price equilibrium) of suppliers under the uniform renewable-generation distribution. For the pure price equilibrium, 	it is straightforward to calculate the equilibrium revenue when there is $p_1=p_2=\bar{p}$  when  $D\geq \sum_i y_i(\bar{p},\varphi_i)$. For the case without pure price equilibrium, i.e., $D< \sum_i y_i(\bar{p},\varphi_i)$, 	 we will characterize the lower support for the mixed price equilibrium and characterize the  equilibrium revenue  based on Theorem \ref{thm:mscdf} and Proposition  \ref{thm:mscomp}.  
	
	We consider  $\varphi_i=1$ and $\varphi_{-i}=0$.   We have the PDF and CDF of the uniform distribution $X_{-i}$ as follows:
	\begin{align}
	f_{-i}= \frac{1}{\bar{X}_{-i}}, ~F_{-i}(x)= \frac{x}{\bar{X}_{-i}}.
	\end{align}
	According to Theorem \ref{thm:quantity}, the weakly dominant bidding quantity strategy is 
	\begin{align}
	&y_i^*=\mathbb{ E }[{X}_i],\\& y_{-i}^*(p_{-i},\varphi_{{-i}})=F_{-i}^{-1}\left(\frac{p_{-i}}{\lambda}\right)=\frac{p_{-i}}{\lambda}\bar{X}_{-i}.
	\end{align}
	
	Next we discuss the case (a) with pure price equilibrium and the case (b) without pure price equilibrium respectively.

	(a) The case with pure price equilibrium:  When $D \geq   \sum_i y_i(\bar{p},\varphi_i)$, both suppliers' bid price $\bar{p}$  and we have
	\begin{align}
	&\pi_i^{RE}(\bm{\varphi})=\bar{p}\mathbb{ E }[X_i],\\ &\pi_{-i}^{RE}(\bm{\varphi})=\pi_{-i}^R(\bar{p},y_{-i}^*(\bar{p},\varphi_{-i}),\bm{\varphi})=\frac{\bar{X}_{-i}}{2\lambda}\bar{p}^2,
	\end{align}
	which leads to the revenue ratio: 
	\begin{align}
	&\frac{\pi_i^{RE}(\bm{\varphi})}{\pi_{-i}^{RE}(\bm{\varphi})}=\frac{\lambda \mathbb{ E }[X_i]}{\mathbb{ E }[X_{-i}] \bar{p}}.
	\end{align}
	If $\mathbb{E}[X_i] \geq \mathbb{E}[X_{-i}]$, then
	\begin{align}
	&\frac{\pi_i^{RE}(\bm{\varphi})}{\pi_{-i}^{RE}(\bm{\varphi})}\geq \frac{\lambda }{ \bar{p}}.
	\end{align}
	
	(b) The case without pure price equilibrium: When $D <   \sum_i y_i(\bar{p},\varphi_i)$, based on the characterization of CDF in Theorem \ref{thm:mscdf},  we discuss the following cases respectively.
	
	\begin{itemize}
		\item Case of $0<D \leq  \mathbb{ E }[X_i]$: According to Theorem \ref{thm:mscdf}, we have the CDF of the mixed equilibrium price over $p\in [l,\bar{p})$ as follows:
		\begin{align}
		&F_i^e(p)= \frac{ \pi_i^R\left(p,\min\{y_{-i}^*(p,\varphi_{-i}),D\},\bm{\varphi}\right)-\pi_{-i}^{RE}(\bm{\varphi})}{\pi_i^R\left(p,\min\{y_{-i}^*(p,\varphi_{-i}),D\},\bm{\varphi}\right)},\label{eq:F1c1}\\
		&F_{-i}^e(p)=\int_l^{\bar{p}} \frac{\pi_{i}^{RE}(\bm{\varphi})}{p^2\cdot \min\{y_{-i}^*(p,\varphi_{-i}),D\}}dp.
		\end{align}

		We can see that  $F_i^e(p)<1 $ over $p\in[l,\bar{p})$ since  $\pi_{-i}^{RE}(\bm{\varphi})>0$.\footnote{Note that   $\pi_{-i}^{RE}(\bm{\varphi})>0$ since the lower support  $l>0$.} According to Proposition \ref{thm:mscomp}, we solve the following equation to derive the equilibrium lower support $l$. 
		\begin{align}
		F_{-i}^e(\bar{p})=\int_l^{\bar{p}} \frac{\pi_i^{RE}(\bm{\varphi})}{p^2\cdot \min\{y_{-i}^*(p,\varphi_{-i}),D\}}dp=1.
		\end{align}
		We discuss the following two cases
		\begin{itemize}
			\item 1) If $D\geq y_{-i}^*(\bar{p},\varphi_{-i})$, we have
			\begin{align}
			&l=\frac{\bar{p}^2\bar{X}_{-i}}{D\lambda}(-1+\sqrt{1+\frac{D^2\lambda^2}{\bar{p}^2\bar{X}_{-i}^2}}).
			\end{align}

	\item 2) If $D< y_{-i}^*(\bar{p},\varphi_{-i})$, we have
		\begin{align}
		&l=\frac{D\lambda}{\bar{X}_{-i}(1+\sqrt{2\frac{D\lambda}{\bar{p}\bar{X}_{-i}}})}.
		\end{align}
			\end{itemize}
		
		We verify that in both cases (1) and (2),  $\min(D,y_{-i}^*(l,\varphi_{-i}))=y_{-i}^*(l,\varphi_{-i})$. According to Lemma \ref{lem:mix}, the  equilibrium revenue of both suppliers will be 
		\begin{align}
		&\pi_i^{RE}(\bm{\varphi})=l\cdot (D,\mathbb{ E }[X_i])=l\cdot D,\\
		&\pi_{-i}^{RE}(\bm{\varphi})=\pi_{-i}^{R}(l,\min(D,y_{-i}^*(l,\varphi_{-i})),\bm{\varphi})=\pi_{-i}^{R}(l,y_{-i}^*(l,\varphi_{-i}),\bm{\varphi})=\frac{\bar{X}_{-i}}{2\lambda}l^2,
		\end{align}
		which leads to the revenue ratio:
		\begin{align}
		&\frac{\pi_i^{RE}(\bm{\varphi})}{\pi_{-i}^{RE}(\bm{\varphi})}=\frac{2\lambda D}{l \bar{X}_{-i} }.
		\end{align}
		In summary, we have
		\begin{align}
		&\frac{\pi_i^{RE}(\bm{\varphi})}{\pi_{-i}^{RE}(\bm{\varphi})}=\left \{
		\begin{aligned}
		&2\sqrt{2\frac{D\lambda}{\bar{p}\bar{X}_{-i}}}+2~, ~~~~~~~~\text{if}~\frac{D\lambda}{\bar{p}\bar{X}_{-i}} < 1,\\
		&2\sqrt{1+\frac{D^2\lambda^2}{\bar{p}^2\bar{X}_{-i}^2}}+2 ,~~~~\text{if}~\frac{D\lambda}{\bar{p}\bar{X}_{-i}} \geq 1.
		\end{aligned}
		\right. 
		\end{align}	
		
		Therefore, when $0<D \leq \mathbb{ E }[X_i]$, we have
		
		\begin{itemize}
			\item  when $0<D \leq \mathbb{ E }[X_i]$, $\frac{\pi_i^{RE}(\bm{\varphi})}{\pi_{-i}^{RE}(\bm{\varphi})}\geq 2$;
			\item  when $D = \mathbb{ E }[X_i]$ and $\mathbb{E}[X_{-i}] =\frac{\bar{X}_{-i}}{2}\leq \mathbb{ E }[X_i]$,  $\frac{\pi_i^{RE}(\bm{\varphi})}{\pi_{-i}^{RE}(\bm{\varphi})}\geq 4$ (due to $\lambda/\bar{p}>1$).
		\end{itemize}
		
		\item {Case of $\mathbb{ E }[X_i]<D<  \sum_i y_i(\bar{p},\varphi_i) $}:  We characterize the revenue ratio between the two suppliers  according to Lemma \ref{lem:mix} as follows.
		\begin{align}
		&\pi_i^{RE}(\bm{\varphi})=l \cdot \min (D,\mathbb{ E }[X_i])=l \cdot \mathbb{ E }[X_i],\\
		&\pi_{-i}^{RE}(\bm{\varphi})=\pi_{-i}^{R}(l,\min(D,y_{-i}^*(l,\varphi_{-i})),\bm{\varphi})\leq \pi_{-i}^{R}(l,y_{-i}^*(l,\varphi_{-i}),\bm{\varphi})=\frac{\bar{X}_{-i}}{2\lambda}l^2.
		\end{align}
		Then, we have 
		\begin{align}
		&\frac{\pi_i^{RE}(\bm{\varphi})}{\pi_{-i}^{RE}(\bm{\varphi})}\geq \frac{l \cdot \mathbb{ E }[X_i]}{ \pi_{-i}^{R}(l,y_{-i}^*(l,\varphi_{-i}),\bm{\varphi})}=\frac{2\lambda \mathbb{ E }[X_i]}{l \bar{X}_{-i} }.
		\end{align}
		If $\mathbb{E}[X_{-i}] \leq \mathbb{ E }[X_i]$, then
		\begin{align}
		\frac{\pi_i^{RE}(\bm{\varphi})}{\pi_{-i}^{RE}(\bm{\varphi})}\geq \frac{\lambda }{l}> \frac{\lambda }{ \bar{p}}.
		\end{align}
		
		Therefore, combining case (a), when  $D >  \mathbb{ E }[X_i]$ and $\mathbb{E}[X_{-i}] \geq \mathbb{ E }[X_i]$, we have
		$\frac{\pi_i^{RE}(\bm{\varphi})}{\pi_{-i}^{RE}(\bm{\varphi})}\geq  \frac{\lambda }{ \bar{p}}.$
		
	\end{itemize}
	
	Finally, combining (a) and Subsection (b), we have Theorem \ref{prop:comparison} proved. \qed
	
		\subsection{Proof of Proposition \ref{prop:positiverev}}
	
	We will discuss the equilibrium revenue with pure price equilibrium and  without pure price equilibrium, respectively.
	\subsubsection{With the pure price equilibrium (i.e., $D\geq \sum_iy_i^*(\bar{p},\varphi_{i})$}: If $\varphi_i=1$, we have
	\begin{align}
	\pi_i^{RE}(\bm{\varphi})&=\bar{p}\mathbb{ E }[X_i]>0.
	\end{align}
	If $\varphi_i=0$, we have
	\begin{align}
	\pi_{i}^{RE}(\bm{\varphi})&= \pi_{i}^R(\bar{p},y_{i}^*(\bar{p},\varphi_{i}),\bm{\varphi})\\
	&=\lambda \int_0^{F_{-i}^{-1}(\frac{\bar{p}}{\lambda})}xf_i(x)dx\\
	&>0.
	\end{align}
	
	\subsubsection{Without the pure price equilibrium (i.e., $D<\sum_iy_i^*(\bar{p},\varphi_{i})$}:If $\varphi_i=1$, due to the lower support $l>0$, we have
	\begin{align}
	\pi_i^{RE}(\bm{\varphi})&=l \min(D,\mathbb{ E }[X_i]) >0.
	\end{align}
	If $\varphi_i=0$, due to the lower support $l>0$, we have
	\begin{align}
	\pi_{i}^{RE}(\bm{\varphi})&= \pi_{i}^R(l,\min(D,y_{i}^*(l,\varphi_{i})),\bm{\varphi})\\
	&>\pi_{i}^R(0,\min(D,y_{i}^*(0,\varphi_{i})),\bm{\varphi})\\
	&=0.
	\end{align}
	
	In conclusion, we have Proposition \ref{prop:positiverev} proved.
	
	\subsection{Proof of Proposition \ref{prop:ns}}
	
	We prove Proposition \ref{prop:ns} by contradiction.
	
	First, we prove $\min_i ~y_i^*(l,\varphi_i)<D$ by contradiction. Suppose that $y_i^*(l,\varphi_i)\geq D$ for both $i=1,2$ and supplier $-i$'s mixed strategy $F_{-i}^e$ has no atom at $\bar{p}$ based on Lemma \ref{lem:mix} (iii). Then, against supplier $-i$'s bidding price $p\in[l,\bar{p})$, according to Proposition \ref{prop:stage3}, supplier $i$'s  selling out electricity quantity at the price $\bar{p}$ is 
	\begin{align}
	{x}_{i}^*(\boldsymbol{p},\boldsymbol{y})=&\min \left\{D-\min \left\{D, y_{-i}^*(p,\varphi_{-i})\right\},y_{i}^*(\bar{p},\varphi_{i})\right\}\\
	=&0.
	\end{align}
	Thus, the equilibrium revenue of supplier $i$ can be characterized as follows 
	\begin{align}
	\pi_i^{RE}(\bm{\varphi})&=\pi_i^{RM}(\bar{p},\mu_{-i}^*,\bm{\varphi})\notag \\&=\bar{p} \int_l^{\bar{p}} 
	{x}_{i}^*(\boldsymbol{p},\boldsymbol{y})\cdot f_{-i}^e(p_{-i})dp_{-i}\\
	&=0. \label{eq:proof_p4}
	\end{align}
	However, at the case of mixed price equilibrium, both suppliers' equilibrium revenue is strictly positive as shown in  Proposition \ref{prop:positiverev}, i.e.,  $\pi_i^{RE}(\bm{\varphi})>0$, which is contradiction to \eqref{eq:proof_p4}. Therefore, we have  $\min_i ~y_i^*(l,\varphi_i)<D$.
	
	Second, we prove $D\leq \sum_i y_i^*(l,\varphi_i)$ by contradiction. Suppose that $D> \sum_i y_i^*(l,\varphi_i)$. Thus, there exists a small  $\varepsilon>0$ such that $D> \sum_i y_i^*(l+\varepsilon,\varphi_i)$ still holds.  Note that $\min_i ~y_i^*(l,\varphi_i)<D$ and we assume that  $ ~y_{-i}^*(l,\varphi_{-i})<D$ without loss of generality. We also let this small $\varepsilon$ satisfy  $ ~y_{-i}^*(l+\varepsilon,\varphi_{-i})<D$.  We can characterize supplier $i$'s  equilibrium revenue using $l$ and $l+\varepsilon$, respectively as follows.
	
	(a) With $l$: 
	\begin{align}
	\pi_i^{RE}(\boldsymbol{\varphi})&=\pi_i^{RM}(l,\mu_{-i}^*,\boldsymbol{\varphi})={l\int_{l}^{\bar{p}} \min(D,y_i^*(l,\varphi_i)) f_{-i}^e(p_{-i})dp_{-i}}.\label{eq:proof_p41}
	\end{align}
	
	(b) With $l+\varepsilon$: 
	\begin{align}
	\pi_i^{RE}(\boldsymbol{\varphi})&=\pi_i^{RM}(l+\varepsilon,\mu_{-i}^*,\boldsymbol{\varphi})\notag \\&={(l+\varepsilon)\cdot\int_{l+\varepsilon}^{\bar{p}} \min(D,y_i^*(l+\varepsilon,\varphi_i)) f_{-i}^e(p_{-i})dp_{-i}}\notag\\&~~~~~~~~~+{(l+\varepsilon)\int_{l}^{l+\varepsilon} \min \left\{D-\min \left\{D, y_{-i}^*(p,\varphi_{-i})\right\},y_{i}^*(l+\varepsilon,\varphi_{i})\right\} \cdot f_{-i}^e(p_{-i})dp_{-i}}\notag \\
	&= {(l+\varepsilon)\int_{l+\varepsilon}^{\bar{p}} \min(D,y_i^*(l+\varepsilon,\varphi_i)) f_{-i}^e(p_{-i})dp_{-i}}+{(l+\varepsilon)\int_{l}^{l+\varepsilon} y_{i}^*(l+\varepsilon,\varphi_{i}) \cdot f_{-i}^e(p_{-i})dp_{-i}} \notag \\
	&>{l\int_{l+\varepsilon}^{\bar{p}} \min(D,y_i^*(l,\varphi_i)) f_{-i}^e(p_{-i})dp_{-i}}+{l\int_{l}^{l+\varepsilon} \min(D,y_i^*(l,\varphi_i))\cdot f_{-i}^e(p_{-i})dp_{-i}} \notag \\
	&={l\int_{l}^{\bar{p}} \min(D,y_i^*(l,\varphi_i)) f_{-i}^e(p_{-i})dp_{-i}}.\label{eq:proof_p42}
	\end{align}
	
	We see that \eqref{eq:proof_p41} and \eqref{eq:proof_p42} contradict with each other. Therefore, we have  $D\leq \sum_i y_i^*(l,\varphi_i)$.
	
	In conclusion, we have Proposition \ref{prop:ns} proved.\qed
	
\vspace{5mm}

\section{Appendix: Proofs of Stage \uppercase\expandafter{\romannumeral1}}\label{appendix:proofstage1}

\subsection{Proof of Theorem \ref{thm:stoeq}}
We prove Theorem  \ref{thm:stoeq} based on  Definition \ref{defi:stoeq} for the storage-investment equilibrium. We first discuss the pure storage-investment equilibrium and then discuss the  mixed storage-investment equilibrium.

First, for the pure price equilibrium, we use the example of the $\text{S}_0\text{S}_0$ case. If the $\text{S}_0\text{S}_0$ case is an equilibrium, each supplier will not be better off if  he deviates to investing in storage, i.e., 
	\begin{align}
	\pi_i^{\text{S}_1\text{S}_0|Y}-C_i\leq 	\pi_i^{\text{S}_0\text{S}_0},\forall i=1,2
	\end{align}
	Therefore,   $C_i\in [\pi_i^{\text{S}_1\text{S}_0|\text{Y}}-\pi_i^{\text{S}_0\text{S}_0},+\infty)$, for both $i=1,2$.
 Similarly, we can derive the conditions for the $\text{S}_1\text{S}_0$ case and the $\text{S}_1\text{S}_1$ case to be the  equilibrium, respectively.
 
 Second, if there is no pure storage-investment equilibrium, we can always compute the mixed storage-investment equilibrium\cite{gamex}. Supplier $i$ invests in the storage with probability $pr_i^s$ and does not invest in storage with with probability $pr_i^n$, where $pr_i^s+pr_i^n=1$. We construct the following set of linear equations as follows to compute $pr_i^s$ and $pr_i^n$\cite{gamex}.
 \begin{equation}
 \left \{
 \begin{aligned}
 &pr_i^s+pr_i^n=1,\forall i=1,2,\\
 &pr_{-i}^s\cdot(\pi_i^{\text{S}_1\text{S}_1}-C_i)+pr_{-i}^n\cdot(\pi_i^{\text{S}_1\text{S}_0|\text{Y}}-C_i)=pr_{-i}^s\cdot\pi_i^{\text{S}_1\text{S}_0|\text{N}}+pr_{-i}^n \cdot \pi_i^{\text{S}_0\text{S}_0},\forall i=1,2.
 \end{aligned} \label{eq:proof_thmstage1}
 \right.
 \end{equation}
By solving \eqref{eq:proof_thmstage1}, we can obtain  $pr_i^s$ and $pr_i^n$ for both $i=1,2$, which is the mixed storage-investment equilibrium. \qed

\subsection{Proof of Proposition \ref{prop:stocost} }
We prove Proposition \ref{prop:stocost} based on Theorem \ref{thm:stoeq}.

Note that $ \pi_i^{\text{S}_1\text{S}_0|\text{Y}}-\pi_i^{\text{S}_0\text{S}_0}$ is bounded for both $i=1,2$. Thus, there always exists $C_i^{\text{S}_0\text{S}_0}$ such that $C_i^{\text{S}_0\text{S}_0}>\pi_i^{\text{S}_1\text{S}_0|\text{Y}}-\pi_i^\text{$\text{S}_0\text{S}_0$}$ for each $i=1,2$. According to Theorem \ref{thm:stoeq}, the  $\text{S}_0\text{S}_0$ case will be the storage-investment equilibrium, which is also unique. \qed

\subsection{Proof of Proposition \ref{prop:stodemandl} }
We prove Proposition \ref{prop:stodemandl} based on the storage-investment-equilibrium shown in  Theorem \ref{thm:stoeq} and suppliers' equilibrium revenue in the case $\text{S}_0\text{S}_0$ shown in  Proposition \ref{prop:pureprice}. We will show that if the demand $D^{m,t}\leq \min_i \mathbb{ E }[X_i^{m,t}]$, the condition $C_i\in [0, \pi_i^{\text{S}_1\text{S}_1}-\pi_i^{\text{S}_1\text{S}_0|\text{N}}]$, for both $i=1,2$ cannot be satisfied.

  According  to Proposition \ref{prop:pureprice}, in  the  $\text{S}_0\text{S}_0$ case, if the demand $D^{m,t}\leq \min_i \mathbb{ E }[X_i^{m,t}]$, then both suppliers' revenue is zero. Therefore,  if the demand  $0<D^{m,t}\leq \min_i \mathbb{ E }[X_i^{m,t}]$ for any $m$ and $t$, we have
\begin{align}
\pi_i^{\text{S}_1\text{S}_1}=0,\forall i=1,2.
\end{align}
However, according to Proposition \ref{prop:positiverev}, we have that $\pi_i^{\text{S}_1\text{S}_0|\text{N}}>0$ always holds. Therefore, if the demand  $0<D^{m,t}\leq \mathbb{ E }[X_i^{m,t}]$ for any $m$ and $t$, we have
\begin{align}
\pi_i^{\text{S}_1\text{S}_1}-\pi_i^{\text{S}_1\text{S}_0|\text{N}}<0,\forall i=1,2.
\end{align}
Based on the condition of $\text{S}_1\text{S}_1$ being the equilibrium in Theorem \ref{thm:stoeq}, the  $\text{S}_1\text{S}_1$ case cannot be a pure equilibrium if $\pi_i^{\text{S}_1\text{S}_1}-\pi_i^{\text{S}_1\text{S}_0|\text{N}}<0,\forall i=1,2$.\qed

\subsection{Proof of Proposition \ref{prop:stodemandh} }

We will prove Proposition \ref{prop:stodemandh} based on Theorem \ref{thm:stoeq}. The key is to show $\pi_i^{\text{S}_1\text{S}_0|\text{Y}}-\pi_i^{\text{S}_0\text{S}_0}=\pi_i^{\text{S}_1\text{S}_1}-\pi_i^{\text{S}_1\text{S}_0|\text{N}}>0$ for both $i=1,2$.

When $D^{m,t}\geq D^{m,t,th}=\max( \sum_i y_i^{m,t*}(\bar{p},1),$ $ \sum_i y_i^{m,t*}(\bar{p},0))$, there exists the pure price equilibrium $p_1=p_2=\bar{p}$ for each type of subgame in Stage \uppercase\expandafter{\romannumeral2} according to Proposition \ref{prop:pureprice}. Therefore, for both $i=1,2$,
\begin{align}
\pi_i^{\text{S}_0\text{S}_0}=\pi_i^{\text{S}_1\text{S}_0|\text{N}}&=\mathbb{ E }_{m,t}[\pi_{i}^{R,m,t}(\bar{p},y_{i}^*(\bar{p},\varphi_{i}),\bm{\varphi})], ~\text{where}~\sum_i\varphi_i=0\\
&=\mathbb{ E }_{m,t}[\lambda \int_0^{y_i^{m,t*}(\bar{p},0)}xf_i^{m,t}(x)dx]\label{eq:proof_p81}\\
&=\mathbb{ E }_{m,t}[\bar{p}y_i^{m,t*}(\bar{p},0)-\lambda \int_0^{y_i^{m,t*}(\bar{p},0)}F_{i}^{m,t}(x)dx],
\end{align}
which has been shown in \eqref{eq:proof_mcom1}. Furthermore, we also have
\begin{align}
\pi_i^{\text{S}_1\text{S}_1}=\pi_i^{\text{S}_1\text{S}_0|\text{Y}}&=\mathbb{ E }_{m,t}[\pi_{i}^{R,m,t}(\bar{p},y_{i}^*(\bar{p},\varphi_{i}),\bm{\varphi})], ~\text{where}~\sum_i\varphi_i=2\\
&=\mathbb{ E }_{m,t}\left[\bar{p} y_i^{m,t*}(\bar{p},1)\right]\label{eq:proof_p82}\\
&=\mathbb{ E }_{m,t}[\bar{p}\bar{X}_i-\bar{p} \int_0^{\bar{X}_i}F_{i}^{m,t}(x)dx].
\end{align}
Thus, we have
\begin{align}
\pi_i^{\text{S}_1\text{S}_0|\text{Y}}-\pi_i^{\text{S}_0\text{S}_0}=\pi_i^{\text{S}_1\text{S}_1}-\pi_i^{\text{S}_1\text{S}_0|\text{N}}&=\mathbb{E}_{m,t}[\bar{p} y_i^{m,t*}(\bar{p},1)-\lambda \int_{0}^{y_i^{m,t*}(\bar{p},0)}xf_i^{m,t}(x)dx] \label{eq:proof_p83}\\
&\triangleq C_i^{th}.
\end{align}
which is based on \eqref{eq:proof_p81} and \eqref{eq:proof_p82}. Note that $C_i^{th}>0$ always holds as implied in \eqref{eq:proof_rev}.
 
 According to Theorem \ref{thm:stoeq}, if $C_i\leq C_i^{th}$, then supplier $i$ will invest in storage (i.e., $\varphi_{i}^*=1$) while if $C_i> C_i^{th}$, then supplier $i$ will not invest in storage (i.e., $\varphi_{i}^*=0$). \qed

\subsection{Proof of Proposition \ref{prop:stoprofit} }

Suppliers always have strictly positive profit at the storage-investment equilibrium  because the without-storage supplier can always have  positively revenue in the cases of $\text{S}_1\text{S}_0$ and $\text{S}_0\text{S}_0$ according to Proposition \ref{prop:positiverev}. We show it as follows.

\begin{itemize}
	\item If  the $\text{S}_0\text{S}_0$ case is the equilibrium, both suppliers get  strictly positive profit (with zero storage investment cost) according to Proposition \ref{prop:positiverev}.
	\item If  the $\text{S}_1\text{S}_0$ case is the equilibrium, the without-storage  suppliers get  strictly positive profit (with zero storage investment cost) according to Proposition \ref{prop:positiverev}. If the with-storage supplier gets non-positive profit, he can always deviate to not investing in storage,  which leads to the  case $\text{S}_0\text{S}_0$ and    brings him strictly positive profit.
	\item If  the $\text{S}_1\text{S}_1$ case is the equilibrium and one supplier gets non-positive profit, he can always deviate to not investing in storage, which leads to the  case $\text{S}_1\text{S}_0$ and brings him strictly positive profit.	
\end{itemize}

In summary, suppliers always have strictly positive profits at the storage-investment equilibrium.\qed

\vspace{5mm}
\section{Appendix: Proofs of oligopoly model} \label{appendix:proofoligopoly}

\subsection{Proof of Proposition \ref{prop:purepricennn}}
 
 This proof can follow the same procedure in the proof of Proposition \ref{prop:pureprice} by verifying the pure price equilibrium according to the definition of the Nash equilibrium. Towards this end, note that for supplier $i$ with or without storage, the revenue function $\pi_i^R\left(p_i, x_i^*(\bm{p},\bm{y}),\bm{\varphi}\right)$ is strictly increasing   with respect to both the price $p_i$  and the selling quantity $x_i^*(\bm{p},\bm{y})$ that is in the range $[0, y_i^*(p_i,\varphi_i)]$ (without considering the other supplier's coupled decisions).  We will discuss the three cases, respectively.
\subsubsection{ The case of $D \geq \sum_{i\in \mathcal{I}} y_i^*(\bar{p},\varphi_i)$} We can prove that when
$D \geq \sum_i y_i^*(\bar{p},{\varphi}_i)$,  $p_i=\bar{p}$ is a pure price equilibrium . Also,  this pure price equilibrium is unique. This proof can follow the same procedure in the Section \ref{appendix:proofstage2}.B.1.a. of the  proof of Proposition \ref{prop:pureprice}. The intuition is that when the demand is larger than the maximum bidding quantity, if any supplier deviates to a lower price, his selling quantity cannot be increased, which leads to a lower revenue.

\subsubsection{ The case of   $D\leq \sum_{i\in\mathcal{U}} y_i^*(\bar{p},\varphi_i)-y_j^*(\bar{p},\varphi_i)$ for any  $j\in\mathcal{ U}$} We first prove  by the  definition of the Nash equilibrium  that when  $D\leq \sum_{i\in\mathcal{U}} y_i^*(\bar{p},\varphi_i)-y_j^*(\bar{p},\varphi_j)$ for any  $j\in\mathcal{ U}$, there exists a pure  price equilibrium $p_i^*=0$  with an equilibrium revenue $\pi_i^{RE}=0$, for any $i\in \mathcal{ I}$. Then, note that this equilibrium is not unique, but we show that suppliers always get zero revenue at any equilibrium.

First,  we prove the  pure  price equilibrium $p_i^*=0$. We assume that $p_i^*=0,\forall i \in \mathcal{ I}$.  We will discuss  two cases of with-storage supplier and without-storage supplier, respectively.

(a) For a supplier $j\in \mathcal{U}$ who invests in storage, if he deviates to a higher price $p_j'>0$, the demand that he gets is the following.
\begin{align}
\min\left(D-\min(D,\sum_{i\in\mathcal{I}\backslash j} y_i^*(0,\varphi_i)),y_j^*(p_j',\varphi_j)\right ),~j\in \mathcal{ U}.\label{eq:proofget}
\end{align}
Note that according to  Theorem \ref{thm:quantity}, we have  $y_k^*(0,\varphi_k)=0,\forall k\in \mathcal{ V}$. Also, we have $y_k^*(p_k,\varphi_k)=\mathbb{ E }[X_k],\forall k\in \mathcal{ U}$. Therefore, 
\begin{align}
\eqref{eq:proofget}=
\min\left(D-\min(D,\sum_{i\in\mathcal{U}} y_i^*(\bar{p},\varphi_i)-y_j(\bar{p},\varphi_j)),y_j^*(p_j',\varphi_j)\right ),~j\in \mathcal{ U},
\end{align}
which is  zero since $D\leq  \sum_{i\in\mathcal{U}} y_i^*(\bar{p},\varphi_i)-y_j^*(\bar{p},\varphi_j),\forall j\in\mathcal{ U}$. Therefore, if  this supplier deviates to a higher price, his revenue will be still zero.

(b) For a supplier $j\in \mathcal{ V}$ who does not invest in storage, if he deviates to a higher price $p_j'>0$, the demand that he gets is 
\begin{align}
&\min\left(D-\min(D,\sum_{i\in\mathcal{I}\backslash j} y_i^*(0,\varphi_i)),y_j^*(p_j',\varphi_j)\right ), ~j\in \mathcal{ V},\\
=&\min\left(D-\min(D,\sum_{i\in\mathcal{U}} y_i^*(\bar{p},\varphi_i)),y_j^*(p_j',\varphi_j)\right ), ~j\in \mathcal{ V}
\end{align}
which is still  zero since $D\leq  \sum_{i\in\mathcal{U}} y_i^*(\bar{p},\varphi_i)$. Therefore, if  this supplier deviates to a higher price, his revenue will be still zero. In conclusion,   the bidding price $p_i^*=0,\forall i \in \mathcal{ I}$ is an equilibrium where no supplier will deviate.

Second, note that the equilibrium here is not unique, however, each supplier always gets zero revenue at any equilibrium. We show this by contradiction  as follows.  If supplier $k$ gets positive revenue, it means that his bidding price  and his obtained demand are both positive. We assume that a set of suppliers $\mathcal{P}$ bid the price $p>0$ the same as this supplier $k$. We denote the set of suppliers whose prices are lower than $p$ as $\mathcal{PL}$  and the set of suppliers whose prices are higher than $p$ as $\mathcal{PH}$.\footnote{Note that $\mathcal{PL}$ and $\mathcal{PH}$  can be both empty sets}  Since this supplier gets positive demand, it means 
 \begin{align}
\sum_{i\in \mathcal{P}} y_i^*({p_i,\varphi_i}) \leq D-\sum_{i\in \mathcal{PL}} y_i^*({p_i,\varphi_i}), \label{eq:poofol}
 \end{align}
 or
  \begin{align}
0<\ D-\sum_{i\in \mathcal{PL}} y_i^*({p_i,\varphi_i})<\sum_{i\in \mathcal{P}} y_i^*({p_i,\varphi_i}).  \label{eq:poofoh}
 \end{align}
 \begin{itemize}
 	\item  Case \eqref{eq:poofoh} and $|\mathcal{P}|\geq 2$: At least one of suppliers in $\mathcal{P}$ can decrease his price by a sufficiently positive value, which can increase his obtained demand and increase his revenue. This shows that this case cannot be one equilibrium.
 	\item  Case \eqref{eq:poofoh}; $|\mathcal{P}|=1$ and $p<\bar{p}$: This supplier can increase his price by a small positive value (which makes the bidding price smaller than the lowest bidding price in set $\mathcal{ PH}\bigcup \bar{p}$), which will not decrease his obtained demand. Thus, this deviation increases his revenue and this case cannot be one equilibrium.
 	\item  Case \eqref{eq:poofoh}; $|\mathcal{P}|=1$ and $p=\bar{p}$: Due to \eqref{eq:poofoh}, we have  $\sum_{i\in \mathcal{PL}} y_i^*({p_i,\varphi_i})< D$. Note that the set $\mathcal{PL}$ contains all the suppliers except the single supplier $k$. Thus, there alyways exists $j\in \mathcal{ U }$ such that $\sum_{i\in\mathcal{U}} y_i^*(\bar{p},\varphi_i)-y_j^*(\bar{p},\varphi_j)<\sum_{i\in \mathcal{PL}} y_i^*({p_i,\varphi_i})< D$, which  contradicts the condition $D\leq  \sum_{i\in\mathcal{U}} y_i^*(\bar{p},\varphi_i)-y_j^*(\bar{p},\varphi_j),\forall j\in\mathcal{ U}$.  This case is impossible. 
 	\item Case  \eqref{eq:poofol} and $p<\bar{p}$: any supplier in $\mathcal{P}$ can always increase his price by a small positive value (which makes the bidding price smaller than price cap $\bar{p}$) without  decreasing his obtained demand, which increases his revenue.  This shows that this case cannot be one equilibrium.
 	\item Case  \eqref{eq:poofol} and $p=\bar{p}$:  Due to  \eqref{eq:poofol}, we have  $\sum_{i\in \mathcal{ I}} y_i^*({p_i,\varphi_i}) \leq D$, which contradicts the condition $D\leq  \sum_{i\in\mathcal{U}} y_i^*(\bar{p},\varphi_i)-y_j^*(\bar{p},\varphi_j),\forall j\in\mathcal{ U}$. Thus, this case is impossible. 
 \end{itemize}
Therefore, we can draw the conclusion that at any equilibrium, suppliers  get zero revenue.

 \subsubsection{The case that there exists $j\in\mathcal{ U}$ such that  $\sum_{i\in\mathcal{U}} y_i^*(\bar{p},\varphi_i)-y_j^*(\bar{p},\varphi_j)<D < \sum_{i\in \mathcal{I}} y_i^*(\bar{p},\varphi_i)$} In this case, there is no pure price equilibrium. This proof can follow the similar procedure in the Section \ref{appendix:proofstage2}.B.1.b of the  proof of Proposition \ref{prop:pureprice}.	We can discuss three cases: (i) all the suppliers bid zero prices;  (ii) suppliers' bidding prices are all equal and positive.  (iii)  suppliers' bidding prices are not equal for all the suppliers. We show that all theses cases cannot be the pure price  equilibrium.
 
First,  for case (i) , at least one supplier $j$ (i.e., the $j$ satisfying $\sum_{i\in\mathcal{U}} y_i^*(\bar{p},\varphi_i)-y_j^*(\bar{p},\varphi_j)<D$) who invests in storage can increase his price, and  he will  get positive demand. This increases his revenue and shows that case (i) cannot be an equilibrium.

Second, for case (ii), we can discuss two conditions $\sum_{i\in \mathcal{ I}}y_i^*(p_i,\varphi_i) \leq D$ and  $\sum_{i\in \mathcal{ I}}y_i^*(p_i,\varphi_i) > D$, which is the same Section \ref{appendix:proofstage2}.B.1.b. For $\sum_{i\in \mathcal{ I}}y_i^*(p_i,\varphi_i) \leq D$, any supplier can always increase  his price without  decreasing his obtained demand, which increases his revenue. For  $\sum_{i\in \mathcal{ I}}y_i^*(p_i,\varphi_i) > D$, at least one supplier can always reduce his price by a sufficiently small positive value, which can increase his demand and increase his revenue. Thus, case (ii)  can not be an equilibrium.

Third, for case (iii), we denote the set of suppliers with the lowest bidding prices $p$ among all the suppliers as $\mathcal{ L }$.
Similarly, we discuss two conditions $\sum_{i\in \mathcal{ L}}y_i^*(p_i,\varphi_i) \leq D$ and  $\sum_{i\in \mathcal{ L}}y_i^*(p_i,\varphi_i) > D$. For $\sum_{i\in \mathcal{ L}}y_i^*(p_i,\varphi_i) \leq D$, any supplier can always increase  his price by a small positive value (which makes the bidding price smaller than the second lowest price) without  decreasing his obtained demand, which increases his revenue. Thus, this case cannot be an equilibrium. For  $\sum_{i\in \mathcal{ L}}y_i^*(p_i,\varphi_i) > D$, there are three possibilities.
\begin{itemize}
	\item The lowest price $p>0$ and $|\mathcal{ L }\mid=1$:  This supplier can increase his price by a small positive value (which makes the bidding price smaller than the second lowest bidding price), which will not decrease his obtained demand. Thus, it  increases his revenue and this case cannot be one equilibrium.
	\item The lowest price $p>0$ and $|\mathcal{ L }\mid \geq 2$: 	At least one of suppliers in $\mathcal{L}$ can decrease his price by a sufficiently small positive value, which can increase his obtained demand and increase his revenue. This shows that this case cannot be one equilibrium.
	\item The lowest price $p=0$: In this case, all the suppliers have zero revenue, and $\sum_{i\in \mathcal{ L}}y_i^*(0,\varphi_i) > D$. Note that demand $D$ also satisfies   $\sum_{i\in\mathcal{U}} y_i^*(\bar{p},\varphi_i)-\max_j y_j^*(\bar{p},\varphi_i)<D$,  $j\in\mathcal{ U }$. We denote $\arg \max_{j \in \mathcal{ U }} y_j^*(\bar{p},\varphi_j)=j^*$. Thus, there are two possibilities that lead to  make this: 
	\begin{itemize}
		\item $j^*\in \mathcal{ L }$: The supplier $j^*$ can increase his zero price to a positive price (which is smaller than the second lowest price) and get positive demand since $\sum_{i\in\mathcal{L}\setminus j^*} y_i^*(0,\varphi_i)\leq \sum_{i\in\mathcal{U}} y_i^*(\bar{p},\varphi_i)- y_{j^*}^*(\bar{p},\varphi_i)<D$. This increases supplier $j^*$'s revenue.
		\item  $j^*\notin \mathcal{ L }$: Any supplier $k$  in $\mathcal{ L }$ can increase his zero price to a positive price (which is smaller than the second lowest price) and get positive demand since  $\sum_{i\in\mathcal{L}\setminus k} y_i^*(0,\varphi_i)<\sum_{i\in\mathcal{U}} y_i^*(\bar{p},\varphi_i)- y_{j^*}^*(\bar{p},\varphi_i)<D$. This increases supplier $k$'s revenue.
	\end{itemize}
	Therefore, the case  that  the lowest price $p=0$ cannot be one equilibrium. Combining the case  $p=0$ and $p>0$,  the condition $\sum_{i\in \mathcal{ L}}y_i^*(p_i,\varphi_i) > D$ is not an equilibrium.

\end{itemize}

 Combining cases (i)-(iii), we show that all theses cases cannot the equilibrium. Thus, there is no pure price equilibrium  if there exists $j\in\mathcal{ U}$ such that  $\sum_{i\in\mathcal{U}} y_i^*(\bar{p},\varphi_i)-y_j^*(\bar{p},\varphi_j)<D < \sum_{i\in \mathcal{I}} y_i^*(\bar{p},\varphi_i)$}. Finally, we have Proposition \ref{prop:purepricennn} proved. \qed

\subsection{Proof of Proposition \ref{lem:mixnnn}}
We first show the existence of mixed price equilibrium and then prove the positive revenues for all the suppliers in the mixed price  equilibrium.

\subsubsection{Existence of mixed price equilibrium} This result can be derived from Theorem 5 \cite{dasgupta1986existence}.

\subsubsection{Positive revenue}
Note that the case that there exists $j\in\mathcal{ U}$ such that  $\sum_{i\in\mathcal{U}} y_i^*(\bar{p},\varphi_i)-y_j^*(\bar{p},\varphi_j)<D < \sum_{i\in \mathcal{I}} y_i^*(\bar{p},\varphi_i)$ is equivalent to the case  $\sum_{i\in\mathcal{U}} y_i^*(\bar{p},\varphi_i)-\max_{j \in \mathcal{ U }}y_j^*(\bar{p},\varphi_j)<D < \sum_{i\in \mathcal{I}} y_i^*(\bar{p},\varphi_i)$. We will first prove by contradiction that for  supplier $n$ with  $n=\arg \max_{i\in \mathcal{ U }}  y_i^*(\bar{p},\varphi_i)$, his equilibrium revenue is positive. Then, we prove that other suppliers except supplier $n$ also have the positive revenues. We denote the support of supplier $i$'s mixed price strategy  as $\mathcal{SP}_i$.

First, we will prove that for supplier $n$, his revenue equilibrium  $\pi_n^{RE}>0$. We prove this by contradiction. We assume that supplier $n$'s equilibrium revenue $\pi_n^{RE}=0$, and  discuss two cases.
\begin{itemize}
	\item For each supplier $j\neq n$, the support  $\mathcal{SP}_j$ only contains 0, which means each supplier $j\neq i$ has the pure price strategy $p_j=0$: Then, for supplier $n$, he can always set a pure price $p_n>0$ to achieve  positive demand and get positive revenue since  $\sum_{i\in\mathcal{U}} y_i^*(\bar{p},\varphi_i)-y_n^*(\bar{p},\varphi_n)<D $, which contradicts the assumption that $\pi_n^{RE}=0$.
	\item In all the suppliers except $n$, there exists at least one supplier $k$ such that  $\mathcal{SP}_k$ contains  positive price $p_k>0$:  For all the suppliers whose  supports contain  positive prices (except $n$), we denote the set of those suppliers as $\mathcal{PS}$. For any supplier $k\in\mathcal{PS}$,  we choose one positive price $p_k \in \mathcal{SP}_k$. Thus, supplier $n$ can always choose a pure price strategy $0<p_n<\min_{k\in\mathcal{PS}} p_k$, such that he can get positive demand and positive revenue with a positive probability.  This contradicts the assumption that $\pi_n^{RE}=0$.
\end{itemize}
 Thus, we can have the conclusion that at the equilibrium,  supplier $n$'s revenue $\pi_n^{RE}> 0$. This also implies that for supplier $n$, his support $\mathcal{SP}_n$ does not contain zero.
 
 Second, we will prove that for any supplier $j \neq n$, his equilibrium revenue is positive. We assume that supplier $j$'s equilibrium revenue  $\pi_j^{RE}=0$.  Note that among the suppliers except $j$, there exists at least one supplier $n$ such that  $\mathcal{SP}_n$ contains  positive price $p_n>0$.  For all the suppliers (except $j$) whose  supports contain  positive prices, we denote the set of those suppliers as $\mathcal{PS}'$. For any supplier $k\in\mathcal{PS}'$,  we choose one positive price $p_k \in \mathcal{SP}_k$. Thus, supplier $j$ can always choose a pure price strategy $0<p_j<\min_{k\in\mathcal{PS}'} p_k$, such that he can get positive demand and positive revenue with a positive probability.  Therefore,  at the equilibrium, supplier $j$'s revenue cannot be zero.
 
 Therefore, based on above discussions, we have that all the suppliers have the positive revenues in the case of   $\sum_{i\in\mathcal{U}} y_i^*(\bar{p},\varphi_i)-\max_{j\in\mathcal{ U}}y_j^*(\bar{p},\varphi_j)<D < \sum_{i\in \mathcal{I}} y_i^*(\bar{p},\varphi_i)$. \qed

\subsection{Proof of Proposition \ref{prop:stocostnnn}}
The proof follows  the definition of Nash equilibrium.

It is straightforward that the benefit brought by investing storage for supplier $i$ is bounded. At the case $\mathcal{S}^{\mathcal{ U}|\mathcal{ V}}$, we denote the equilibrium profit of supplier $i$ as $\Pi_i^*(\mathcal{S}^{\mathcal{ U}|\mathcal{ V}})$ and the expected  equilibrium  revenue (scaled in one hour) over the investment horizon as  $\pi_i^{REE}(\mathcal{S}^{\mathcal{ U}|\mathcal{ V}})$. For any case $\mathcal{S}^{\mathcal{ U}|\mathcal{ V}}$,   one without-storage supplier $i$ has the profit $\Pi_i^*(\mathcal{S}^{\mathcal{ U}|\mathcal{ V}})=\pi_i^{REE}(\mathcal{S}^{\mathcal{ U}|\mathcal{ V}}),i\in\mathcal{ V}$ at the equilibrium.
 However, if he deviates to investing in storage, he has the profit $\Pi_i^*(\mathcal{S}^{\mathcal{ U}\bigcup i|\mathcal{ V}\setminus i})=\pi_i^{REE}(\mathcal{S}^{\mathcal{ U }\bigcup i|\mathcal{ V}\setminus i})-C_i$. 
 Thus, for $i\in \mathcal{V}$, we have 
 \begin{align}
&\Pi_i^*(\mathcal{S}^{\mathcal{ U}\bigcup i|\mathcal{ V}\setminus i})-\Pi_i^*(\mathcal{S}^{\mathcal{ U}|\mathcal{ V}})\\=&\pi_i^{REE}(\mathcal{S}^{\mathcal{ U }\bigcup i|\mathcal{ V}\setminus i})-\pi_i^{REE}(\mathcal{S}^{\mathcal{ U}|\mathcal{ V}})-C_i.
 \end{align}
 
  Note that $\pi_i^{REE}(\mathcal{S}^{\mathcal{ U }\bigcup i|\mathcal{ V}\setminus i})-\pi_i^{REE}(\mathcal{S}^{\mathcal{ U}|\mathcal{ V}})$ is bounded for any $\mathcal{S}^{\mathcal{ U}|\mathcal{ V}}$. If the storage cost $C_i>C_i^{no}$, where  $C_i^{no}$ is the maximum value of   $\pi_i^{REE}(\mathcal{S}^{\mathcal{ U }\bigcup i|\mathcal{ V}\setminus i})-\pi_i^{REE}(\mathcal{S}^{\mathcal{ U}|\mathcal{ V}})$  over all the cases $\mathcal{S}^{\mathcal{ U}|\mathcal{ V}}$, then this supplier $i\in \mathcal{V}$ will not deviate to investing  in storage in any case of $\mathcal{S}^{\mathcal{ U}|\mathcal{ V}}$. Thus, no supplier investing in storage is the unique equilibrium.
  
 \qed

\subsection{Proof of Proposition \ref{prop:stodemandlnnn}}
The proof follows the definition of Nash equilibrium.

Note that in the subgame $S^{\mathcal{ U}|\mathcal{ V}}$, when $0<D^{m,t}\leq \min_{j\in \mathcal{ U}} (\sum_{i\in\mathcal{U}} y_i^*(\bar{p},\varphi_i)-y_j^*(\bar{p},\varphi_j))$  for any $t$ and $m$, each supplier has zero revenue for any $t$ and $m$  as shown in Proposition  \ref{prop:purepricennn}. Thus, for each supplier $i\in \mathcal{ I}$, his expected equilibrium revenue $\pi_i^{REE}(S^{\mathcal{ U}|\mathcal{ V}})=0$.  Then, for supplier $j\in\mathcal{ U}$ who invests in storage, his profit is $\pi_i^{REE}(S^{\mathcal{ U}|\mathcal{ V}})-C_i<0$ since $C_i>0$.  Therefore, this supplier $i$  can always deviate to not investing storage which leads to a nonnegative profit. This shows that when $0<D^{m,t}\leq \min_{j\in \mathcal{ U}} (\sum_{i\in\mathcal{U}} y_i^*(\bar{p},\varphi_i)-y_j^*(\bar{p},\varphi_j))$, the  case  $S^{\mathcal{ U}|\mathcal{ V}}$ (i.e., suppliers of set $\mathcal{ U}$  investing in storage and suppliers of set $\mathcal{V}$  not investing in storage) cannot be a pure storage-investment equilibrium.
\qed

\subsection{Proof of Proposition \ref{prop:stodemandhnnn}}
The intuition of this proposition is that when the demand $D$ is sufficiently large, there is not competition between suppliers and they make decisions of storage investment independently.

As implied in Proposition  \ref{prop:purepricennn}, when demand $D^{m,t}\geq\sum_{i\in \mathcal{I}}y_i^{m,t*}(\bar{p},\varphi_i)$ in subgame $\mathcal{S}^{\mathcal{ U }|\mathcal{ V}}$, each supplier $i$ can bid the price cap $\bar{p}$ to get his biding quantity  $y_i^{m,t*}(\bar{p},\varphi_i)$. For convenience, at hour $t$ of month $m$,  we denote the  bidding quantity of supplier $i$ at price cap $\bar{p}$ in subgame  $\mathcal{S}^{\mathcal{ U }|\mathcal{ V}}$ as $y_i^{m,t*}(\bar{p},\varphi_i| \mathcal{S}^{\mathcal{ U }|\mathcal{ V}})$. We also denote the set of all the subgames as $\mathcal{ S}^{\Omega}$. Thus, if the demand  $D^{m,t}\geq\max_{\mathcal{S}^{\mathcal{ U }|\mathcal{ V}} \in \mathcal{ S}^{\Omega}}\sum_{i\in \mathcal{I}}y_i^{m,t*}(\bar{p},\varphi_i| \mathcal{S}^{\mathcal{ U }|\mathcal{ V}})\triangleq D^{m,t,th'}$ for any $t$ and $m$,  then  each supplier $i$ can bid the price cap $\bar{p}$ to get his biding quantity  $y_i^{m,t*}(\bar{p},\varphi_i)$ in any subgame for any $t$ and $m$. This  leads to the revenue $\pi_{i}^{R,m,t}(\bar{p},y_{i}^{m,t*}(\bar{p},\varphi_{i}),\bm{\varphi})$ that can be directly calculated based on supplier $i$'s parameter. In this case, we have the following.

\begin{itemize}
	\item If supplier invests in storage, i.e., $\varphi_{i}=1$, his equilibrium revenue is 
	\begin{align}
\mathbb{ E }_{m,t}[\pi_{i}^{R,m,t}(\bar{p},y_{i}^*(\bar{p},1),\bm{\varphi})]=
	&=\mathbb{ E }_{m,t}\left[\bar{p} y_i^{m,t*}(\bar{p},1)\right], \label{eq:nearf1}
	\end{align}
	which has been shown in \eqref{eq:proof_p82}.
	\item  If supplier does not invest in  storage i.e., $\varphi_{i}=0$, his equilibrium revenue is 	
	\begin{align}
\mathbb{ E }_{m,t}[\pi_{i}^{R,m,t}(\bar{p},y_{i}^*(\bar{p},0),\bm{\varphi})]=\mathbb{ E }_{m,t}[\lambda \int_0^{y_i^{m,t*}(\bar{p},1)}xf_i^{m,t}(x)dx],  \label{eq:nearf2}
	\end{align}
	which has been shown in \eqref{eq:proof_p81}.

\end{itemize}

We compared \eqref{eq:nearf1} and \eqref{eq:nearf2}, and we characterize $C_i^{th'}$ the same as  \eqref{eq:proof_p83} as follows. 
\begin{align}
\mathbb{E}_{m,t}[\bar{p} y_i^{m,t*}(\bar{p},1)-\lambda \int_{0}^{y_i^{m,t*}(\bar{p},0)}xf_i^{m,t}(x)dx] 
\triangleq C_i^{th'}.
\end{align}
\qed

\subsection{Proof of Proposition \ref{prop:stoprofitnnn}}
We prove this by contradiction and discuss a total of three cases.

\begin{itemize}
	\item If one supplier does not invest in storage and gets zero profit (note that a without-storage supplier always has nonnegative profits),  it only means the demand lies in the condition  $D\leq \sum_{i\in\mathcal{U}} y_i^*(\bar{p},\varphi_i)-y_j^*(\bar{p},\varphi_j),\forall j\in \mathcal{ U }$ as shown in Proposition  \ref{prop:purepricennn} and Proposition \ref{lem:mixnnn}, where all the suppliers get zero revenues in the local energy market. This state is not stable because  the with-storage supplier  gets negative profit and  he can always choose not to invest in storage, which increases his profit.
	\item If one supplier invests in storage and gets negative profit,  he can always choose not to invest in storage, which increases his profit. Thus, this case cannot be an equilibrium.
	\item If one supplier invests in storage and gets zero profit,  it  means the demand cannot lie in the condition  $D\leq \sum_{i\in\mathcal{U}} y_i^*(\bar{p},\varphi_i)-y_j^*(\bar{p},\varphi_j),\forall j\in \mathcal{ U }$ (otherwise, this supplier will get negative profit) as shown in Proposition  \ref{prop:purepricennn} and Proposition \ref{lem:mixnnn}. This state is not stable since this supplier can further choose not to invest ins storage, where the demand still cannot satisfy $D\leq \sum_{i\in\mathcal{U}} y_i^*(\bar{p},\varphi_i)-y_j^*(\bar{p},\varphi_j),\forall j \in \mathcal{ U }$. This leads to a positive revenue, i.e., the positive profit for this supplier.

\end{itemize}

In summary, any supplier always has strictly positive profits at the storage-investment equilibrium.\qed

\bibliographystyle{IEEEtran}
\bibliography{storage}
%
%
%

\end{document}